\documentclass[11pt]{article}
\pdfoutput=1
\usepackage{hyperref}
\usepackage{amsmath,amstext}
\usepackage{makeidx}         
\usepackage{graphicx}        
\usepackage{multicol}        
\usepackage[bottom]{footmisc}

\usepackage{color}					 
\usepackage{soul} 					 

\renewcommand{\vec}[1]{\ensuremath{\text{\boldmath $#1$\unboldmath}}}

\newcommand{\enm}{\ensuremath}
\newcommand{\C}{{\enm{ \mathcal C }}}

\newcommand{\Lag}{{\enm{ \mathcal L }}}
\newcommand{\Bez}{{\enm{ \mathcal B }}}
\newcommand{\mvx}{\vec{x}}
\newcommand{\mvX}{\vec{X}}

\newcommand{\mvn}    {\ensuremath{{\bf{n}}}}
\newcommand{\mvt}    {\ensuremath{{\bf{t}}}}
\newcommand{\mvp}    {\ensuremath{{\bf{p}}}}
\newcommand{\mvq}    {\ensuremath{{\bf{q}}}}
\newcommand{\mvc}    {\ensuremath{{\bf{k}}}}

\renewcommand{\o}{\ensuremath{p}}
\newcommand{\phik}[1]{\ensuremath{{\mathcal L}^{(#1)}}}
\newcommand{\bez}[2]{\ensuremath{\mathcal{B}_{#1}^{(#2)}}}
\newcommand{\ap}[1]{(#1)}
\newcommand{\cro}[1]{\left[#1\right]}
\newcommand{\beqn}[1]{\begin{equation}\label{#1}}
\newcommand{\eeqn}{\end{equation}}
\def\k{\ensuremath{k}}			

\newcommand{\Comb}[2]{{{#2}\choose {#1}}}
\def\a{\ensuremath{\xi}} 		
\newcommand{\matB}[1]{{\enm{ \vec T_{\Bez\to\Lag}^{\,(#1)} }}}
\newcommand{\RA}{{\enm{ \vec\xi }}}

\graphicspath{{./}{../figures/}}

\makeindex             

\begin{document}

\title{Optimizing the geometrical accuracy of curvilinear meshes}
\author{Thomas Toulorge$^{a,b}$, Jonathan Lambrechts$^{a,b}$ and Jean-Fran{\c c}ois Remacle$^a$\\
\small $^a$ Universit\'e catholique de Louvain, Institute of Mechanics, Materials and Civil\\
\small Engineering (iMMC), B\^atiment Euler, Avenue Georges Lema\^itre 4,\\
\small 1348 Louvain-la-Neuve, Belgium\\
\small $^b$ Fonds National de la Recherche Scientifique, rue d'Egmond 5, 1000 Bruxelles, Belgium
}

\maketitle

\begin{abstract}
This paper presents a method to generate valid high order meshes 
with optimized geometrical accuracy. The high order meshing procedure
starts with a linear mesh, that is subsequently curved without taking
care of the validity of the high order elements. An optimization procedure
is then used to both untangle invalid elements and optimize the
geometrical accuracy of the mesh. Standard measures of the distance
between curves are considered to evaluate the geometrical accuracy in
planar two-dimensional meshes, but they prove computationally too costly
for optimization purposes. A fast estimate of the geometrical accuracy,
based on Taylor expansions of the curves, is introduced. An unconstrained
optimization procedure based on this estimate is shown to yield significant
improvements in the geometrical accuracy of high order meshes, as measured
by the standard Haudorff distance between the geometrical model and the mesh.
Several examples illustrate the beneficial impact of this method on CFD
solutions, with a particular role of the enhanced mesh boundary smoothness.
\end{abstract}

\section{Introduction}
The development of high-order numerical technologies for engineering
analysis has been underway for many years now. For example, Discontinuous
Galerkin methods (DGM) have been thoroughly studied in the literature,
initially in a theoretical context \cite{dg:book}, and now from the
application point of view \cite{adigma, idihom}. Compared to standard
second-order-accurate numerical schemes, high-order methods exhibit
superior efficiency in problems with high resolution requirements,
because they reach the required accuracy with much coarser grids.

However, many contributions have pointed out that the accuracy of these methods
can be severely hampered by a too crude discretization of the
geometry~\cite{bassi, bernard, toulorge}. It is now widely accepted
that linear geometrical discretizations may annihilate the benefits of
high-order schemes in cases featuring curved geometries, that is, in
most cases of engineering and scientific interest.

This problem has motivated the development of methods for the generation
of high-order meshes, in which curvilinear elements are meant to provide
sufficient geometrical accuracy on the boundary. Elements are then most
often defined in a Lagrangian manner by a set of high-order nodes.
Until now, efforts have mostly been targeted at ensuring the validity of
the mesh. Indeed, the
naive approach consisting in simply curving the boundaries of a linear
mesh to match the geometry often results in tangled elements
\cite{jcp2013}. The curvature of the boundary must somehow be
``propagated'' into the domain for all elements to be valid. In the case
of locally structured meshes, such a situation can be avoided by means of
an efficient isoparametric technique~\cite{moxey2015}. For unstructured
meshes, untangling procedures based on topological
operations~\cite{dey1999, luo2004automatic, sahni2010},
mechanical analogies~\cite{xie, abgrall2012, perssonperaire} or
optimization procedures~\cite{gargallo2013, jcp2013} have been proposed.

Although the improved representation of the geometry of the domain is the
prime motivation for the use of high-order meshes, only few authors
have taken into consideration the quality of the geometrical
approximation. In the literature, the limited work dedicated to this topic
has focused on placing adequately the high-order nodes when curving the
mesh boundaries. Simple techniques include interpolating them between the
first-order boundary nodes in the parametric space describing the
corresponding CAD entity~\cite{dey1999,gargallo2013} or projecting them
on the geometry from their location on the straight-sided element.
More sophisticated procedures have also been proposed. In
Ref.~\cite{xie}, the high-order nodes on boundary edges
are interpolated in the physical space through a numerical
procedure involving either the CAD parametrization (in the
case of a mesh edge assigned to an edge of the geometric
model), or an approximation of the geodesic connecting the
two first-order vertices (in the case of an edge located
on a 3D surface). Nodes located within surface elements are
obtained through a more sophisticated version of this
procedure. Instead of interpolating, Sherwin and
Peir\'o~\cite{sherwin2002} use a mechanical analogy with
chains of springs in equilibrium that yields the adequate
node distribution along geometric curves and geodesics for
edge nodes. Two-dimensional nets of springs provide the
appropriate distribution of surface element nodes.

This paper presents a method that makes it possible to build
geometrically accurate curvilinear meshes. Unlike previous work
reported in the literature, the representation of the model by
the mesh is formally assessed by measuring \emph{distances} between
the geometric model and the corresponding high-order mesh boundary,
or by evaluating a fast estimate of the geometrical error. The aim of
the method is to minimize this geometrical error through the use of
standard optimization algorithms. Although most of the paper deals
with two-dimensional meshes, it is shown that the approach can
easily be extended to three spatial dimensions.

Consider a model entity $\C$ and the mesh entity $\C_m$ that is meant
to approximate $\C$. The first questions that arise are how to define a
proper distance $d(\C,\C_m)$ between $\C$ and $\C_m$, and how to compute
this distance efficiently. Two main definitions for such a distance have
been proposed in the computational geometry literature, namely the
Fr\'echet distance and the Hausdorff distance. 

In this context  of curvilinear meshing, distances $d(\C,\C_m)$ that
are computed are usually small in comparison with the typical
dimension of either $\C$ or $\C_m$. Consequently, $d(\C,\C_m)$ has to
be computed with high accuracy.  In this paper, we show that computing
standard distances between the mesh and the geometry may be too expensive
for practical computations. Alternative measures of distance are
presented, that are both fast enough to compute and sufficiently accurate.
An optimization procedure is then developed to drastically reduce the
model-to-mesh distance while enforcing the mesh validity.

The paper is organized as follows. In Section~\ref{sec:distance},
the problem of defining and computing a proper model-to-mesh
distance is examined. The mesh optimization procedure is described
in Section~\ref{sec:opti}. Section~\ref{sec:examples} illustrates
the method with examples, and the extension of the approach to three
dimensions is presented in Section~\ref{sec:3d}. Conclusions are drawn
in Section~\ref{sec:conclusions}.

\section{Model-to-Mesh Distance}\label{sec:distance}

\subsection{Distance Between Curves}

\subsubsection{Setup}

Consider the following planar parametric curve
$$\C \equiv \{\eta\in[\eta_0,\eta_p] \mapsto \mvx(\eta) \in R^2\}$$ 
and the following $p+1$ successive points on $\C$ 
$$\mvx_i = \mvx(\eta_i),~~ \mbox{with}~~\eta_0 <\eta_1<\eta_2 \dots < \eta_{p-1}<\eta_p.$$
A curvilinear mesh edge $\C_m$ is defined as the Lagrange approximation of
$\C$ at order $p$
\begin{equation}\label{eq:lag1d}
\C_m \equiv \left\{\xi \in [0,1] \mapsto \mvx_m(\xi)= \sum_{i=0}^{p}
\phik{\o}_i(\xi)\;\in R^2\right\}.
\end{equation}
In \eqref{eq:lag1d}, $\phik{\o}_i(\xi)$ is the $i$th Lagrange
polynomial of order $p$.

\begin{figure}[h!]
 \begin{center}
 \includegraphics[width=0.5\textwidth]{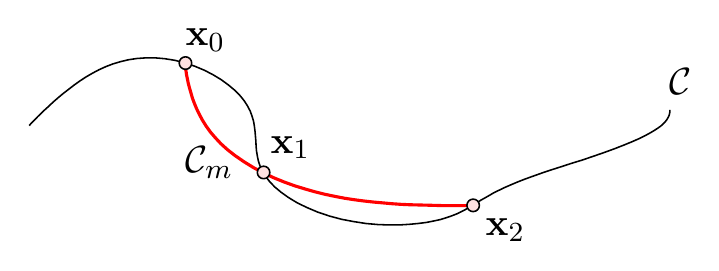}
\end{center}
\caption{Typical setup: a model edge $\mathcal C$ and a quadratic mesh
  edge ${\mathcal C}_m$.\label{fig:geom1db}}
\end{figure}

Curves $\C$ and $\C_m$ that are both bounded
by the vertices $\mvx_0$ and $\mvx_p$ and coincide at least at the
$p+1$ Lagrange points $\mvx_i,~i=0,\dots,p$ (see Figure \ref{fig:geom1db}). 

\subsubsection{Formal Definitions of Distance}

Define $\alpha(t)$ (resp. $\beta(t)$) as an arbitrary continuous
nondecreasing function from $t \in [0,1]$ 
onto $\eta \in [\eta_0,\eta_p]$ (resp. $ \xi \in [0,1]$). The {\bf Fr\'echet distance}
between $\C$ and $\C_m$ is defined as 
$$d_{\mathrm{F}} (\C,\C_m) = \inf_{\alpha, \beta}\,\,\max_{t \in [0,1]} \|\mvx_m(\beta(t))-\mvx(\alpha(t))\|.$$
There is a standard interpretation of the Fr\'echet distance. Consider
a man is walking with a dog on a leash. The man is walking on the one
curve and the dog on the other one. Both may vary their speed, as
$\alpha$ and $\beta$ are arbitrary. Backtracking  is not allowed which
implies that $\alpha$ and $\beta$ are non-decreasing.
Then, the Fr\'echet distance between the curves is the minimal length of a
leash that is necessary. 

The {\bf Hausdorff distance} between $\C$ and $\C_m$ is the smallest value
$d$ such that every point of $\C$ has a point of $\C_m$ within distance $d$
and every point of $\C_m$ has a point of $\C$ within distance
$d$~\cite{rote1991}. It is formally defined as
\begin{eqnarray}\label{eq:haus}
   d_{\mathrm H}(\C,\C_m) = \max\{\,\sup_{\eta \in [\eta_0,\eta_p]} \inf_{\xi
      \in [0,1]} \|\mvx_m(\xi)-\mvx(\eta)\|,\, \nonumber\\
     \sup_{\xi \in [0,1]} \inf_{\eta \in [\eta_0,\eta_p]} \|\mvx_m(\xi)-\mvx(\eta)\|\,\}\mbox{.} \nonumber
\end{eqnarray}

\begin{figure}[h!]
 \begin{center}
 \includegraphics[width=0.3\textwidth]{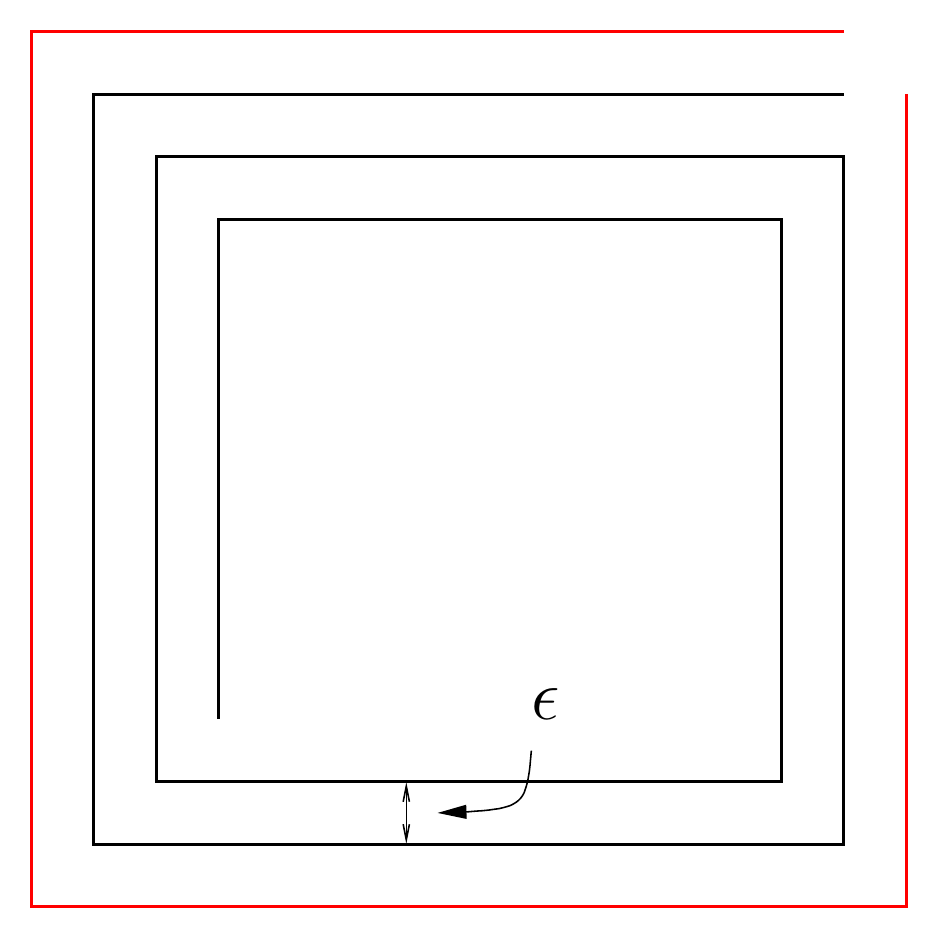}
\end{center}
\caption{Two curves that can be made arbitrary close when $\epsilon \rightarrow 0$ in terms of their Hausdorff distance
(that is exactly $3\epsilon$) but not in terms of their Fr\'echet distance that remains finite and equal to 
the diagonal of the square. \label{fig:close}}
\end{figure}

Not only  $d_H(\C,\C_m) \leq d_F(\C,\C_m)$, 
but the Fr\'echet distance between two curves can be arbitrarily 
larger than their Hausdorff distance. The Fr\'echet distance is usually considered as a 
more reliable measure of similarity between curves. Figure \ref{fig:close} shows
two curves that can be made arbitrary ``Hausdorff-close'' while being quite dissimilar.

The definition of the Hausdorff and Fr\'echet measures, that involves
infima and suprema over infinite sets of parametrizations, makes it
difficult to devise algorithms for computing these distances between
arbitrary curves. However, an alternative that can lead to practical
algorithms is to calculate the Hausdorff and Fr\'echet distances between
polygonal approximations of the curves under consideration, as explained
in the next Section.

\subsection{Distance Between Polygonal Curves}

\subsubsection{Optimal Sampling of Curves}\label{sec:sampling}

Let us first consider the problem of approximating an arbitrary curve
by a polygonal curve. In order to maximize the efficiency of the distance
computation, it is necessary to find a polygonal curve that contains
as few vertices as possible and still approximates the original curve with
sufficient accuracy.

\begin{figure}[h!]
 \begin{center}
 \includegraphics[width=0.4\textwidth]{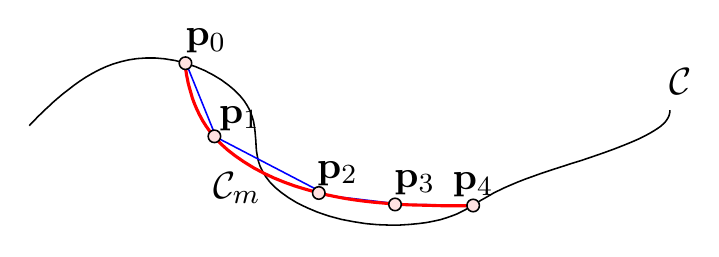}
\end{center}
\caption{A model edge $\mathcal C$, a mesh
  edge ${\mathcal C}_m$ and a polygonal curve M (in blue).\label{fig:geom1dc}}
\end{figure}

Assume $m+1$ points $\mvp_i = \mvx_m(\xi_i),~i\in [0,m],$ that are sampled
on $\C_m$. This defines a polygonal curve $M$ formed of $m$ segments for
which segment $i$ goes from $\mvp_i$ to $\mvp_{i+1}$ (see
Fig.~\ref{fig:geom1dc}). Let us do the same with $\C$ and define a
polygonal curve $N$ composed of $n+1$ points $\mvq_i,~i\in [0,n].$

The goal is to find sampling points on the original curve ${\mathcal C}_m$
(resp. $\mathcal C$) in such a way that the distance between the polygonal
curve $M$ (resp. $N$) and ${\mathcal C}_m$ (resp. $\mathcal C$) is smaller
than a threshold distance $\epsilon$.

A parametrization of ${\mathcal C}_m$ is given by \eqref{eq:lag1d} and a
parametrization of $\mathcal C$ is usually available from the CAD model. A
possible sampling strategy for these curves consists in starting with the
segment $[\mvx_0, \mvx_p]$ and refining recursively the discretization with
points distributed uniformly in the parameter space, until the desired
accuracy is reached. However, $\mathcal C$ is often a B\'ezier or a
rational B\'ezier spline. As a polynomial curve, ${\mathcal C}_m$ is also
a particular case of a B\'ezier curve. For such curves, de Casteljau's
algorithm provides a more efficient way to refine the discretization and
control the geometrical accuracy at the same time.

\begin{figure}[h!]
 \begin{center}
 \includegraphics[width = 6cm]{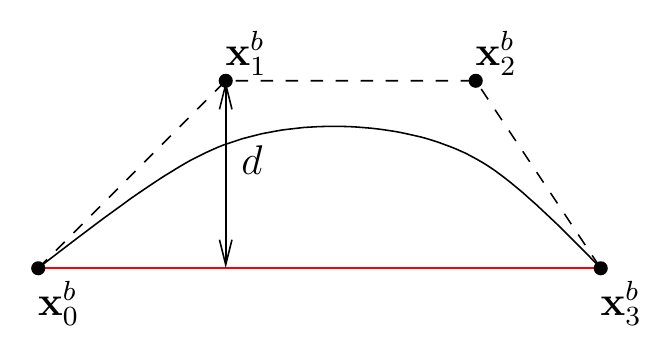}
\end{center}
\caption{A cubic B\'ezier curve, its control polygon (dashed lines)
 and the coarsest approximation of the curve in red. \label{fig:decas}}
\end{figure}

Consider the Bernstein basis polynomials of degree $\o$:
\beqn{e:bezFunc}
\bez\k\o(\a) = \Comb\k\o\,(1-\a)^{\o-\k}\,\a^\k\qquad(\a\in[0,1]\ ;\ \k=0,...,\o)
\eeqn
where $\Comb\k\o = \frac{\o!}{\o!(\o-\k)!}$ is the binomial coefficient.
Since Lagrange and Bernstein polynomials span the same function space,
we can re-write \eqref{eq:lag1d} as a B\'ezier curve
\begin{equation}\label{eq:bez1d}
\mvx_m(\xi)= \sum_{i=0}^{\o}
\bez i\o(\a)\;\mvx^b_i,~~~\xi \in [0,1]
\end{equation}
where the $\mvx^b_i$'s are the control points of the B\'ezier curve, that
form a control polygon (see Figure \ref{fig:decas}). The control points
$\mvx^b_i$'s can be computed from the node locations $\mvx_i$'s by means
of a transformation matrix $\matB\o$: 
{\def\RA{\xi}
\[ \matB\o =
  \cro{\begin{array}{ccc}
      \bez0\o\ap{\RA_0} & \hdots & \bez\o\o\ap{\RA_0}\\
      \bez0\o\ap{\RA_1} & \hdots & \bez\o\o\ap{\RA_1}\\
      \vdots & \ddots & \vdots\\
      \bez0\o\ap{\RA_\o} & \hdots & \bez\o\o\ap{\RA_\o}
\end{array}}.
\]}

\begin{figure}[h!]
 \begin{center}
 \includegraphics[width = 6cm]{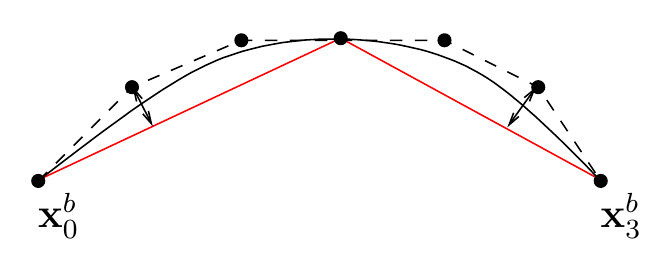}
\end{center}
\caption{A first application of de Casteljau's algorithm. The polygonal
approximation in red gets closer to the control polygons (and thus to
the B\'ezier curves).\label{fig:decas2}}
\end{figure}

A classical way to optimally sample a B\'ezier curve is to use
de Casteljau's algorithm. A first approximation of the B\'ezier 
curve is constructed as the single line segment between 
$\mvx^b_0$ and $\mvx^b_{\o}$ (red line segment in Figure~\ref{fig:decas}). 
The distance $d$ between this single segment and the control polygon
is an upper bound of the distance between the curve and the segment
because of the convex hull property.
If needed, the curve is then split into two sub-curves using de
Casteljau's algorithm, each of them coinciding exactly with the
original curve. This argument is applied  recursively (see Figure
\ref{fig:decas2}) to every sub-curve where the distance between the
control polygon and the corresponding segment is less than $\epsilon$.
The extremities of the sub-curves at the finest level are thus the
vertices $\mvp_i$ (resp. $\mvq_i$) of the polygonal approximation for
${\mathcal C}_m$ (resp. $\mathcal C$).

\begin{figure}
\begin{center}
\includegraphics[width=0.475\textwidth]{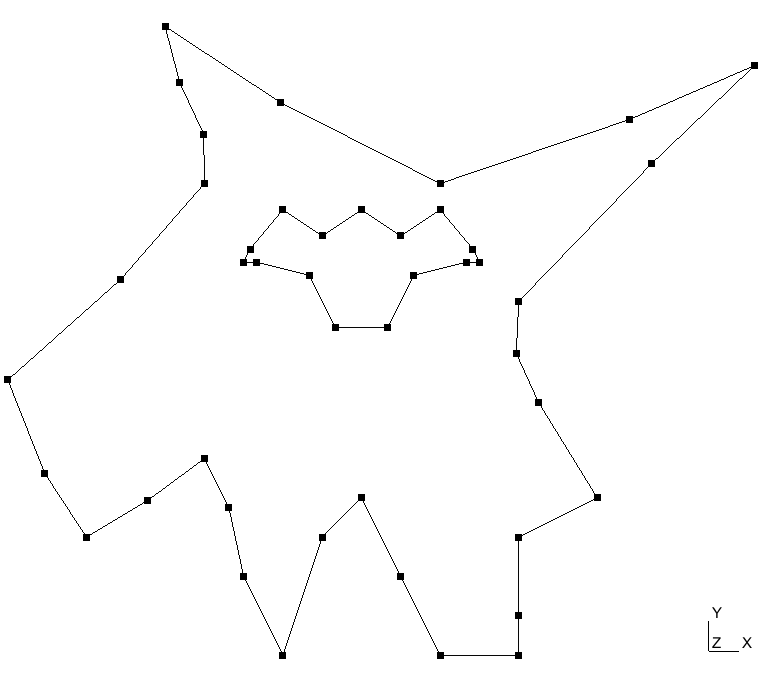}
\includegraphics[width=0.475\textwidth]{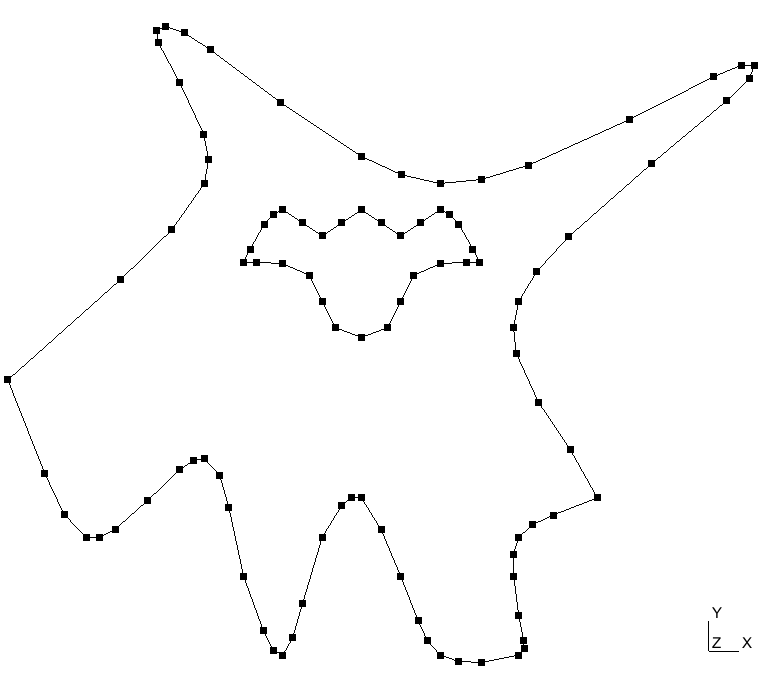}\\
\includegraphics[width=0.475\textwidth]{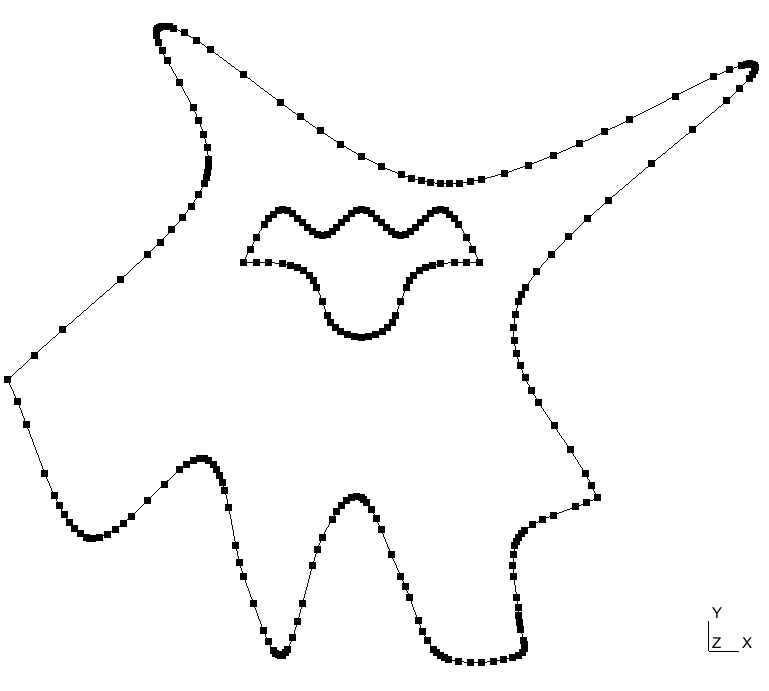}
\includegraphics[width=0.475\textwidth]{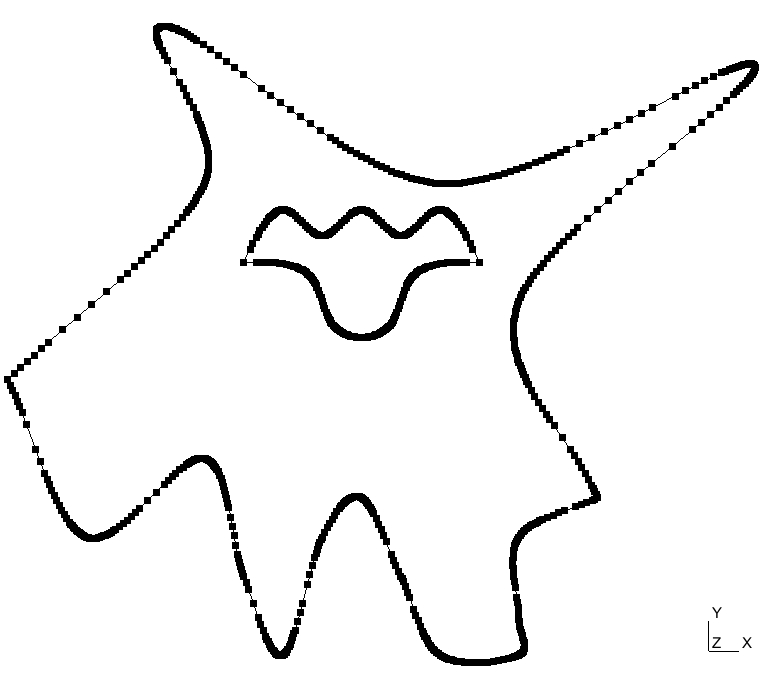}
\end{center}
\caption{Adaptive subdivision of a 2D boundary with 
$\epsilon = 5 \times 10^{-2}$  (88 points), 
$\epsilon = 10^{-2}$  (196 points), 
$\epsilon = 10^{-3}$   (554 points), 
$\epsilon = 10^{-4}$  (858 points) \label{fig:blob}}
\end{figure}

A similar algorithm can be applied to rational B\'ezier splines. 
Most of the CAD entities can be casted as rational B\'ezier splines so
that this optimal subdivision can be applied to most of the curves
that are present in CAD models. As an example, Figure \ref{fig:blob}
presents the subdivision of  a boundary using different values of the
threshold parameter $\epsilon$. For the unusual cases involving
non-standard parametrizations, the generic recursive sampling algorithm
can be used.

\subsubsection{Discrete distances between polygonal curves}

The simplest way of approximating the distances
$d_{\mathrm{H}} (\C,\C_m)$ and $d_{\mathrm{F}} (\C,\C_m)$ is to compute
the \emph{discrete} Hausdorff and Fr\'echet distances $\delta_H(M,N)$ and
$\delta_F(M,N)$, i.e. the Hausdorff and Fr\'echet distances restricted
to discrete point sets. 

Computing $\delta_H(M,N)$
consist essentially in computing both Voronoi diagrams of the
$\mvp_i$'s and of the $\mvq_i$'s and to find Voronoi cells
of $\mvp_i$ that contains $\mvq_j$ and Voronoi cells
of $\mvq_j$ that contains $\mvp_i$. 
Computing $\delta_H(M,N)$  requires ${\mathcal O} (m+n) \log(n+m)$
operations.  In our implementation, we do not
explicitly construct the Voronoi diagram. An approximate nearest
neighbor algorithm \cite{arya1998optimal} is used to locate closest
points, leading to the same algorithmic complexity.  
The following result holds
\[
d_H (M,N) \leq \delta_H(M,N) \leq d_H(M,N) + \max \{D(M),D(N)\}.
\]
where $D(M)$ (resp. $D(N)$) is the maximum distance between two successive
points in $M$ (resp. $N$). In Section~\ref{sec:sampling}, the accuracy
$\epsilon$ of the polygonal representation is supposed to be much smaller
than the actual distance between the CAD and the mesh in such a way that 
\[
|d_H(M,N) - d_H(\C,\C_m)| = {\mathcal O} (\epsilon).
\]
Consequently, the discrete distance is accurate enough if and only if 
\[
\max \{D(M),D(N)\} < \epsilon,
\]
which implies that the approximate nearest neighbor algorithm should not
be applied directly to the vertices of $M$ and $N$, but to sets of points
sampled at interval $\epsilon$ on these polygonal curves instead. Note
that it is still useful to construct the optimal polygonal approximations
$M$ and $N$ and to sample these rather than obtaining directly a dense
sampling of $\C$ and $\C_m$, because evaluating a point of a B\'ezier curve
is computationally much more intensive than evaluating a point of a line
segment. Unfortunately, the number of points to be submitted to the
approximate nearest neighbor algorithm is clearly a problem in our
context, where a relative accuracy of  $10^{-6}$ is often required.

The discrete Fr\'echet distance $\delta_F(M,N)$ \cite{eiter1994computing}
considers only positions of the ``leash'' where its endpoints are located
at vertices of the two polygonal curves and never in the interior of an edge.
The discrete Fr\'echet distance can be computed in polynomial time, i.e.
in ${\mathcal O}(mn)$ operations, using a simple dynamic programming
algorithm that is described in \cite{eiter1994computing}. It is very
difficult to find a sub-quadratic algorithm that computes the discrete
Fr\'echet distance. We have again
\[
d_F (M,N) \leq \delta_F(M,N) \leq d_F(M,N) + \max \{D(M),D(N)\}.
\]
Computing the discrete Fr\'echet distance may be out of reach if
massive oversampling is applied.

\subsubsection{Direct distances between polygonal curves}

Let us now consider the real Hausdorff and Fr\'echet distances between
polygonal curves, that is, considering the line segments per se and not
only discrete sets of points on them.

The computation of the direct Hausdorff distance between two polygonal
curves is related to the Voronoi diagram of the line segments. 
The distance can only occur at points that are either endpoints of 
line segments or intersection points of the Voronoi diagram of one of 
the sets with a segment of the other. This observation leads us to 
the following quadratic algorithm for computing the direct 
Hausdorff distance $d_H(M,N)$. 
\begin{itemize}
\item Compute the bissector of all possible pairs of segments 
  of polygonal curve $N$. Two line segment have a bisector with 
  up to 7 arcs (lines and parabolas). Store all arcs in a list.
\item Compute the intersections of each arc with $M$.
\item Compute the distance between those intersection points and $N$. 
  The one sided Hausdorff distance is the maximum of those distances.
\end{itemize}
The Voronoi diagram of line segments could be theoretically
computed in ${\mathcal O}((n+m)\log(n+m))$ operations \cite{alt1995computing}. Yet, it involves
the computation of the whole Voronoi diagram. 
To our knowledge, few robust implementations of Voronoi diagrams of
line segments exist
\cite{held2001vroni} and no extension to higher dimensions than two 
has been proposed to date.

It is possible to compute the Fr\'echet distance $d_F(M,N)$ 
between two polygonal curves in  
${\mathcal O}\left(mn \log(mn)\right)$ operations \cite{alt1995computing}. The
algorithm is even more complex that for direct Hausdorff
distance.

\subsection{Geometrical error based on Taylor expansions}

In the above sections, the curves ${\mathcal C}_m$ and ${\mathcal C}$
have been approximated by polygons, so that their relative distance
can be easily and efficiently computed without any assumption about the
curves. Another approach for simplifying the computation of the
geometrical error is to take advantage of the fact that the high-order
nodes $\mvx_i$ defining the mesh edge ${\mathcal C}_m$ are located both
on ${\mathcal C}_m$ and on the model curve ${\mathcal C}$. A Taylor
expansion of the natural parameter in the vicinity of $\mvx_i$ for
each curve then provides an estimation of the geometrical error.

Assume a curve defined by $\mvx(t)$, $t\in[t_1,t_2]$. The curvilinear
abscissa $s(t)$ of a point $\mvx(t)$ of the curve is the length of the
segment defined by parameter range $[t_1,t]$, i.e. the length of the
curve from the origin $\mvx(t_1)$ to $\mvx(t)$:
\[
s(t) = \int_{t_1}^t ~\left\|\mvx_{,t}\right\| dt
\]
We have $ds=\left\|\mvx_{,t}\right\| dt$. The arc length $s(t)$
provides the natural parametrization $\bar{\mvx}(s)$ of the curve:
\[
\bar{\mvx}(s(t)) = \mvx(t),~~t\in[t_1,t_2],
\]
with $\left\|\bar{\mvx}_{,s}\right\| = 1$, where $\bar{\mvx}_{,s}$ is
the derivative of $\bar{\mvx}$ with respect to $s$. The unit tangent
vector to the curve is computed as 
\begin{equation}\label{eq:tangent}
\mvt = \bar{\mvx}_{,s} = {\mvx_{,t} \over \|\mvx_{,t}\|}.
\end{equation}
The curvature vector $\mvc$ of the curve at a point $\mvx$ can be defined
as the amplitude  of  the variations of the unit tangent $\mvt$ along the
curve. The vector  $\mvc = \mvt_{,s}$ is obviously orthogonal to $\mvt$
because $\mvt$'s amplitude is equal to one along $s$. We have
$$\mvc = \bar{\mvx}_{,ss} = \frac{1}{\left\|\mvx_{,t}\right\|^3}
\left(\mvx_{,tt} \left\|\mvx_{,t}\right\| -
    \mvx_{,t} \frac{\mvx_{,t} \cdot \mvx_{,tt}}{\|\mvx_{,t}\|}\right).$$
It is thus possible to approximate the curve with a Taylor expansion
of $\bar{\mvx}$ around $s_0=s(t_0)$ at second order as
\begin{equation}\label{eq:curvature}
\bar{\mvx}(s_0+s) = \bar{\mvx}(s_0) + s\left.\mvt\right|_{s_0} +
\frac{s^2}{2} \left.\mvc\right|_{s_0} + {\mathcal O}(s^3).
\end{equation}

Applying this expansion for a mesh edge ${\mathcal C}_m$ and the
corresponding model curve ${\mathcal C}$, the geometrical error between
both curves can be estimated near each of their common points
$\mvx_i$ as
\[
\delta^i_{T,L} = \left\| h (\mvt_m - \mvt)\right\|
\]
for a linear approximation and
\[
\delta^i_{T,Q} = \left\| h (\mvt_m - \mvt)
+ {h^2 \over 2} (\mvc_m - \mvc)\right\|
\]
for a quadratic approximation. In these expressions, the unit tangent
vector $\mvt_m$ (resp. $\mvt$) and the curvature vector $\mvc_m$ (resp.
$\mvc$) is computed on $\C_m$ and (resp. $\C$) at point $\mvx_i$, and
$h$ is proportional to a ``local edge length'' computed from the Jacobian
of $\C_m$. The derivatives of $\mvx_m$ required to compute $\mvt_m$ and
$\mvc_m$ can be easily obtained from Eq.~\eqref{eq:lag1d}. For the
vectors $\mvt$ and $\mvc$  related to the model curve ${\mathcal C}$,
the derivatives of $\mvx$ are provided by the CAD model. The geometrical
error for the whole mesh edge $\C_m$ can then be computed as
\[
\delta_T = \left(\sum_{i=0}^{p} {d^i_T}^2\right)^\frac{1}{2}.
\]

\subsection{A simple example}

\begin{figure}
\begin{center}
\includegraphics[width=0.75\textwidth]{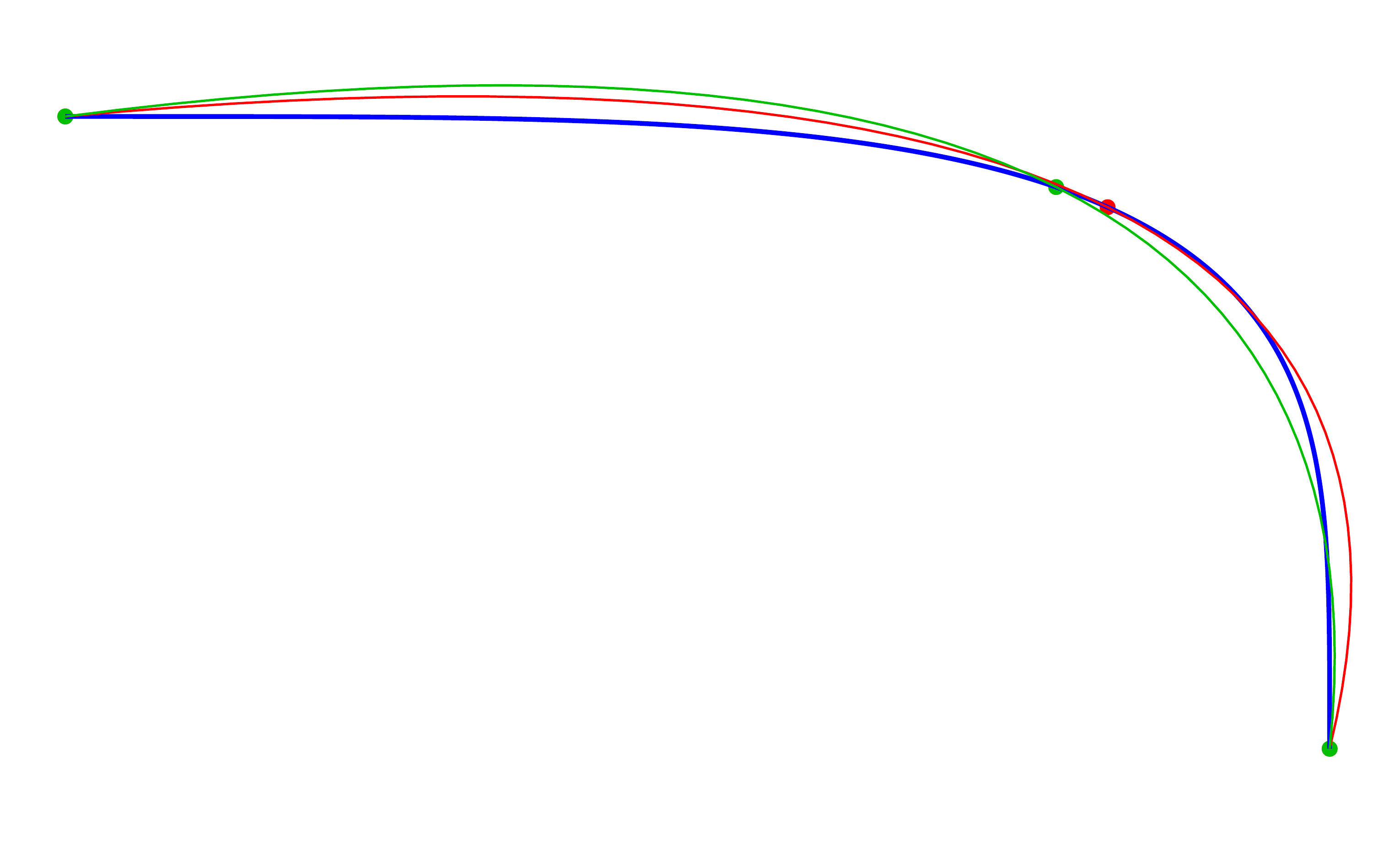}
\end{center}
\caption{Geometrical model representing a Lam\'e curve (thick blue
line) and corresponding quadratic mesh edges minimizing
$\delta_H=\delta_F$ (red thin line) and $\delta_T$ (green thin
line).\label{fig:simpleCurve}}
\end{figure}

In order to illustrate the different estimates of the geometrical error
described above, we consider the simple case of a particular
Lam\'e curve:
\[
y=\frac{1}{2}\left(1-x^4\right)^\frac{1}{4},\;\;x\in[0, 1].
\]
The geometrical model is a B\'ezier spline representing this curve with
negligible error (see Fig.~\ref{fig:simpleCurve}). The spline is
parametrized with a variable $u\in[0, 1]$.

A quadratic mesh edge is built to approximate the model curve. Its end
vertices are fixed at the extremities of the curve (i.e. $u=0$ and
$u=1$), and we explore the values of $\delta_F$,  $\delta_H$ and
$\delta_T$ for different locations of the high-order node on the model
curve. In particular, we consider 100 locations in the range
$u\in[0.2, 0.68]$, where the Jacobian of the edge is positive (i.e.
the mesh is valid).

\begin{figure}
\begin{center}
\includegraphics[width=0.75\textwidth]{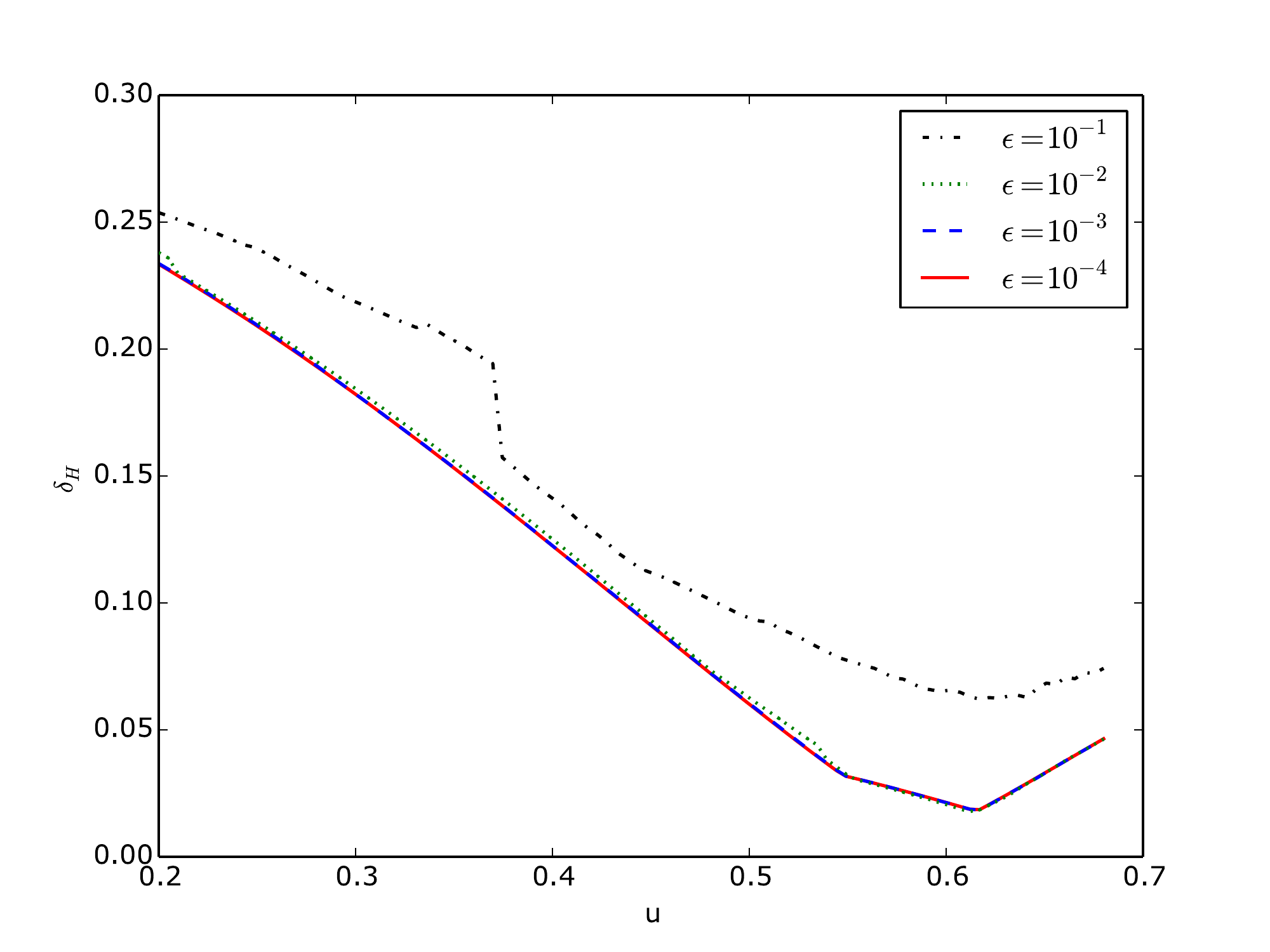}
\end{center}
\caption{Case of a Lam\'e curve: discrete Hausdorff distance
$\delta_H=\delta_F$ for the quadratic mesh edge as a function of
the position $u$ of the high-order node on the geometrical model.
\label{fig:dist_accuracy}}
\end{figure}

The evolution of the discrete Hausdorff distance $\delta_H$ with the
high-order node location is shown in Fig.~\ref{fig:dist_accuracy} for
different values of the accuracy threshold $\epsilon$. A value of
$\epsilon=10^{-3}$ seems to yield sufficient accuracy for this curve.
In this particular case, the discrete Fr\'echet distance $\delta_F$ is
equal to the discrete Hausdorff distance $\delta_H$. 

\begin{figure}
\begin{center}
\includegraphics[width=0.75\textwidth]{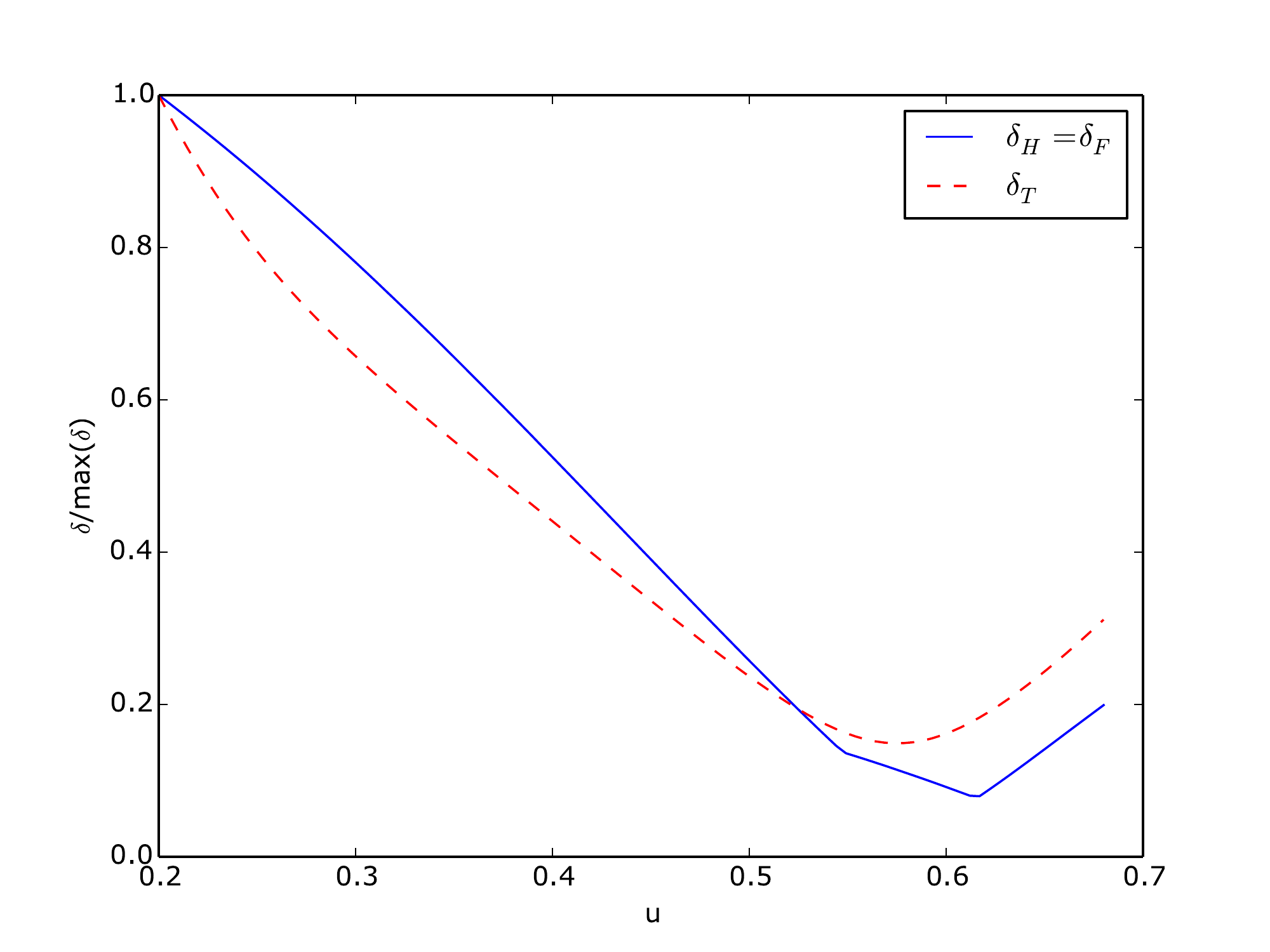}
\end{center}
\caption{Case of a Lam\'e curve: quantities $\delta_H=\delta_F$ and
$\delta_T$, normalized by their maximum value, for the quadratic mesh
edge as a function of the position $u$ of the high-order node on the
geometrical model.\label{fig:scaledDist}}
\end{figure}

Fig.~\ref{fig:scaledDist}
shows the quantities $\delta_H=\delta_F$ and $\delta_T$ normalized by
their respective maximum value, so that they can be compared to each
other. Although both curves are qualitatively similar, they do not
reach a minimum for the same high-order node location ($u=0.617$ for
$\delta_H$ and $u=0.573$ for $\delta_T$), which can be visualized
in Fig.~\ref{fig:simpleCurve}. Moreover, it clearly appears
that the Taylor-based geometrical error $\delta_T$ is a continuously
differentiable function of the high-order node position, while the
the Hausdorff and the Fr\'echet distances $\delta_H$ and $\delta_F$ are
not differentiable everywhere, in particular at their minimum.

The approximate CPU time for one distance evaluation, measured on a
recent laptop computer, is about $3.0\cdot 10^{-2}\;\textrm{s}$ for
$\delta_F$, and $2.4\cdot 10^{-3}\;\textrm{s}$ for $\delta_H$ with
$\epsilon=10^{-3}$, whereas it is only $1.1\cdot10^{-6}\;\textrm{s}$
for $\delta_T$. Given its continuously differentiable nature and
its low computational cost, the Taylor-based geometrical error
estimate $\delta_T$ is clearly much more appropriate  than the other
distances for an optimization procedure such as the one described in
Sec.~\ref{sec:opti}.

\section{Mesh optimization}\label{sec:opti}

In a recent paper \cite{jcp2013}, a technique that allows to untangle high
order/curvilinear meshes is presented. The technique makes use of unconstrained
optimization where element Jacobians are constrained to lie in a prescribed
range through moving log-barriers. The untangling procedure starts from a possibly 
invalid curvilinear mesh and moves mesh vertices with the objective 
of producing elements that all have bounded Jacobians. Bounds on
Jacobians are computed using results of papers \cite{bounds-imr,bounds-jcp}.
In what follows, we extend the optimization procedure in charge of untangling
the invalid elements in order to take into account the geometrical error
$\delta_T$.

The procedure described in Ref.~\cite{jcp2013} consists in solving a
sequence of minimization problems, where the objective function
$f(\mvx_i)$ is composed of two parts ${\mathcal E}$ and
${\mathcal F}\epsilon$:
\[
f={\mathcal E}+{\mathcal F}_\epsilon
\]
Here $\mvx_i$ is the position of node $i$. For a node located on a
boundary, it is possible to work with the parametric coordinate(s) of
the node on the geometrical model entity given by the CAD model. As
the scale of the parametric coordinate can differ significantly from the
scale of the physical coordinates, preconditioning may then be required
for the Conjugate Gradient to converge properly.

The first part ${\mathcal E}$ relies on the assumption that the
method is provided with a straight-sided mesh of high quality. This
mesh has potentially been defined to satisfy multiple criteria, such as
a predetermined size field, or anisotropic adaptation. The conversion
of such meshes to high order is expected to preserve as much as possible
all these features. Therefore, the nodes shall be kept as close as
possible to their initial location in the straight sided mesh. In this
work, the definition of ${\mathcal E}$ is the one of  \cite{jcp2013}
i.e. 
\begin{equation}
\label{eq:energy1}
{\mathcal E}(\mvx_i) = \frac{K_{\mathcal E}}{L^2}
\sum_{i} \left\|\mvx_i-\mvX_i\right\|^2
\end{equation}
with $\mvX_i$ the position of the node $i$ in the straight-sided mesh,
$K_{\mathcal E}$ a non dimensional constant and $L$ a characteristic
size of the problem.

The second part ${\mathcal F}$ of the functional controls the positivity
of the Jacobian. A $\log$ barrier \cite{jcp2013} prevent Jacobians from
becoming too small:
\[
{\mathcal F}_\epsilon(\mvx_i) =
\sum_{e} \sum_{l} F^e_l(\mvx_i,\epsilon)
\]
with $l$ iterating on all coefficients $B^e_l $ of the B\'ezier
expansion of the Jacobian of $e$ and where 
\begin{equation}
\label{eq:barrier1}
F^e_l(\mvx_i,\epsilon) =  \left[\log\left(\frac{B^e_l(\mvx^e) -
\epsilon J^e_0}{J^e_0 - \epsilon J^e_0} \right)\right]^2  +
\left(\frac{B^e_l(\mvx^e)}{J^e_0} -1\right)^2
\end{equation}
is the log barrier function defined in such a way that ${\mathcal F}$
blows up when $B^e_l/J^e_0\rightarrow\epsilon$, but still vanishes when
$B^e_l = J^e_0$. In this expression, $\mvx^e$ is the vector gathering
the positions $\mvx_i$ of all nodes $i$ belonging to element $e$, and
$J^e_0$ is the constant straight sided Jacobian of $e$.

The value of $K_{\mathcal E}$ has little influence on results. The
presence of ${\mathcal E}$ prevents the problem from being under-determined,
and it orients the optimization procedure towards a solution that tends
to preserve the straight-sided mesh, but it is clearly dominated
by  ${\mathcal F}_\epsilon$ when invalid elements exist in the domain.

A Conjugate Gradient algorithm is used to minimize the objective
function $f$ with respect to the node positions $\mvx_i$ for a fixed
value of the log barrier parameter $\epsilon$. A sequence of such
minimization problems is solved, in between which the $\epsilon$
is progressively increased, so that the Jacobian of all elements
is forced to exceed a user-defined target value.

In the present work, the procedure described above is followed in a first
step. In a second step, a similar procedure is used, albeit with an
objective function f taking into account the geometrical model:
\[
f={\mathcal E}+{\mathcal F}_\epsilon+{\mathcal D}_{\epsilon^\prime}
\]

The third part ${\mathcal D}_{\epsilon^\prime}$ of the functional
controls the error $\delta_T$ between the mesh and the geometrical
model. Again, a log barrier is used:
\[
{\mathcal D}_{\epsilon^\prime}(\mvx_i) = K_{\mathcal D}
\sum_{b} D^b(\mvx_i,\epsilon)
\]
with $b$ iterating on all boundary mesh edges and
\[
D^e_l(\mvx_i,\epsilon^\prime) = \left[\log\left(\frac{\delta_T^b(\mvx^b) -
\epsilon \delta_0}{\delta_0 - \epsilon\delta_0}\right)\right]^2  +
\left(\frac{\delta_T^b(\mvx^b)}{\delta_0}\right)^2
\] 
where $\delta_T^b$ is the is the geometrical error $\delta_T$ for the
boundary mesh edge $b$, and $\delta_0$ is a target value for $\delta_T^b$.
The vector $\mvx^b$ collects the positions $\mvx_i$ of all nodes $i$
belonging to the boundary mesh edge $b$. Derivatives of $\delta_T$ with
respect to $\mvx_i$ are computed using finite differences. 

In this second step, ${\mathcal F}_\epsilon$ is used as a fixed log barrier
(constant $\epsilon$) that is meant to prevent the Jacobian of elements to
fall back below the target value reached in the first step. On the contrary,
${\mathcal D}_{\epsilon^\prime}$ is a moving log barrier where the parameter
$\epsilon^\prime$ is iteratively updated to drive the geometrical error
$\delta_T$ towards its target value. The parameter $K_{\mathcal D}$ reflects
the weight given to the geometrical error contribution with respect to the
contribution of the Jacobians. In this work, we typically choose values
around $K_{\mathcal D}=0.1$.

\section{Examples}\label{sec:examples}

\subsection{NACA0012}\label{sec:naca}

We consider the classical geometry of the NACA0012 airfoil with unit chord
length. Sequences of 6 triangular meshes are generated, where the airfoil
is discretized by $4$, $6$, $10$, $18$, $34$ and $66$ elements respectively,
for a total of $244$, $298$, $436$, $546$, $832$ and $1178$ elements in each
mesh. The first sequence consists of linear meshes. Two others are composed
of quadratic meshes resulting from the optimization procedure described in
Section~\ref{sec:opti}: one is optimized for element validity only,
while the other also minimizes the geometrical error. The last two
sequences are made up of cubic meshes generated in the same manner as
the quadratic ones.

\begin{figure}
\begin{center}
\includegraphics[width=0.85\textwidth]{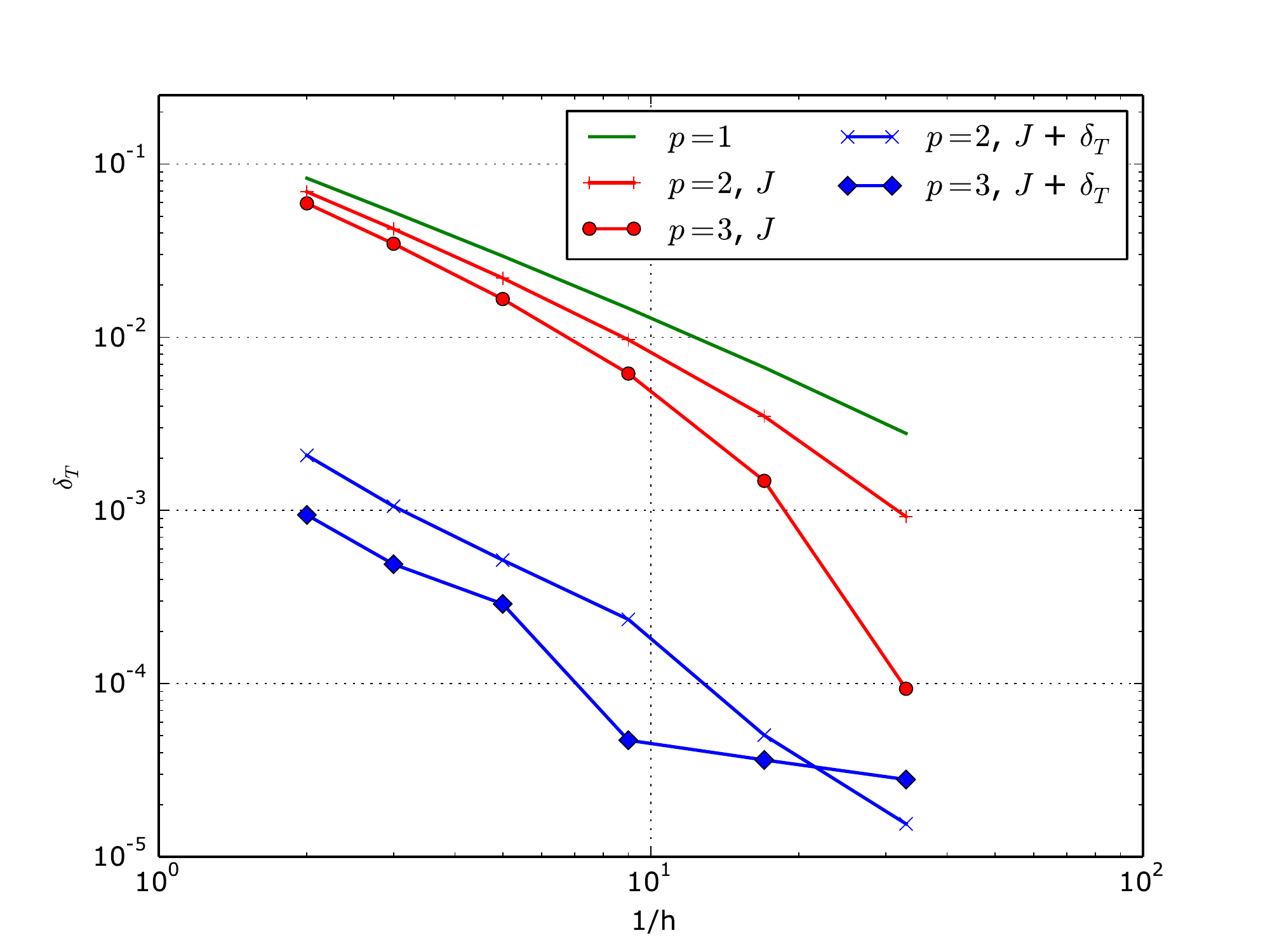}\\
\includegraphics[width=0.85\textwidth]{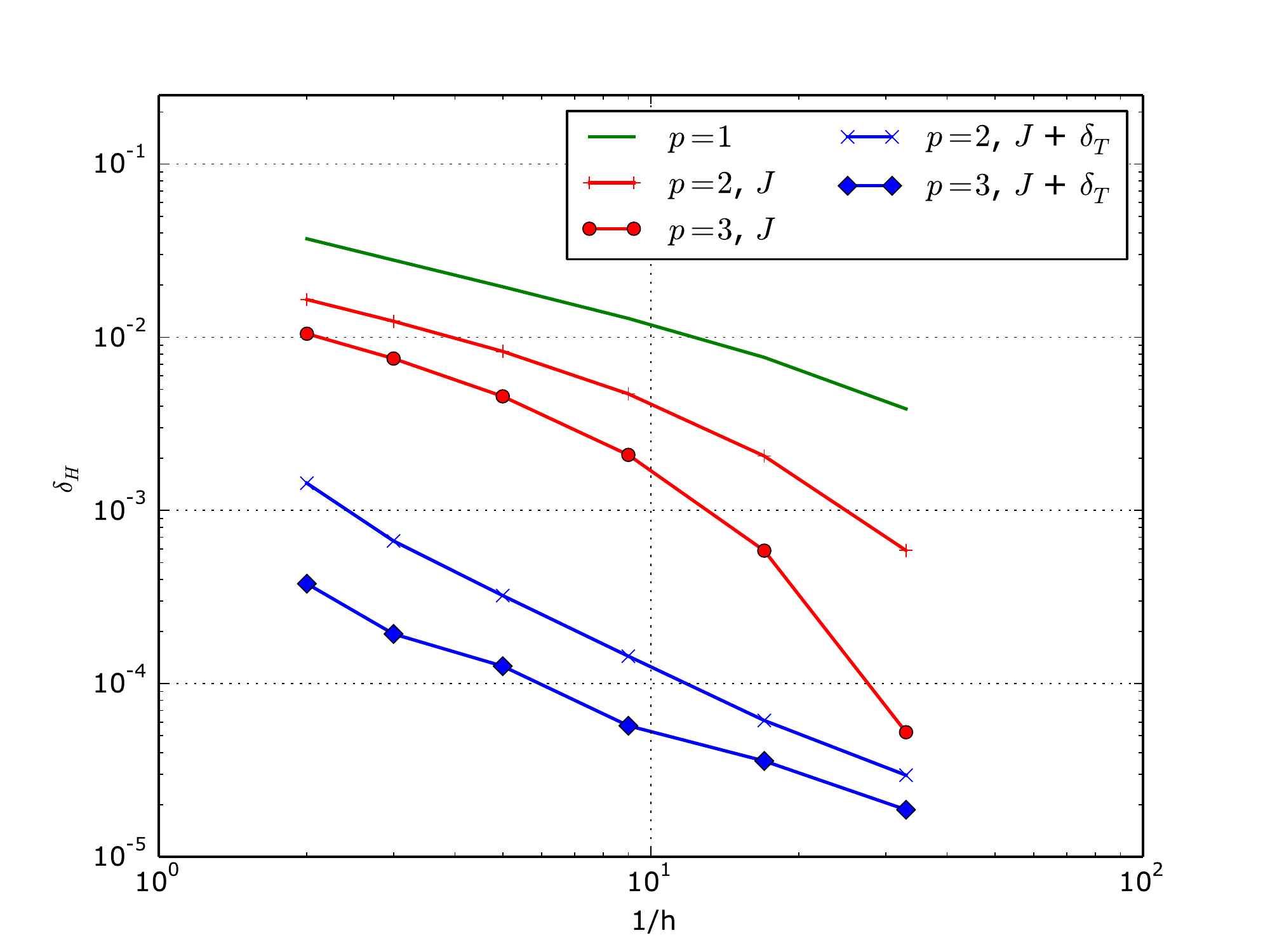}
\end{center}
\caption{Geometrical error $\delta_T$ (top) and model-to mesh Hausdorff
distance $\delta_H$ (bottom) for the NACA0012 profile: linear meshes
($p=1$) as well as quadratic ($p=2$) and cubic ($p=3$) meshes optimized
either for element validity only ($J$) or for both element validity and
geometrical error ($J+\delta_T$).
\label{fig:naca_dist}}
\end{figure}

Figure \ref{fig:naca_dist} shows how the geometrical error $\delta_T$
and the model-to-mesh Hausdorff distance $\delta_H$ evolve with the mesh
size. In meshes optimized for validity only, both $\delta_T$ and $\delta_H$
clearly decrease with decreasing mesh size, but most meshes are
too coarse to yield the optimal convergence rate. However, minimizing
the geometrical error $\delta_T$ reduces both $\delta_T$ and $\delta_H$ of
at least one order of magnitude in most cases. Geometrically-optimized
quadratic meshes are even more accurate than valid cubic meshes. The
geometrical optimization is most beneficial for coarser meshes, which lie
precisely the range of mesh size where refining brings less geometrical
accuracy. For fine cubic meshes, the improvement is less significant:
meshes optimized for validity only converge at near asymptotic rate for
both while $\delta_T$ and $\delta_H$, while the convergence rate with
meshes optimized for both validity and geometrical accuracy is not
improved (or even reduced for $\delta_T$).
Thus, optimizing meshes with respect to $\delta_T$ may be particularly
interesting with numerical schemes of very high order running on coarse
meshes, where it may yield a suitable geometrical approximation of the
model without unnecessary mesh refinement.

\begin{figure}
\begin{center}
\includegraphics[width=0.475\textwidth]{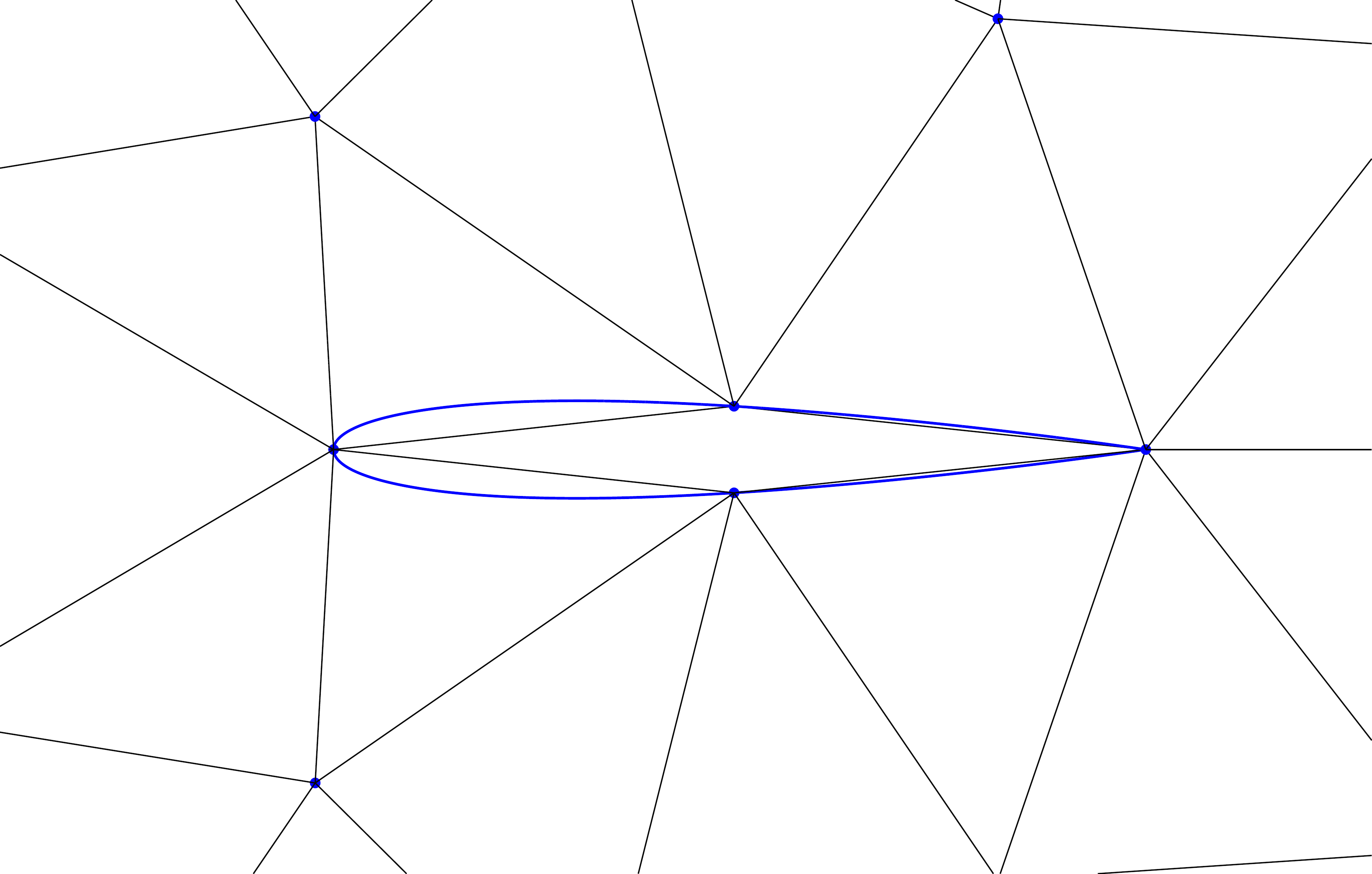}
\\[0.25\baselineskip]
\includegraphics[width=0.475\textwidth]{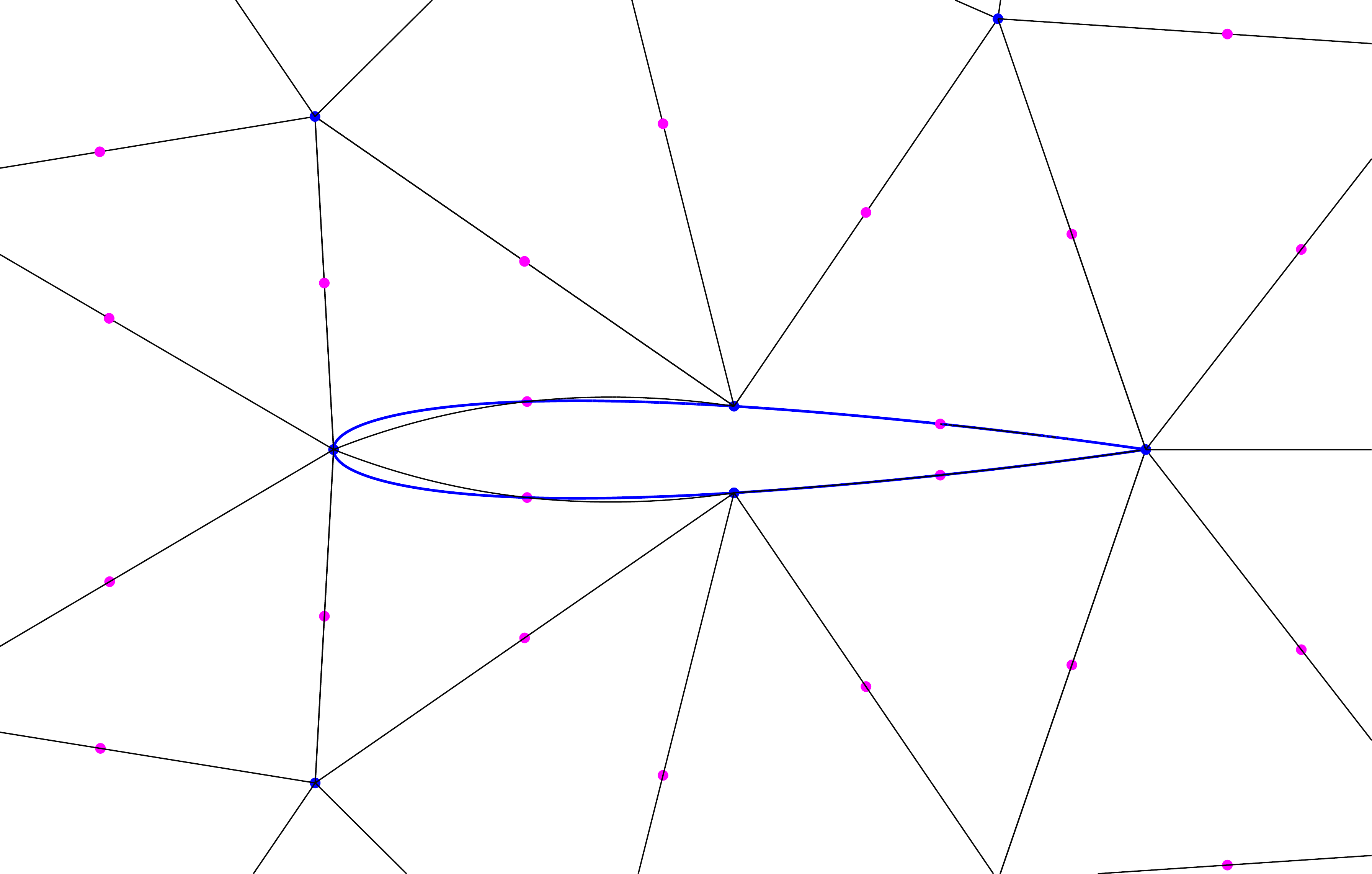}
\includegraphics[width=0.475\textwidth]{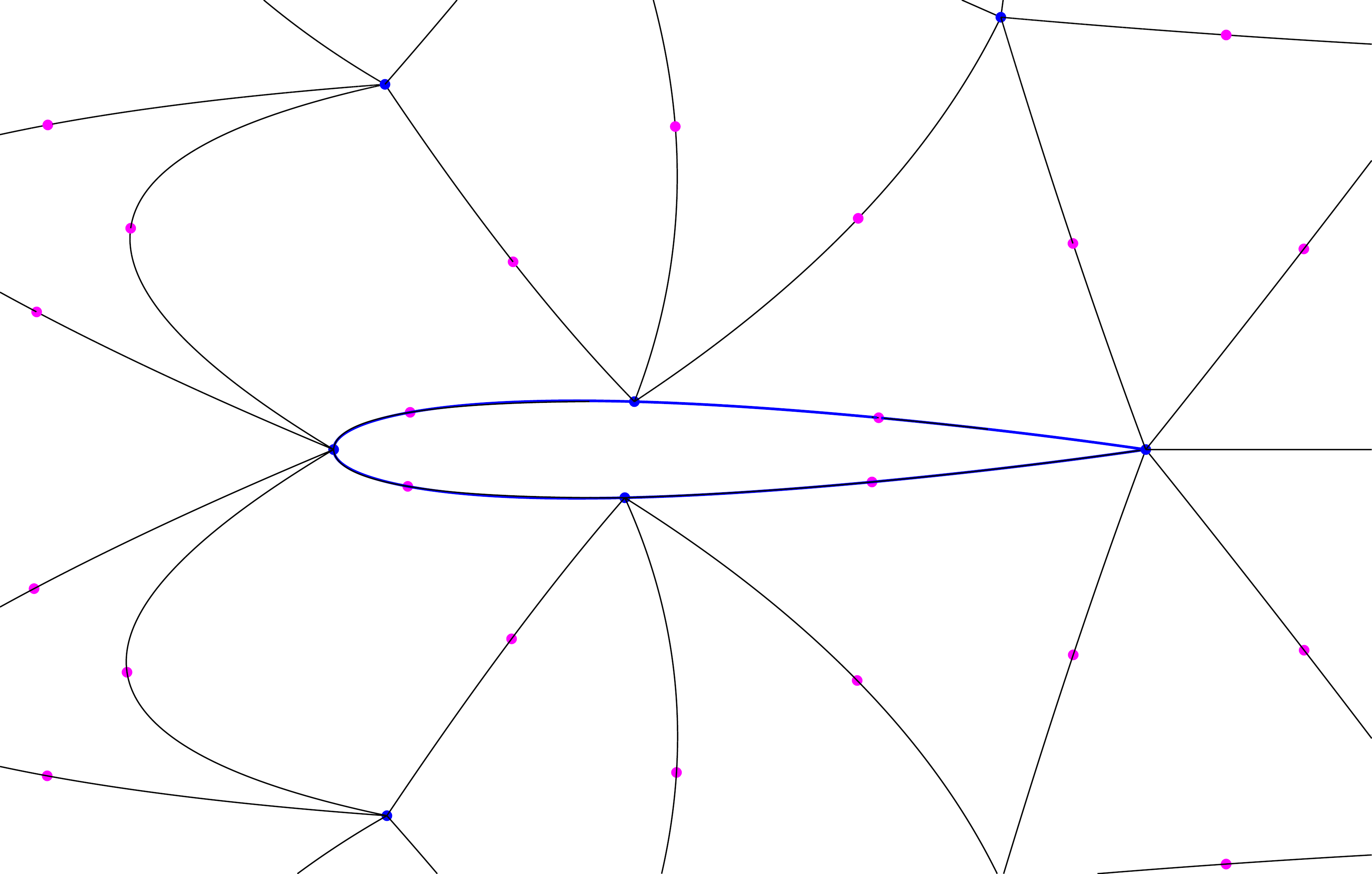}
\\[0.25\baselineskip]
\includegraphics[width=0.475\textwidth]{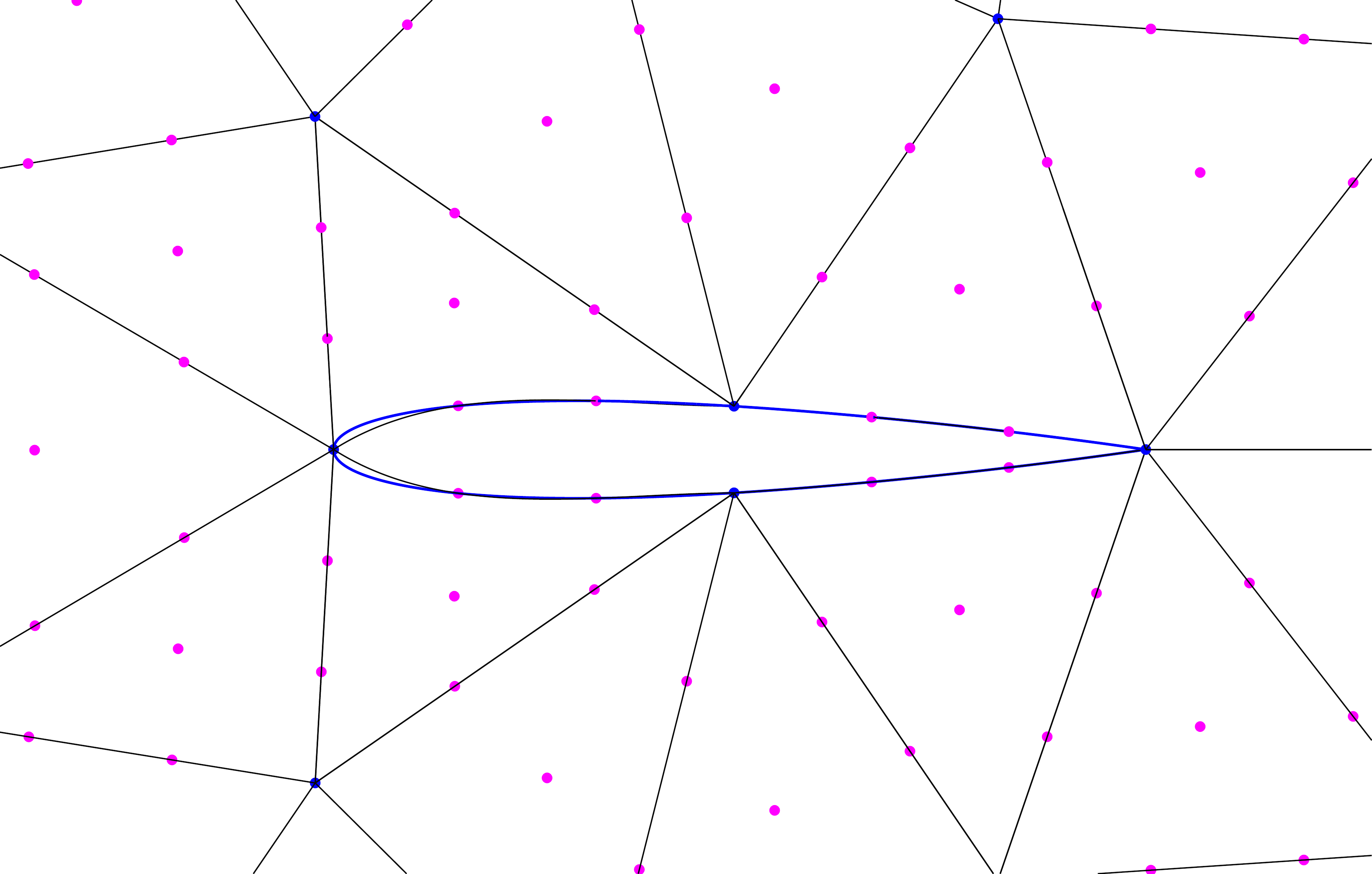}
\includegraphics[width=0.475\textwidth]{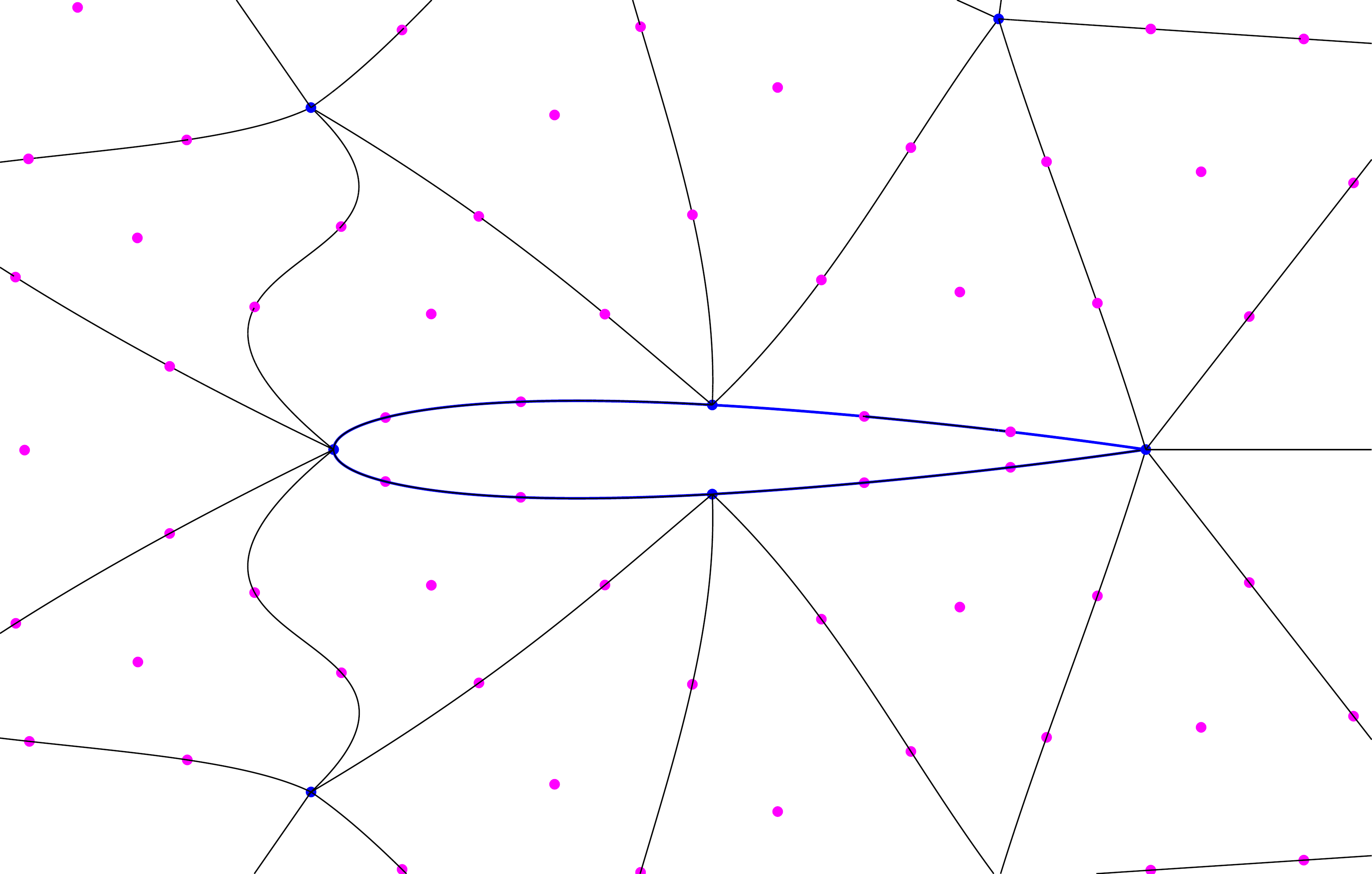}
\end{center}
\caption{Coarsest meshes of the NACA0012 profile. Top: linear mesh.
Center: quadratic meshes optimized for validity only (left) and for
validity as well as geometrical error (right). Bottom: cubic meshes
optimized for validity only (left) and for validity as well as
geometrical error (right).\label{fig:naca_mesh_0}}
\end{figure}

\begin{figure}
\begin{center}
\includegraphics[width=0.475\textwidth]{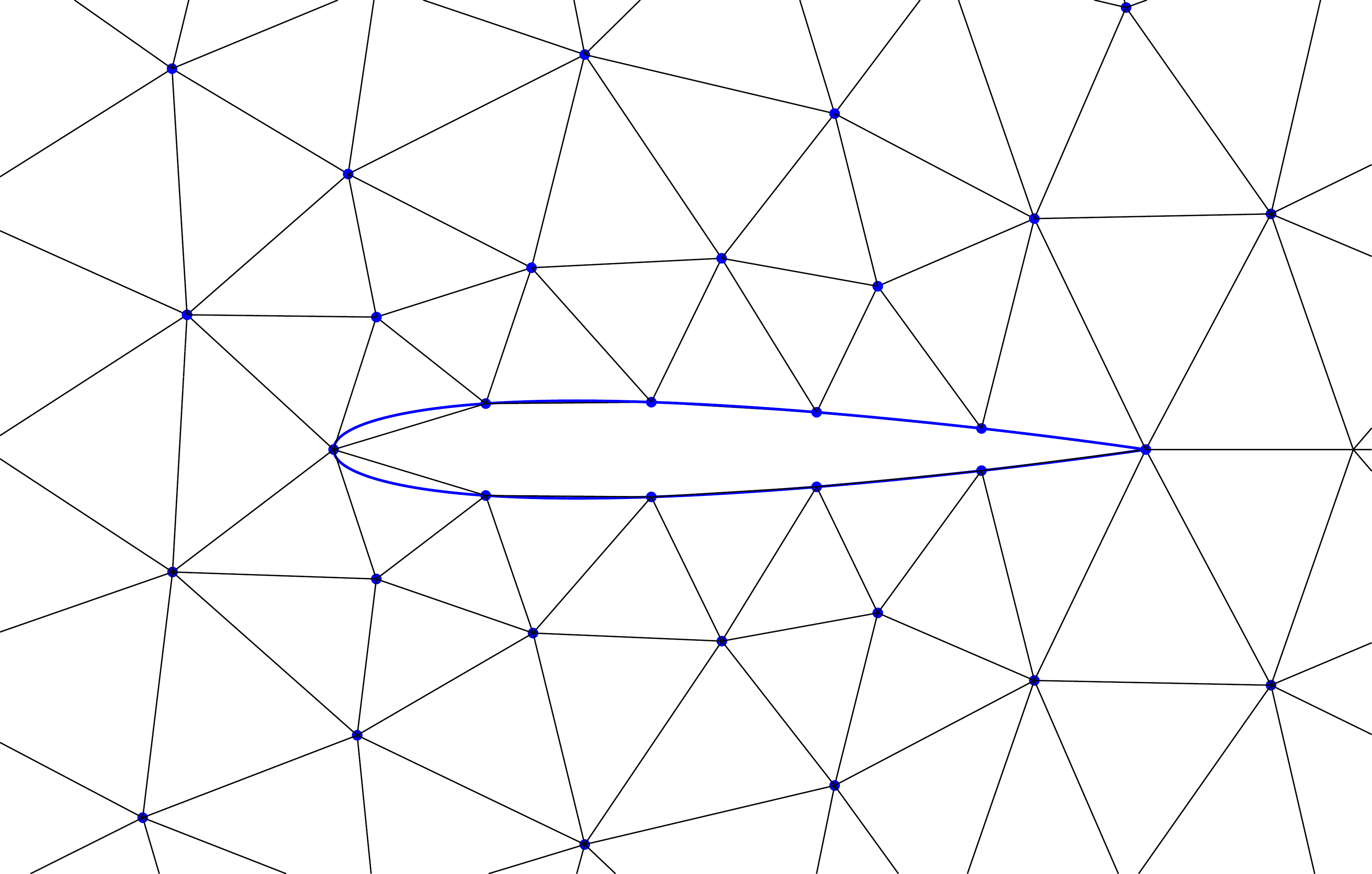}
\\[0.25\baselineskip]
\includegraphics[width=0.475\textwidth]{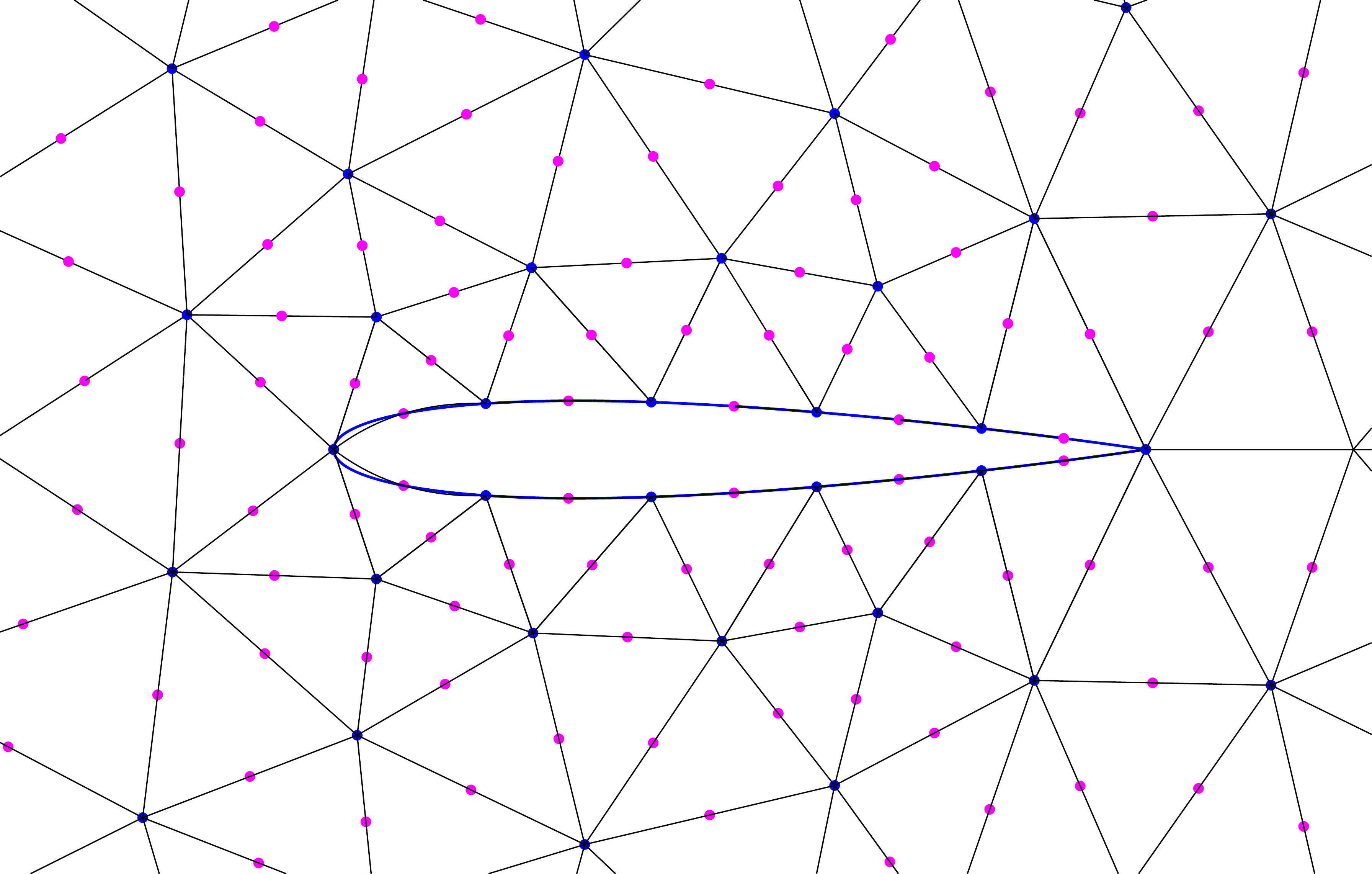}
\includegraphics[width=0.475\textwidth]{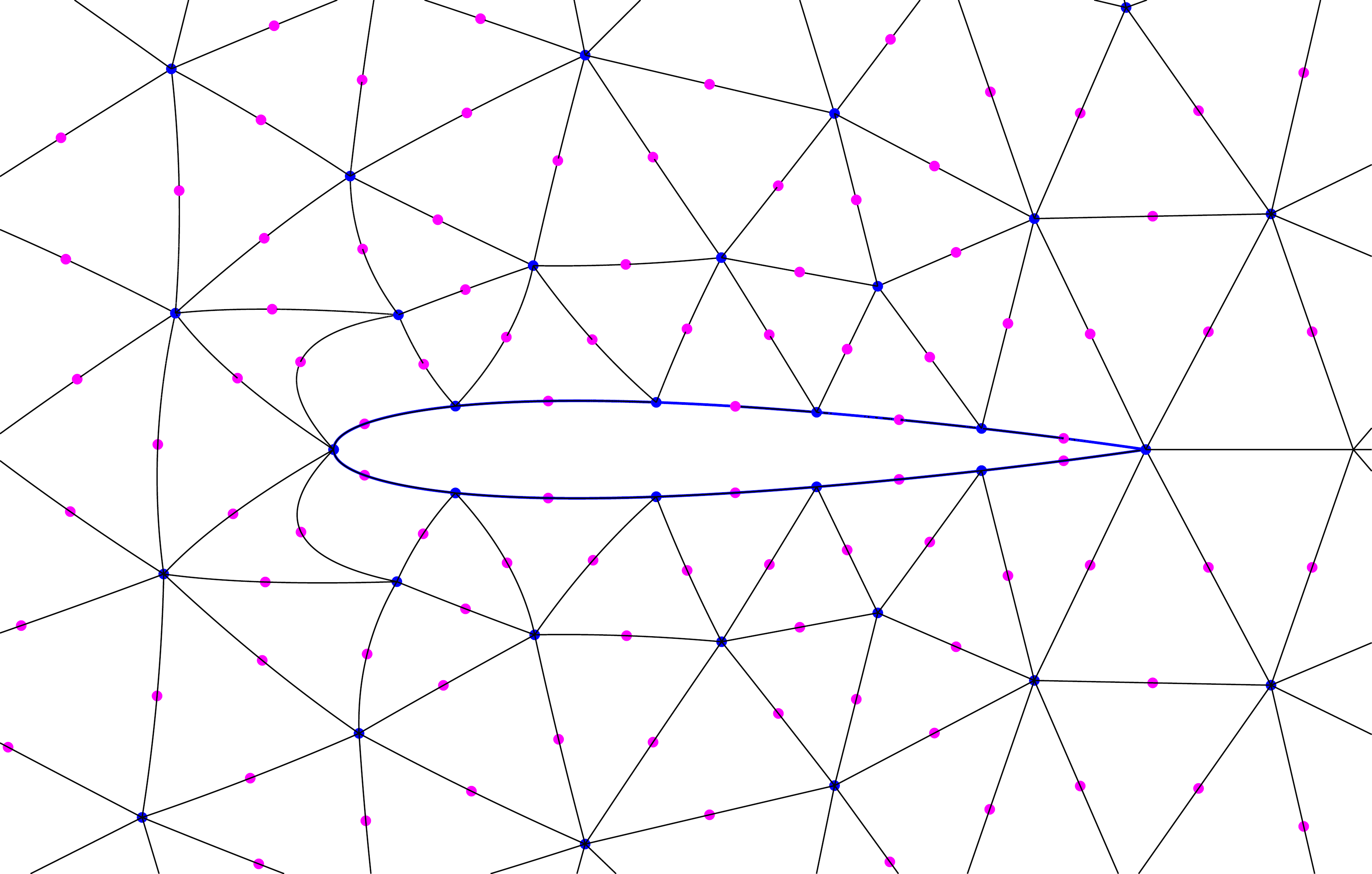}
\\[0.25\baselineskip]
\includegraphics[width=0.475\textwidth]{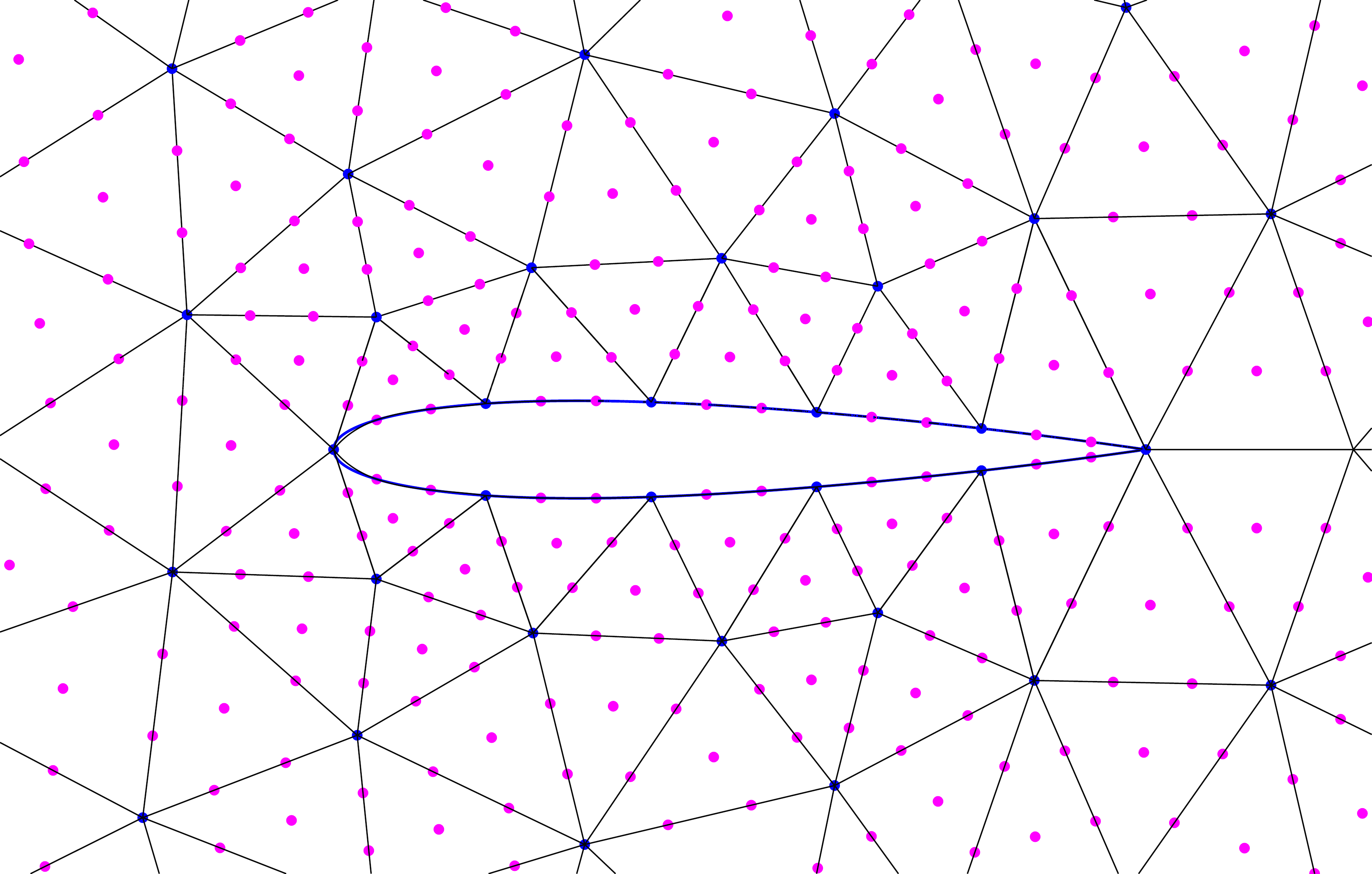}
\includegraphics[width=0.475\textwidth]{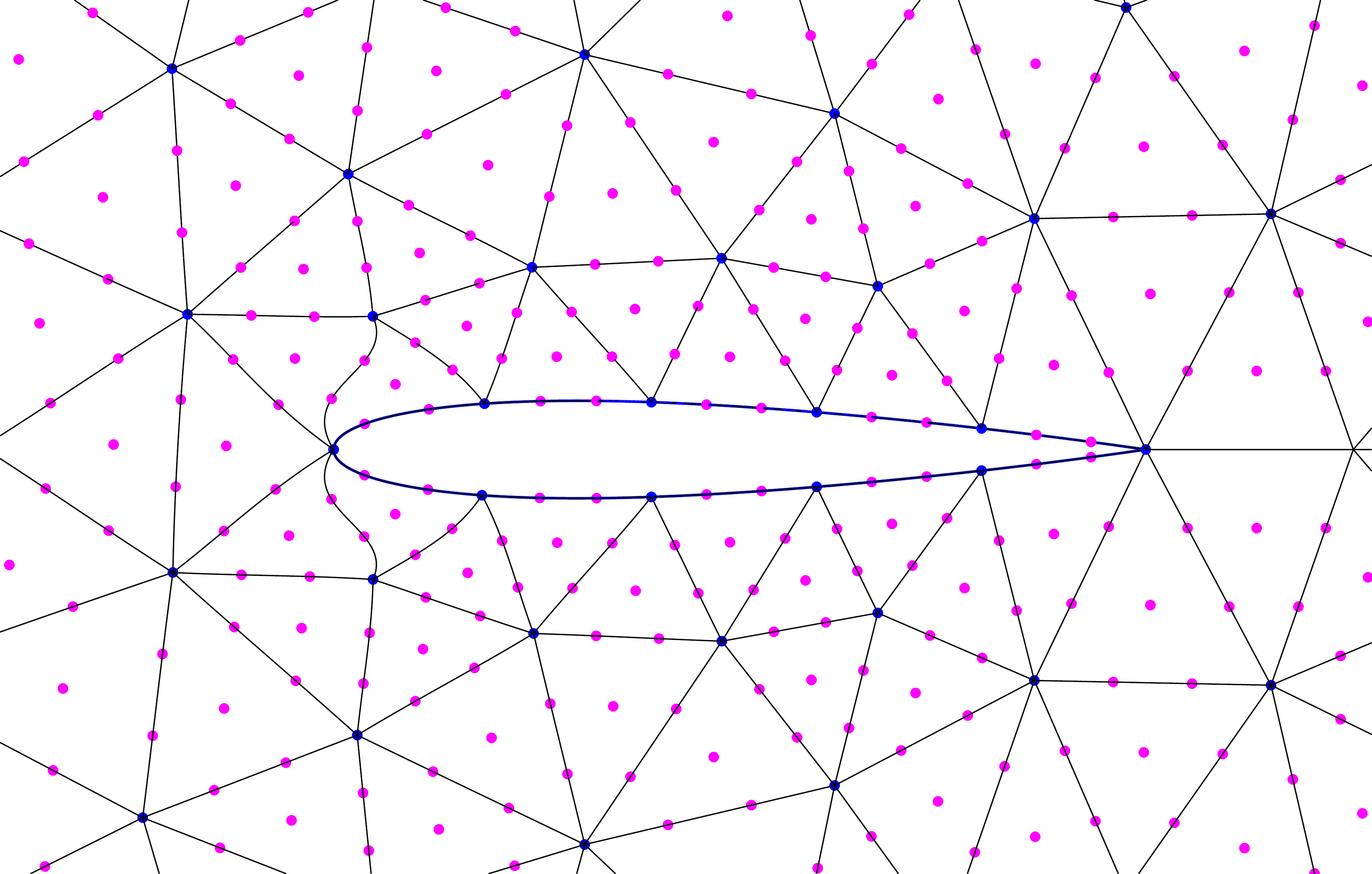}
\end{center}
\caption{Medium-size meshes of the NACA0012 profile. Top: linear mesh.
Center: quadratic meshes optimized for validity only (left) and for
validity as well as geometrical error (right). Bottom: cubic meshes
optimized for validity only (left) and for validity as well as
geometrical error (right).\label{fig:naca_mesh_2}}
\end{figure}

Examples of meshes are shown in Figures~\ref{fig:naca_mesh_0}
and~\ref{fig:naca_mesh_2}. In geometrically-optimized meshes, the
high-order nodes located on the boundary are clearly moved along the CAD
curve to minimize $\delta_T$, while they remain in the middle of corner
nodes when validity only is considered. In coarse meshes, some elements
need to be strongly deformed to satisfy both geometrical and validity
criteria, which may affect the simulations adversely. Indeed, highly
distorted elements are known to harm the accuracy of finite element
approximations~\cite{botti2012influence}. They may also deteriorate the
conditioning of the spatial discretization operator, with negative impact
on time integration~\cite{toulorge2}. Moreover, a correct integration of
polynomial quantities on such elements may require costly higher-order
quadrature rules. Fortunately, the effect is less pronounced in finer
meshes.

In order to illustrate the impact of the geometrical accuracy on
simulations, computations solving the Euler flow around the NACA0012
airfoil at Mach number $M=0.5$ and $3^\circ$ angle-of-attack have been
carried out. A high-order ($p=6$ i.e. sixth-order polynomials)
discontinuous Galerkin scheme was used for the spatial discretization,
and steady-state solutions were obtained through a pseudo-time approach
involving a backward Euler scheme in combination with a Newton-Krylov
solver. Slip wall conditions are imposed on the airfoil and
characteristic-based free-stream boundary conditions are used at the
far-field boundary of the domain.

\begin{figure}
\begin{center}
\includegraphics[width=0.475\textwidth]{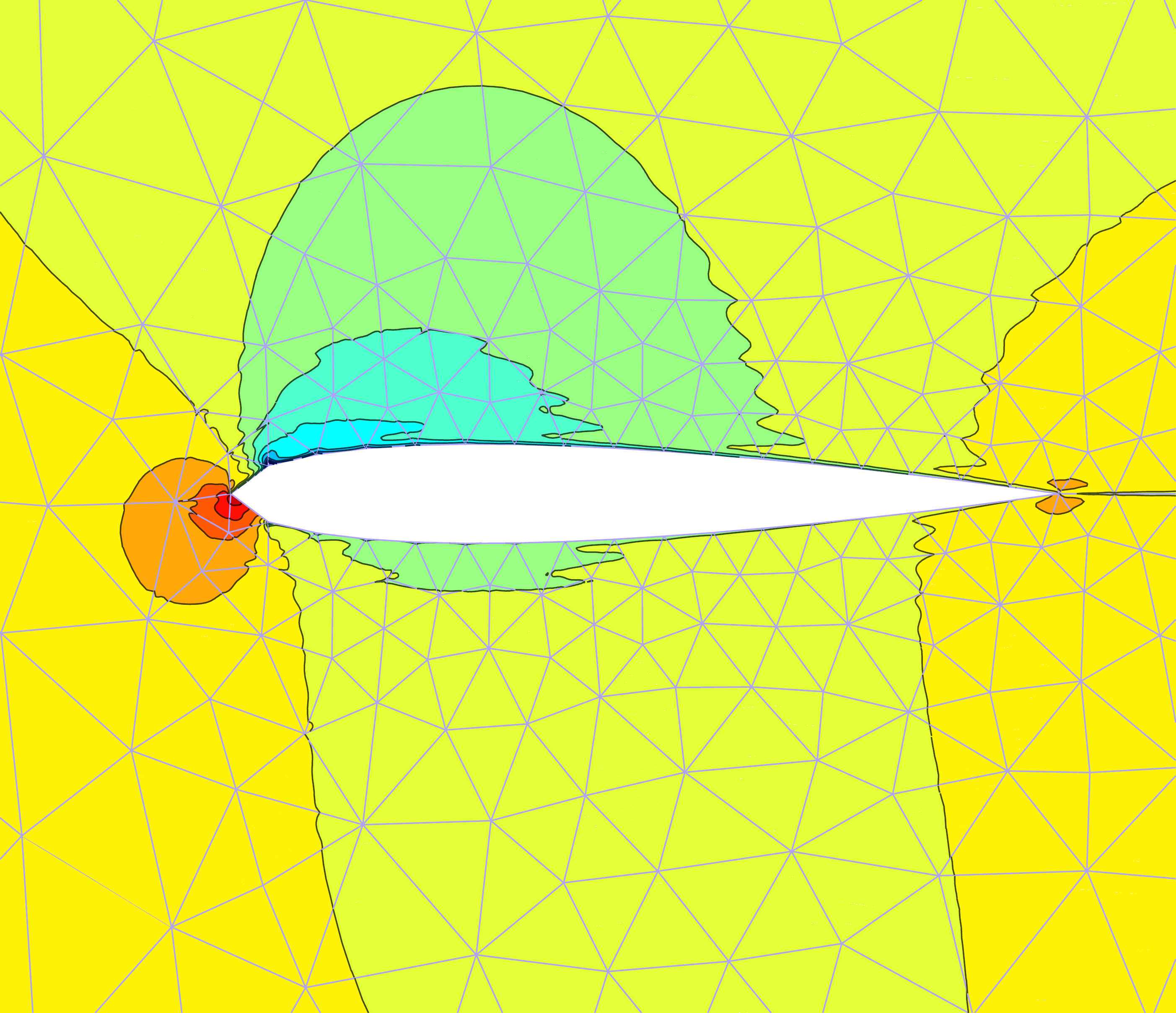}
\includegraphics[width=0.475\textwidth]{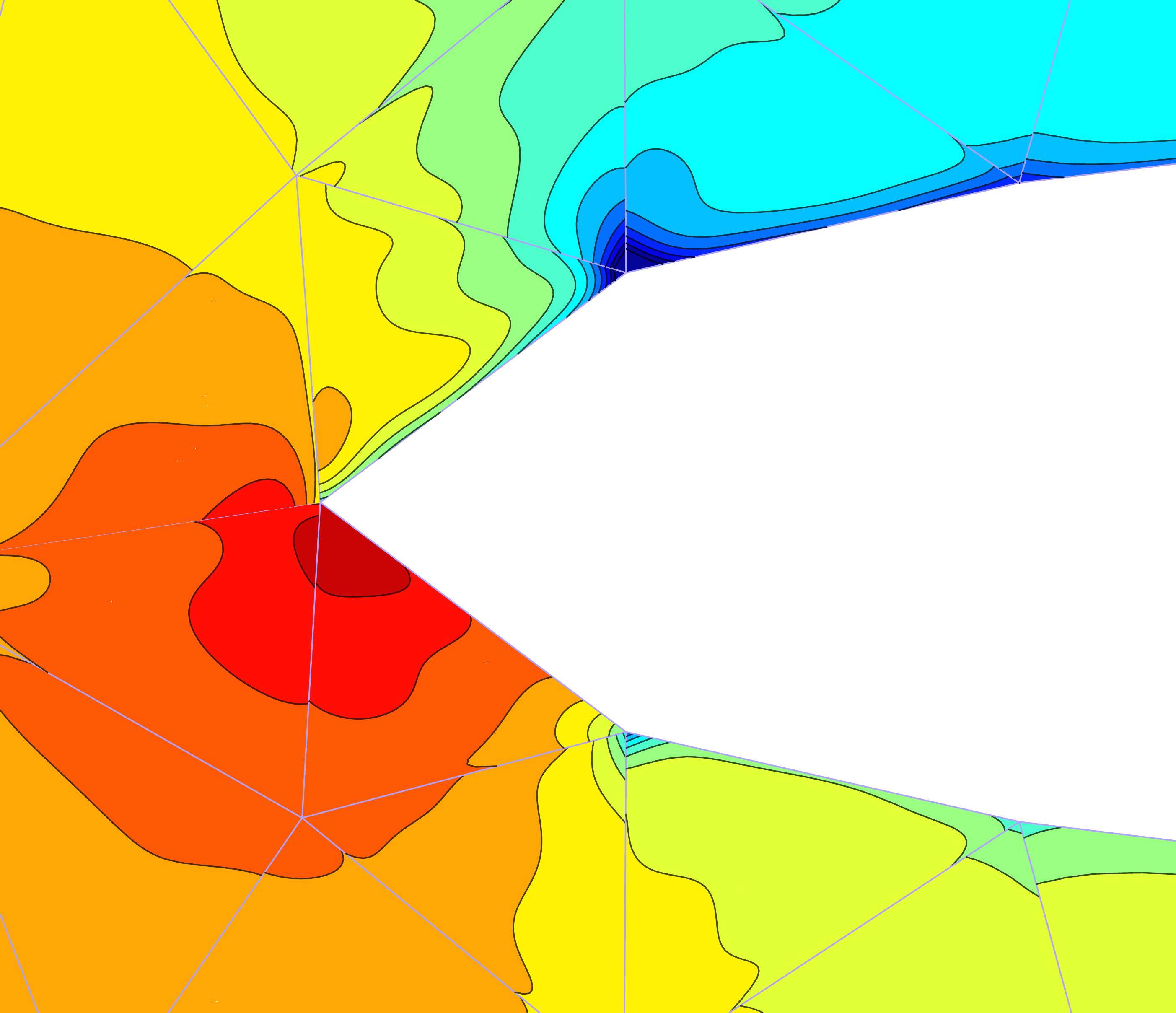}
\\[0.25\baselineskip]
\includegraphics[width=0.475\textwidth]{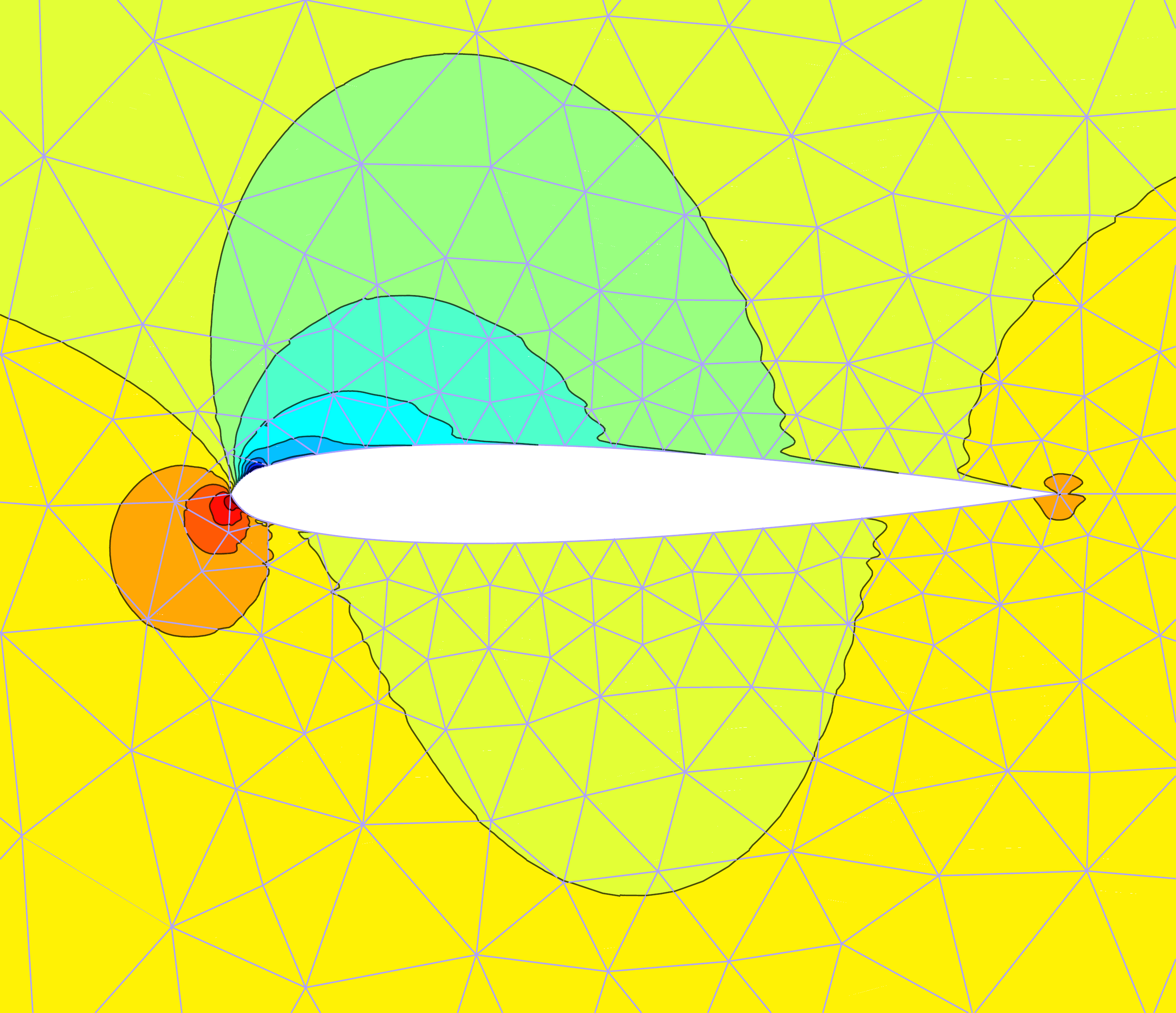}
\includegraphics[width=0.475\textwidth]{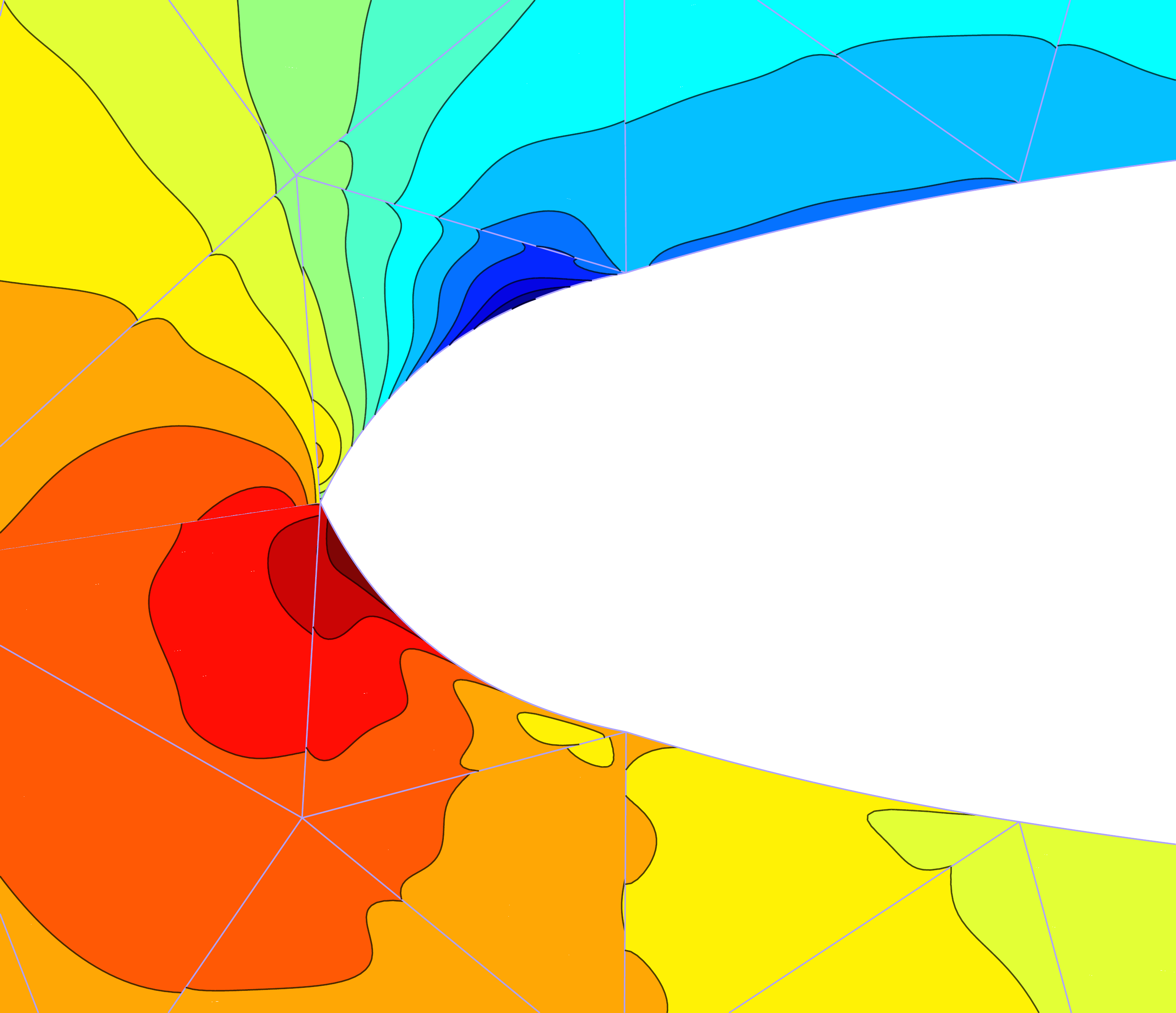}
\\[0.25\baselineskip]
\includegraphics[width=0.475\textwidth]{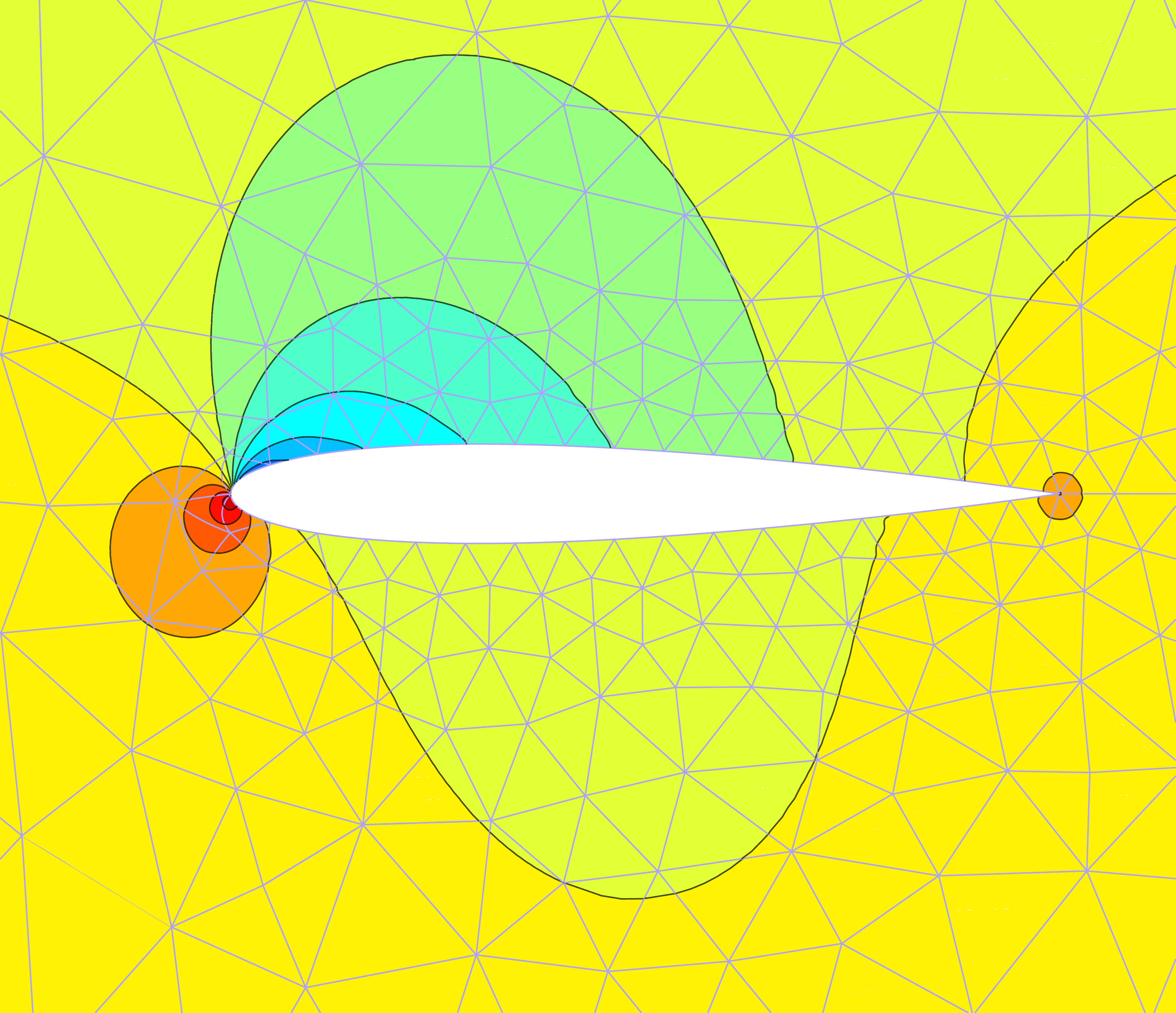}
\includegraphics[width=0.475\textwidth]{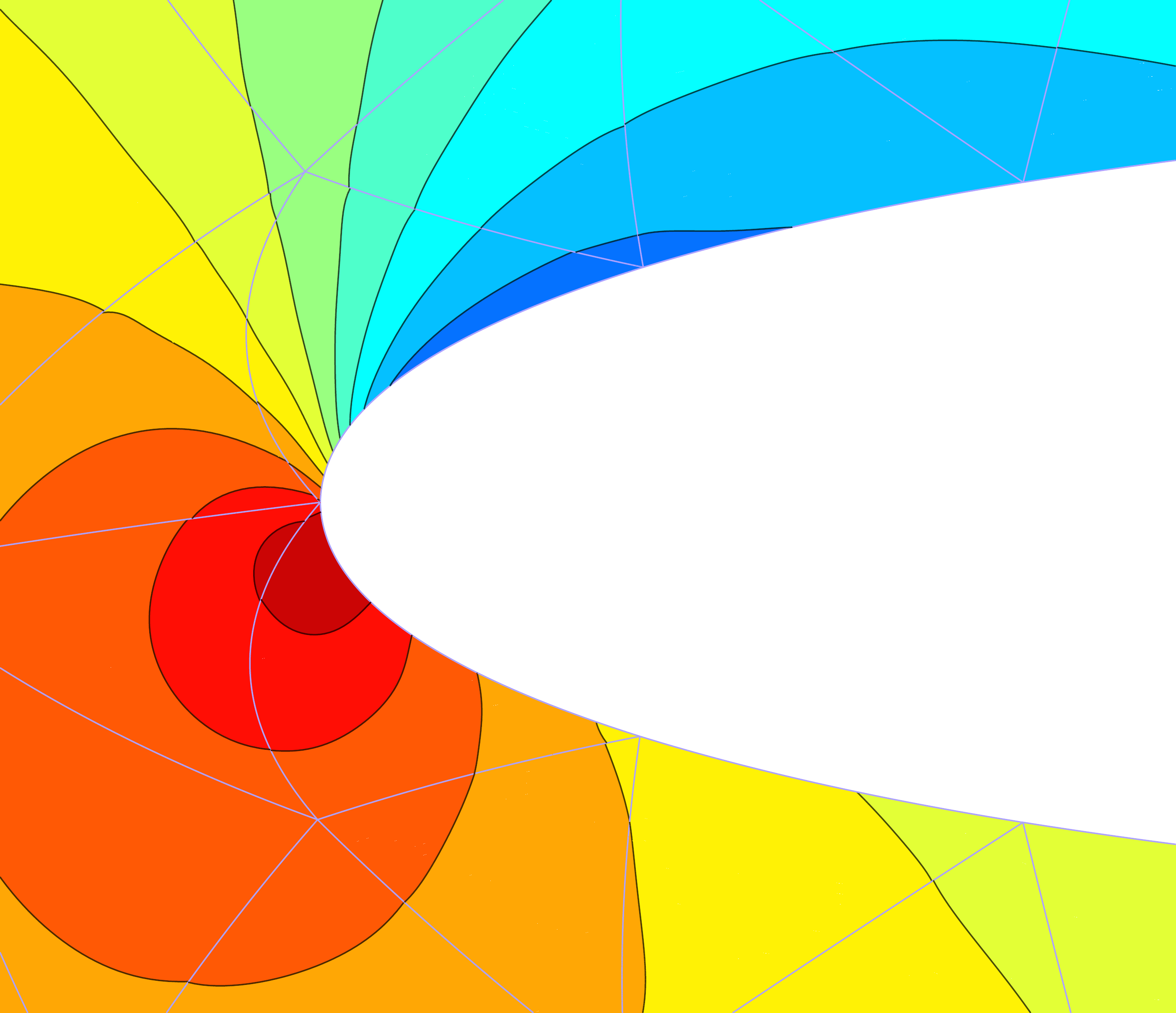}
\end{center}
\caption{Density field for the NACA0012 case: flow around the airfoil
(left column) and zoom around the leading edge (right column). Top row:
results with a linear mesh. Middle row: results with a quadratic mesh
optimized for validity only. Bottom row: results with a quadratic mesh
optimized for both validity and geometrical accuracy.
\label{fig:naca_simu}}
\end{figure}

Results for meshes in which the airfoil is discretized with 34
elements are shown in Figure~\ref{fig:naca_simu}. Unsurprisingly, the
numerical method does not converge properly with the linear mesh, and
the residual cannot be decreased by more than two orders of magnitude.
The density field is clearly different from the expected solution.
With the quadratic mesh optimized for validity only, the airfoil is
represented more accurately, but the corresponding solution still
exhibits spurious oscillations and flow features near corner nodes on
the wall boundary, where the representation of the airfoil is not
smooth. A drop of four orders of magnitude in residual is achieved in
26 pseudo-time iterations. With the geometrically-optimized mesh
however, the computation converges towards the expected smooth
solution in 19 pseudo-time iterations. In this purely inviscid
test case, the increased boundary smoothness resulting from the
minimization of $\delta_T$ is instrumental in converging towards the
exact solution without spurious entropy generation at the boundary.
Moreover, the geometrically-optimized quadratic mesh represents the
model so accurately that it is meaningless to use a higher-order
mesh, as the Hausdorff distance $\delta_H\approx 4\cdot 10^{-5}$ is
probably already lower than the manufacturing tolerance of the
airfoil.

\subsection{Rattray island}

\begin{figure}
\begin{center}
\includegraphics[width=0.4\textwidth]{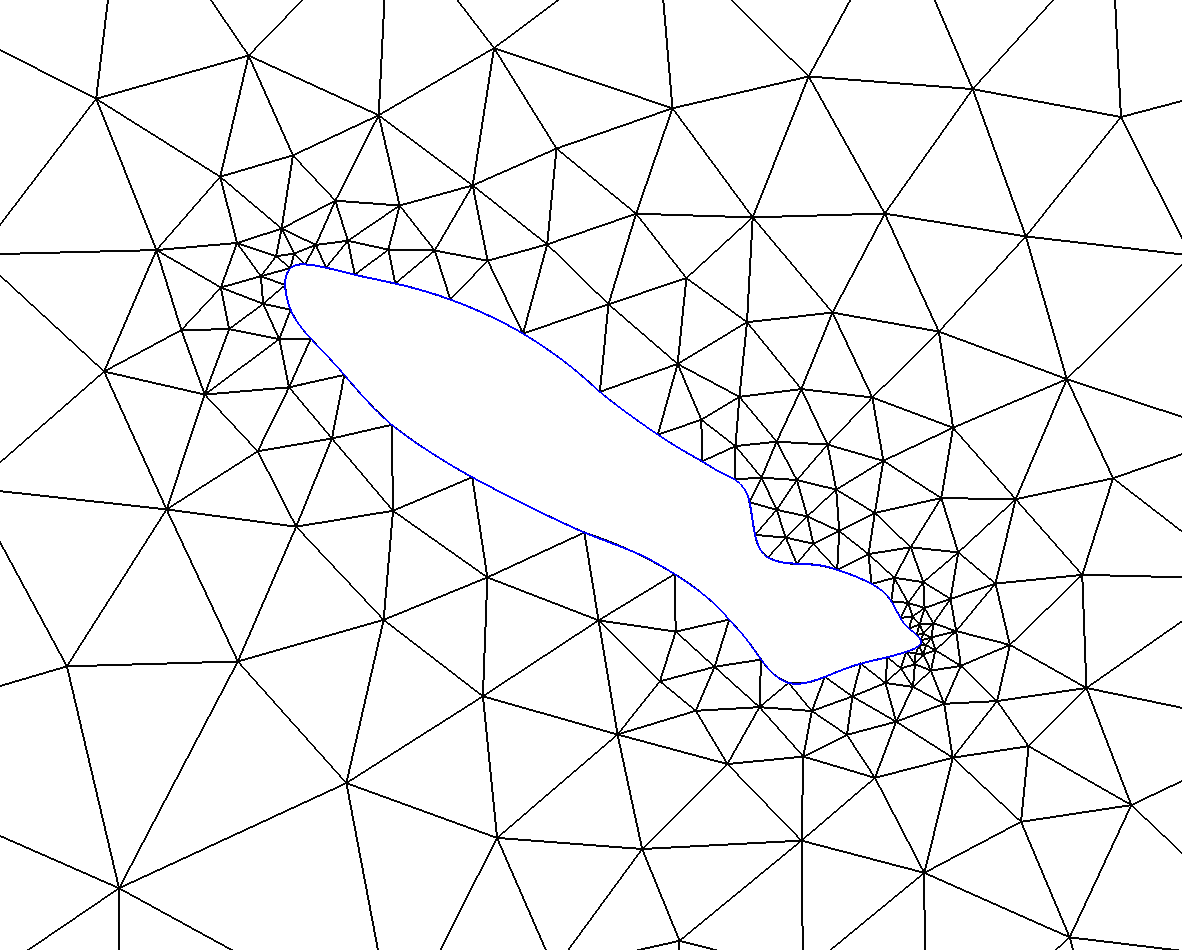}
\\[0.25\baselineskip]
\includegraphics[width=0.4\textwidth]{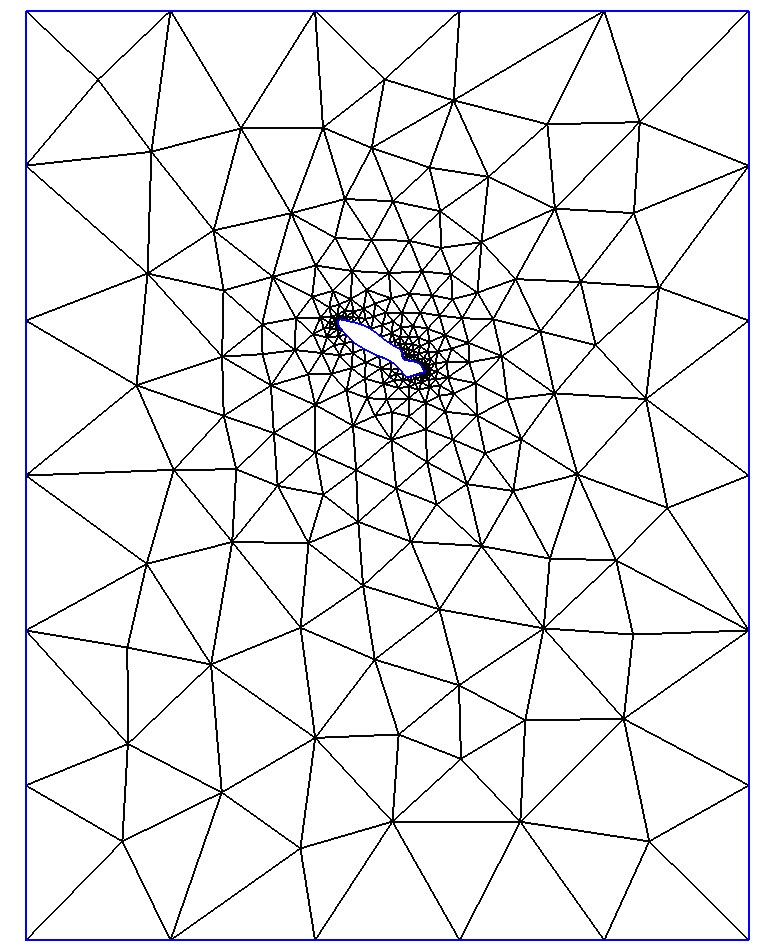}
\includegraphics[width=0.4\textwidth]{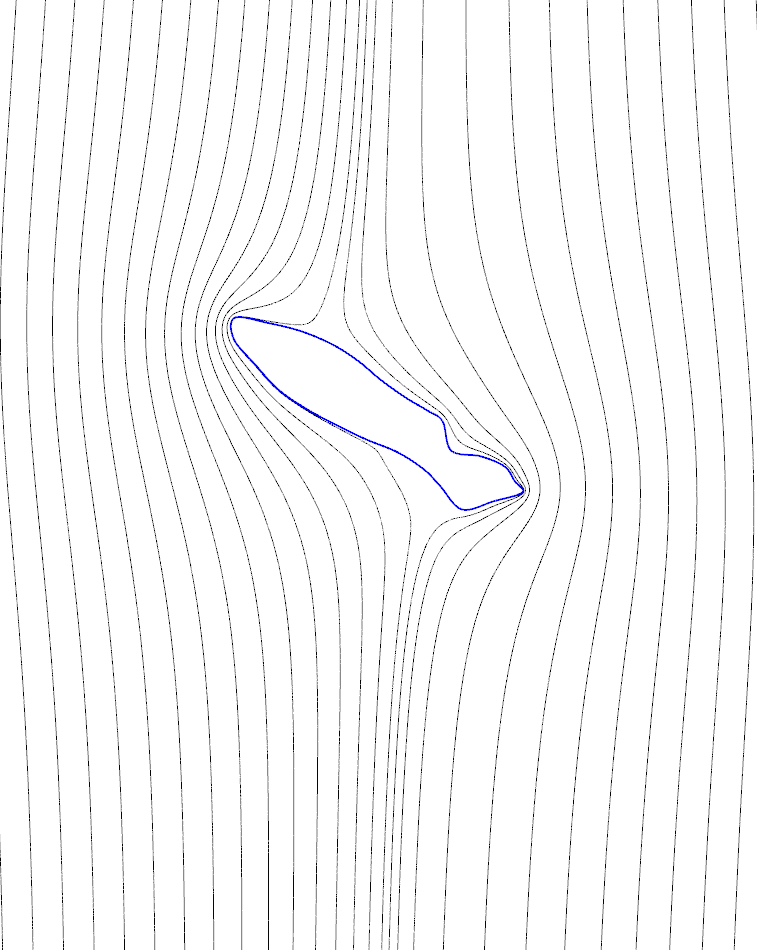}
\end{center}
\caption{Rattray test case: general view of the computational domain
and quadratic mesh (bottom left), zoom on the mesh around the island
(top) and streamlines of the ideal solution in the vicinity of the
island (right).
\label{fig:rattray}}
\end{figure}

We consider now an ocean modelling application focusing on the
Rattray island, that is located in the Great Barrier Reef near
Australia. Simulations have been performed, in which the shallow water
equations are solved without diffusion nor Coriolis force. The
water depth at rest is uniform and equal to $25\,\textrm{m}$. Slip
wall conditions are imposed on the island coast and the lateral sides
of the domain. At the upstream and downstream sides of the domain,
uniform free-stream conditions are prescribed with a velocity
corresponding to a Froude number of $Fr=0.02$, which
is representative of the tidal stream~\cite{wolanski1994}. The
island, that has a length of about $1350\,\textrm{m}$, is oriented
at $60^\circ$ compared to the free stream. In this setup, there
is no source of vorticity, and the ideal solution is a steady
irrotational flow. A view of the domain and the solution is given
in Figure~\ref{fig:rattray}.

\begin{figure}
\begin{center}
\includegraphics[width=0.475\textwidth]{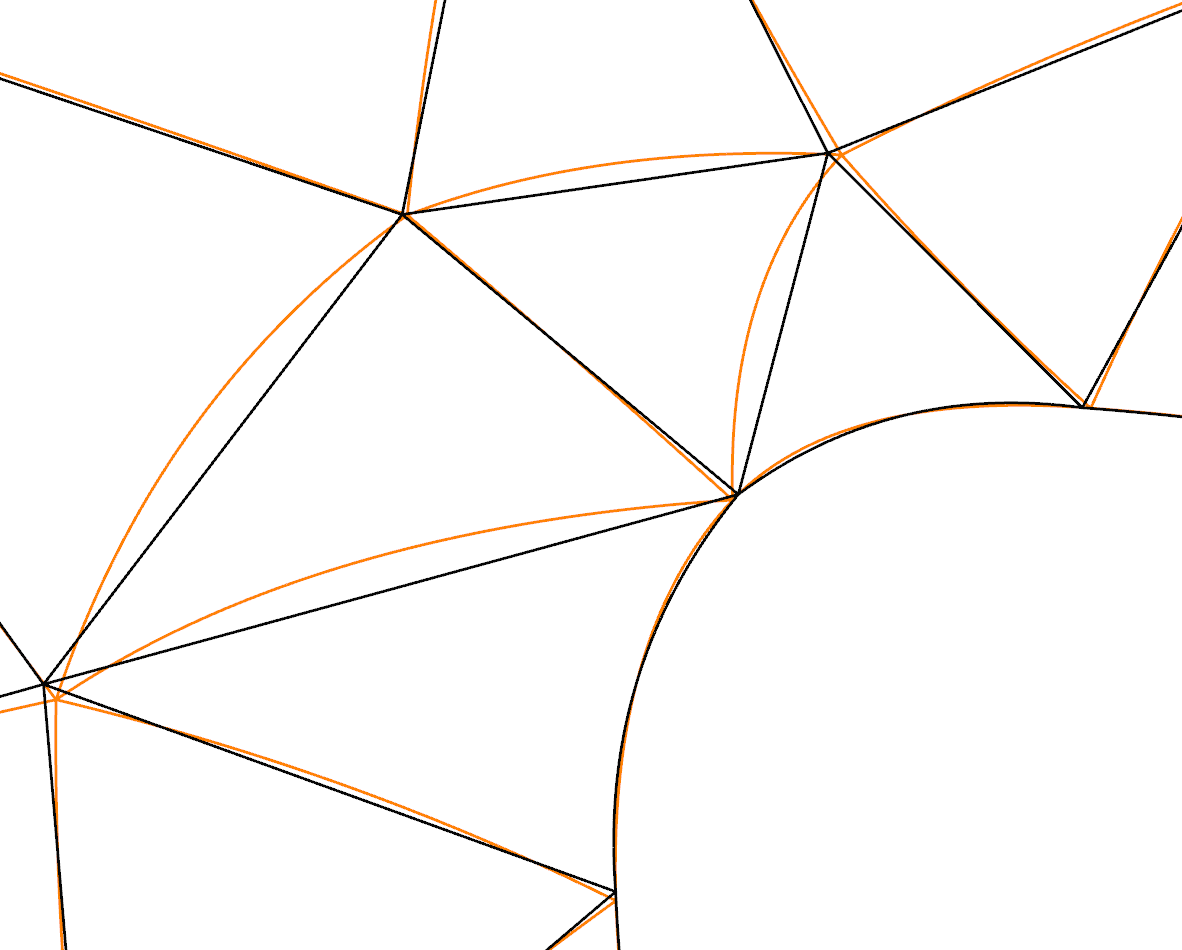}
\includegraphics[width=0.475\textwidth]{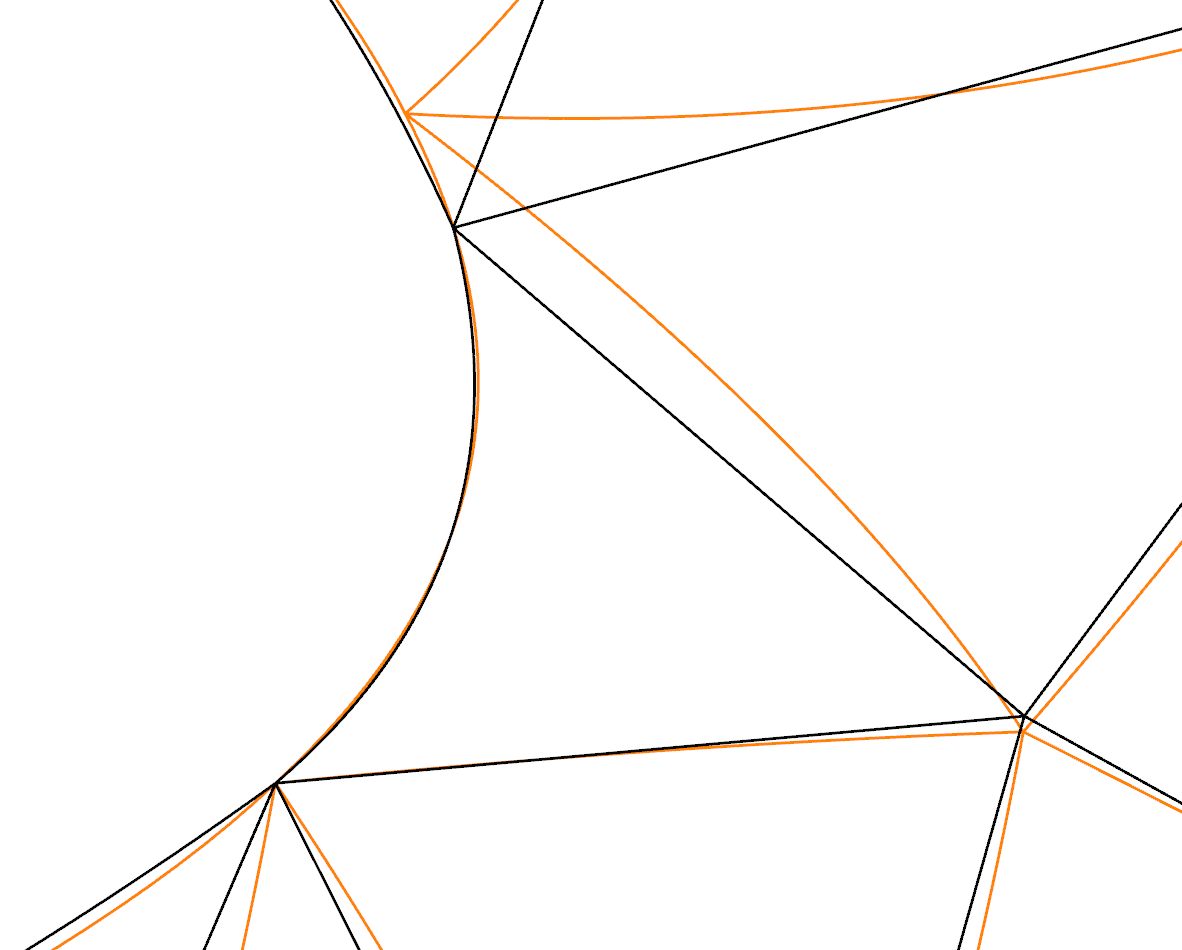}
\end{center}
\caption{Rattray test case: comparison between the original quadratic
mesh (black) and the geometrically-optimized mesh (orange) at the
upstream tip (left) and the downstream tip (right) of the island.
\label{fig:rattray_meshComp}}
\end{figure}

A quadratic mesh of $596$ elements with higher density near curved
boundaries is generated. This mesh is already valid without
optimization. A second mesh is obtained by minimizing $\delta_T$
according to the procedure described in Section~\ref{sec:opti}.
Figure~\ref{fig:rattray_meshComp} shows a comparison of both meshes
at the tips of the island. The mesh boundaries of both meshes look
very similar. Indeed, the model-to-mesh Hausdorff distance
for the original mesh ($\delta_H=1.45\,\textrm{m}$) is not
significantly different from the one for the optimized mesh
($\delta_H=0.66\,\textrm{m}$). This is due to the fact that the
geometry is already well resolved by the unoptimized mesh. However,
a close examination of Figure~\ref{fig:rattray_meshComp} reveals that
the boundary of the unoptimized mesh is not perfectly smooth at
element corners, while the optimized mesh looks better in this respect.

\begin{figure}
\begin{center}
\includegraphics[width=0.475\textwidth]{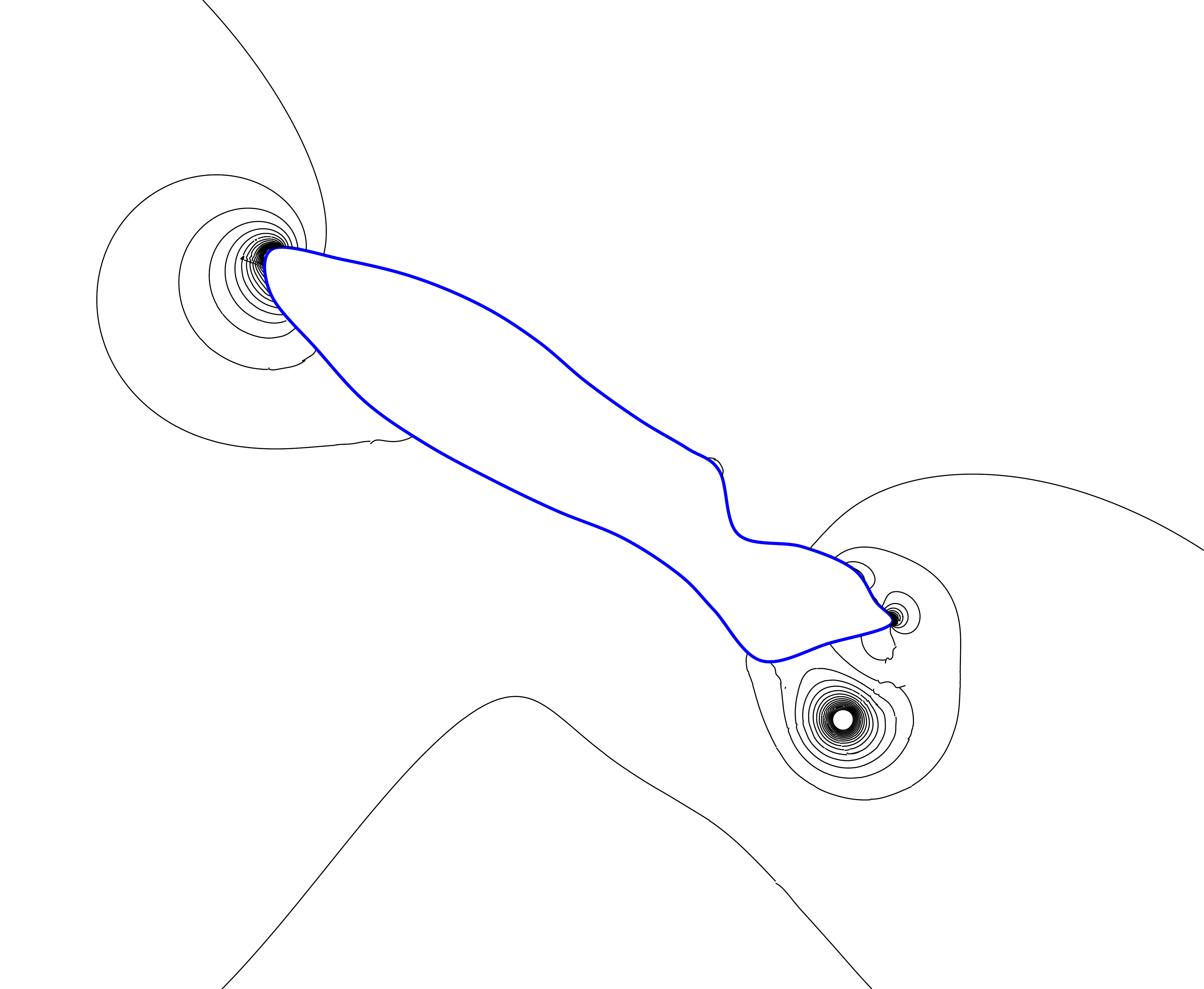}
\includegraphics[width=0.475\textwidth]{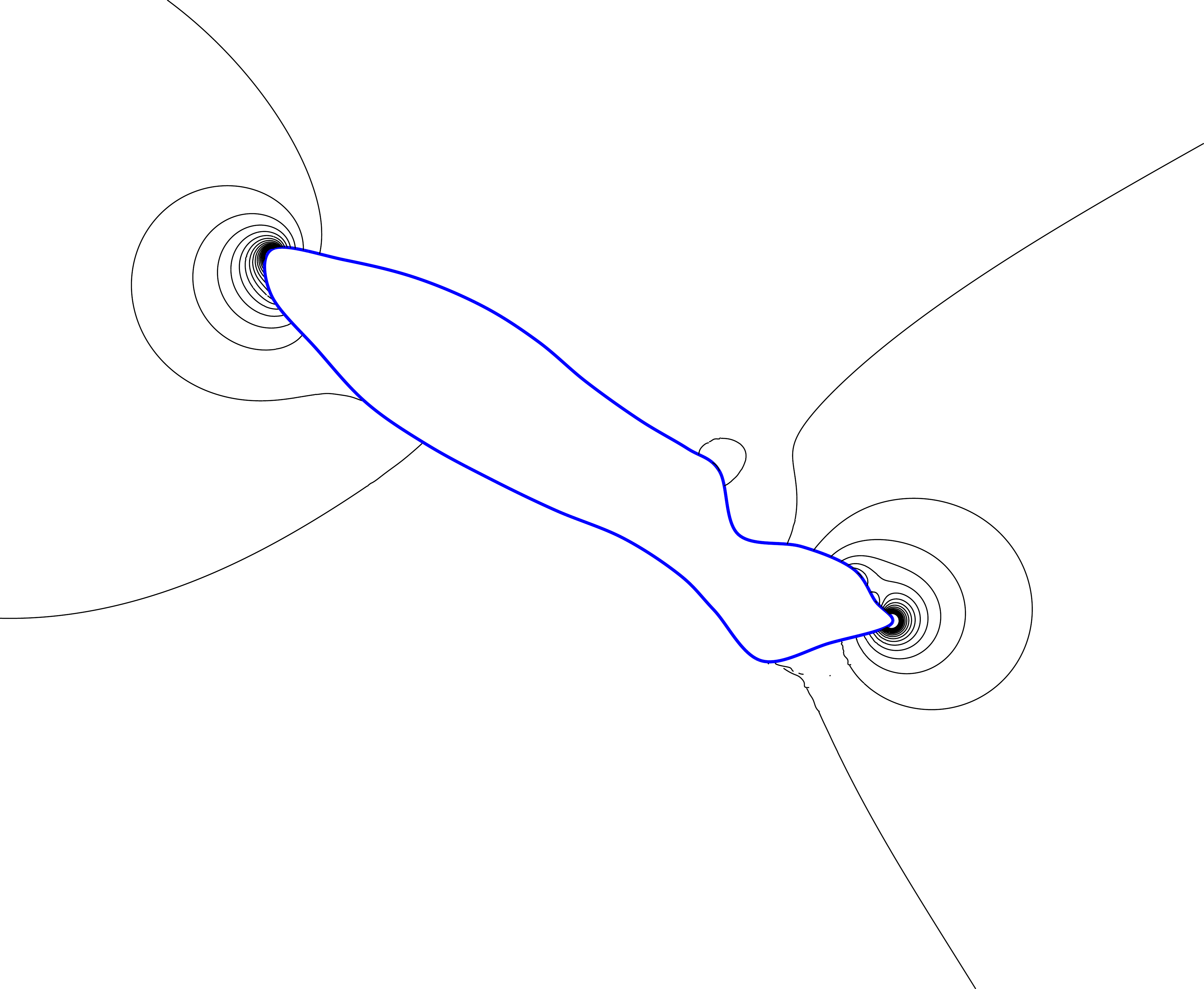}
\\[0.25\baselineskip]
\includegraphics[width=0.475\textwidth]{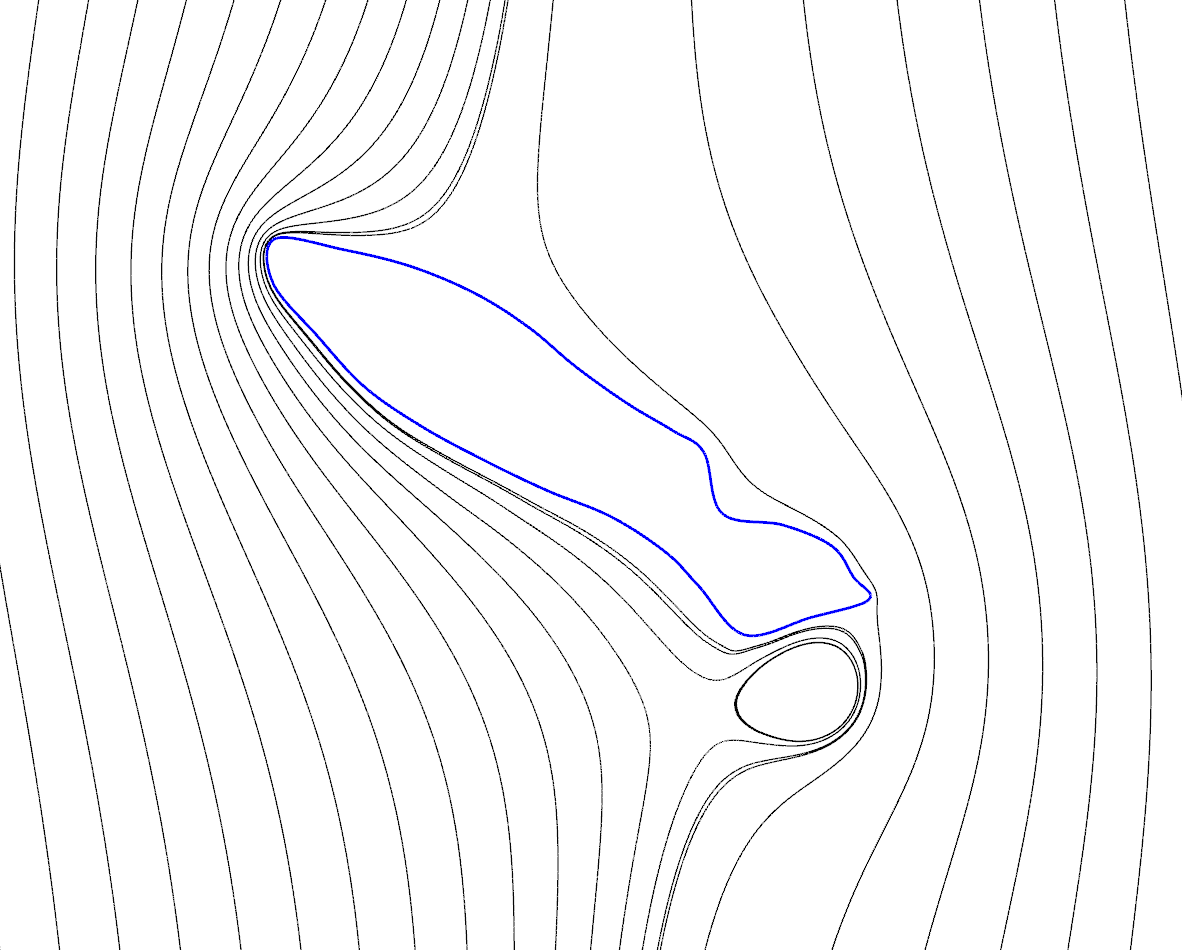}
\includegraphics[width=0.475\textwidth]{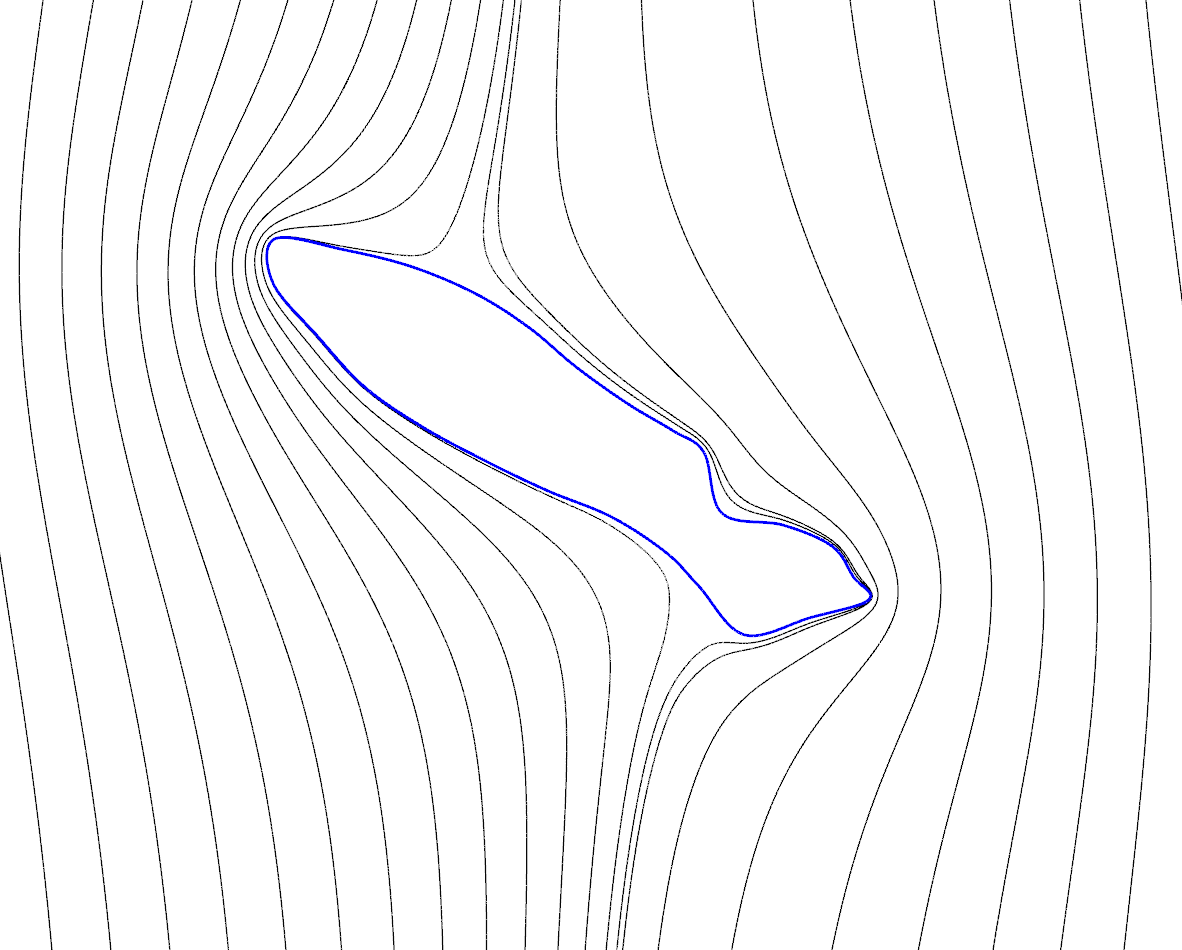}
\end{center}
\caption{Rattray test case: contours of the sea surface elevation
(upper row) and streamlines (bottom row) at $t=2100\,\textrm{s}$.
Results obtained with the original quadratic mesh (left column) and
with the geometrically-optimized mesh (right column).
\label{fig:rattray_solComp}}
\end{figure}

Even though the improvement in the Taylor-based geometrical error
is not necessarily impressive ($\delta_T=2.19\,\textrm{m}$ for the unoptimized
mesh against $\delta_T=0.37\,\textrm{m}$ for the optimized one), the impact
on the solution is important. Simulations were performed with the same
numerical method as in Section~\ref{sec:naca}. Figure~\ref{fig:rattray_solComp}
compares the results between both meshes. With the unoptimized mesh,
vortices are shed from the downstream tip of the island, preventing
the flow from reaching steady-state, while a drop of 4 orders of
magnitude in residuals can be achieved with the optimized mesh,
leading to a nearly potential solution. This test case shows, even
more than in the NACA0012 case, that the gain in boundary smoothness
brought by the geometrical optimization is crucial to obtain the
correct solution in some problems.

\subsection{High-lift airfoil}

\begin{figure}
\begin{center}
\includegraphics[width=0.65\textwidth]{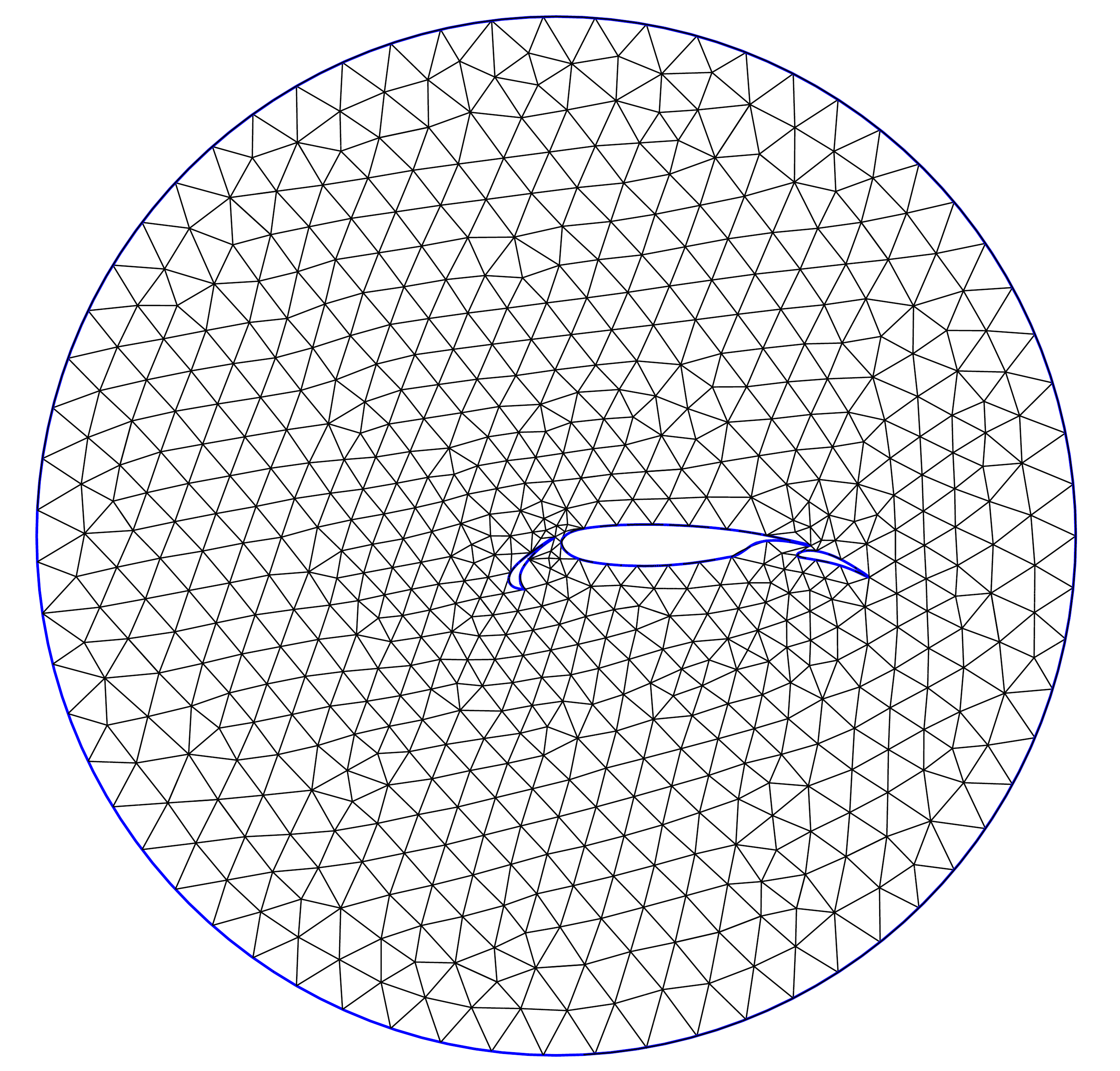}
\end{center}
\caption{Acoustics test case: computational domain and medium-size
quadratic mesh (left).
\label{fig:ra16sc1_domain}}
\end{figure}

In this section, we apply the methods described in
Section~\ref{sec:opti} to an acoustic application involving a
high-lift airfoil. The geometry is a 3-element airfoil based on the
RA16SC1 profile, with the slat and flap deflected by $30^{\circ}$ and
$20^{\circ}$ respectively. The chord of the main element is
$480\,\textrm{mm}$, and the computational domain is a disc of radius
$1\,\textrm{m}$ centered on a point P located close to the trailing edge.
The acoustic excitation consists of a monopole source placed at point P,
with an amplitude of $1\,\textrm{Pa}$ and frequency of
$7816\,\textrm{Hz}$. The computational domain is shown in
Figure~\ref{fig:ra16sc1_domain}.

Simulations are performed with a discontinuous Galerkin nodal scheme
for the space discretization, and the standard fourth-order
four-stage Runge-Kutta integrator in time. Slip wall conditions are
imposed on the airfoil, while characteristic-based non-reflecting
boundary conditions are prescribed in the far field. The monopole
is modeled by a Gaussian pressure perturbation of half-width
$3\,\textrm{mm}$. Simulation are run until a periodic regime is
reached. A reference solution is obtained through a computation on a
fine grid.

\begin{figure}
\begin{center}
\includegraphics[width=0.425\textwidth]{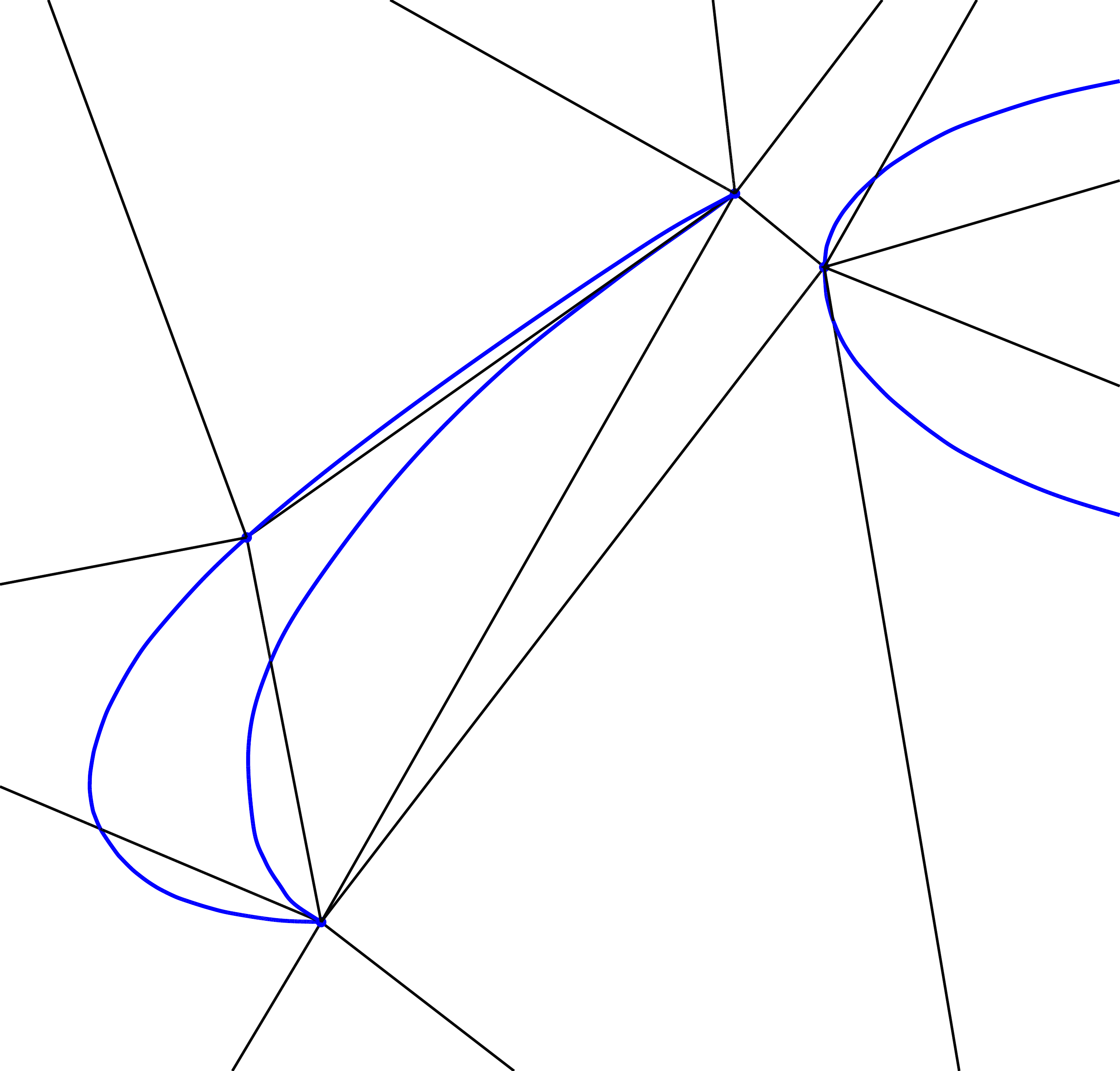}
\includegraphics[width=0.425\textwidth]{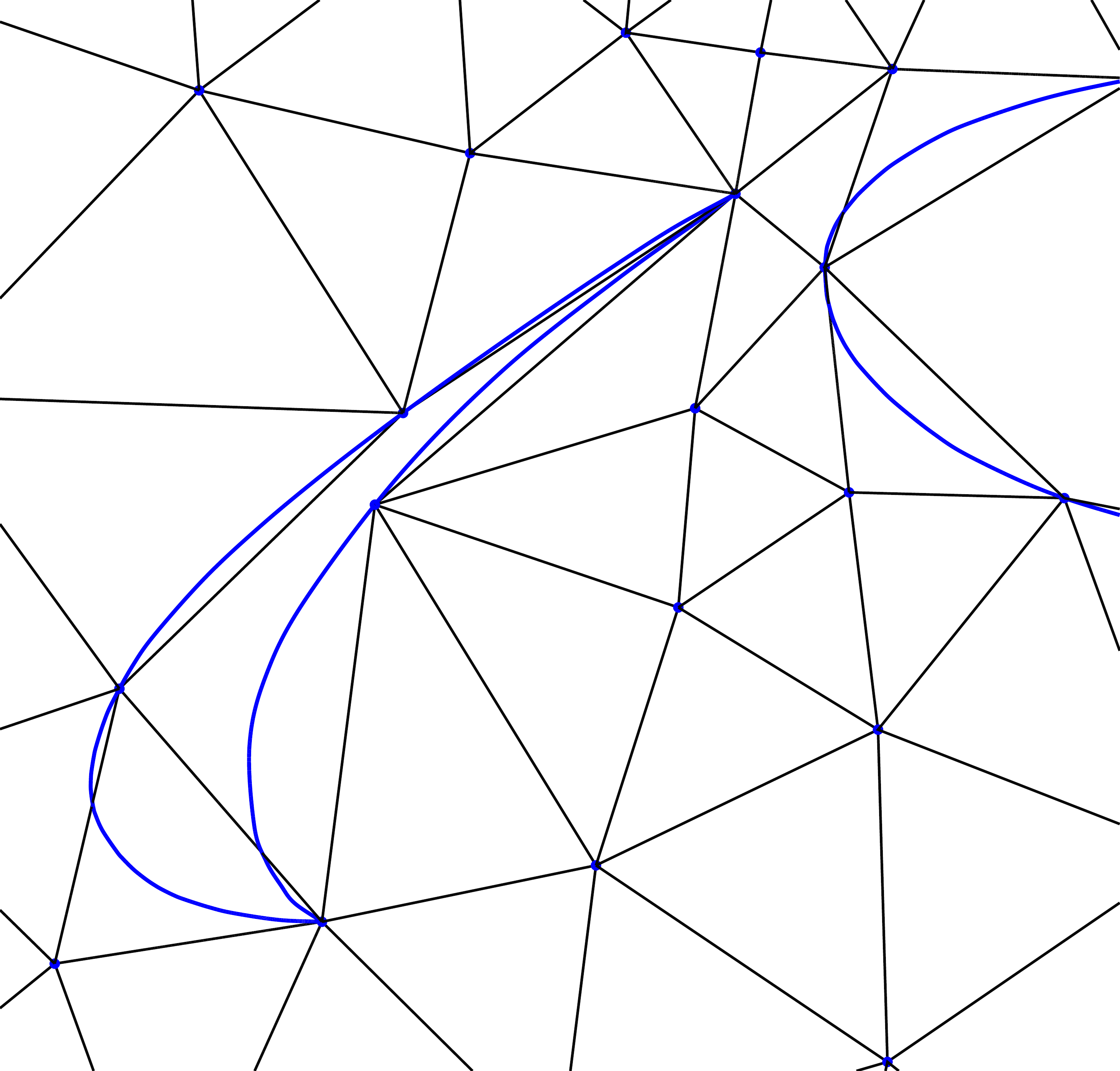}
\\[0.25\baselineskip]
\includegraphics[width=0.425\textwidth]{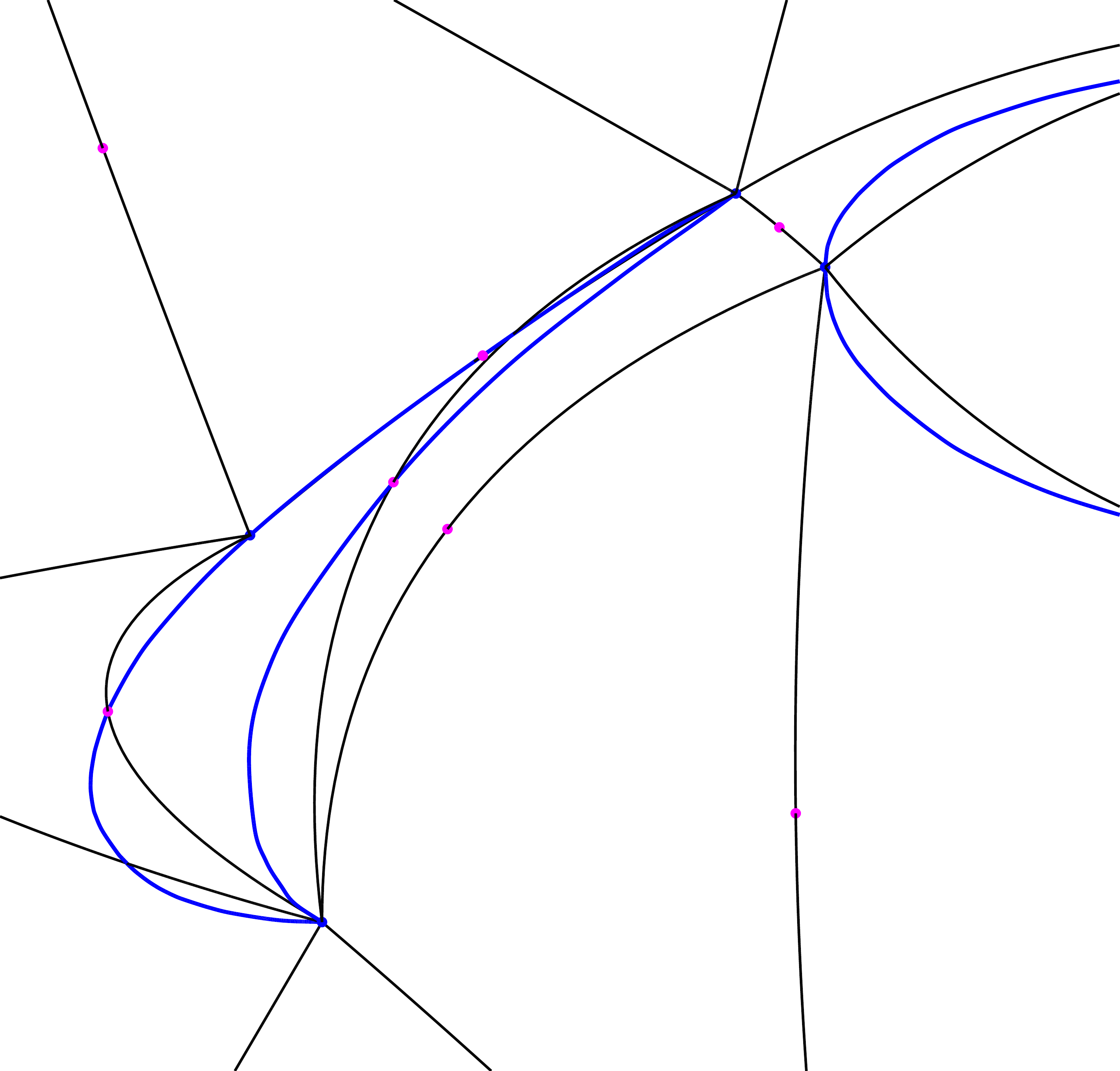}
\includegraphics[width=0.425\textwidth]{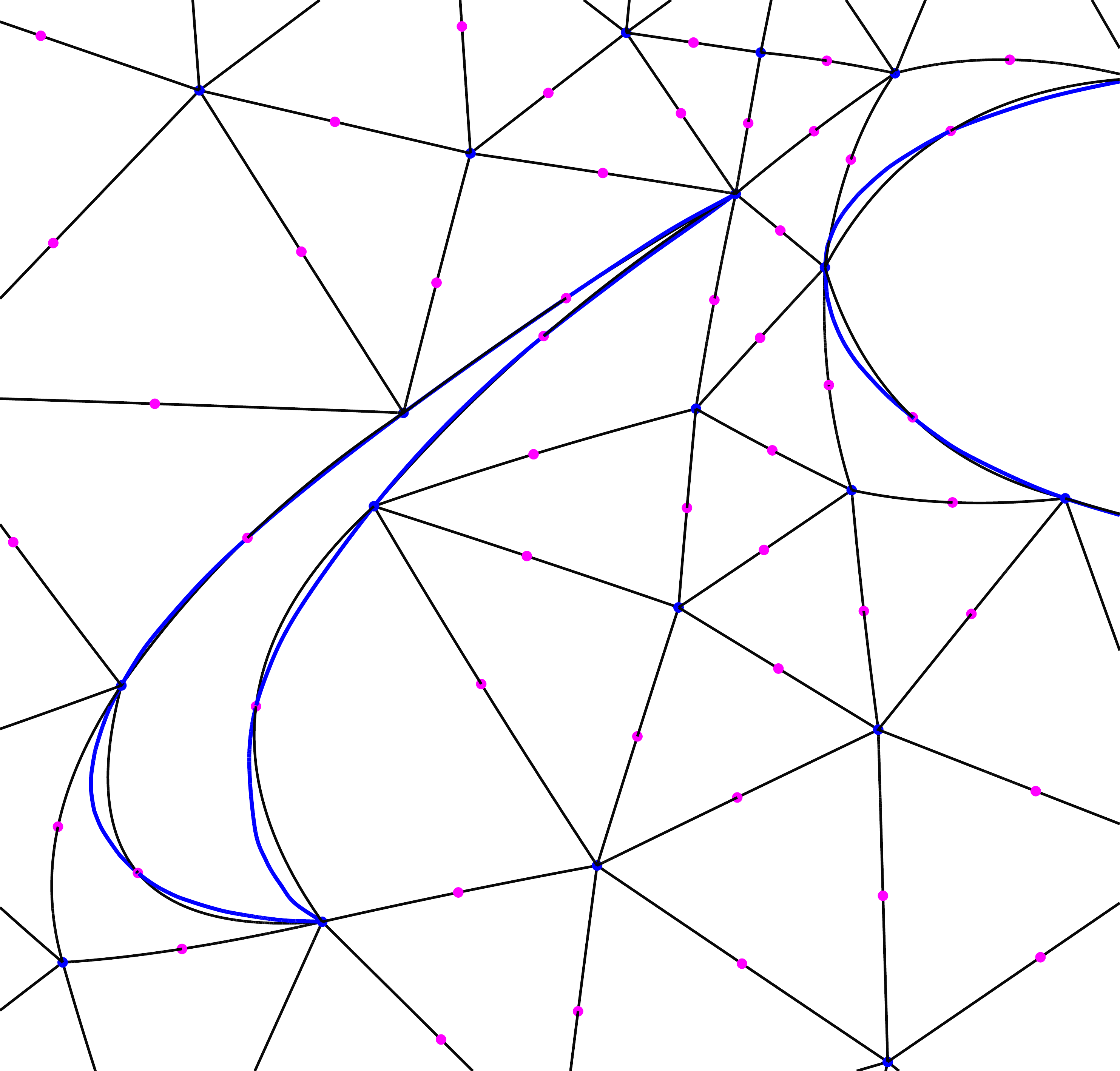}
\\[0.25\baselineskip]
\includegraphics[width=0.425\textwidth]{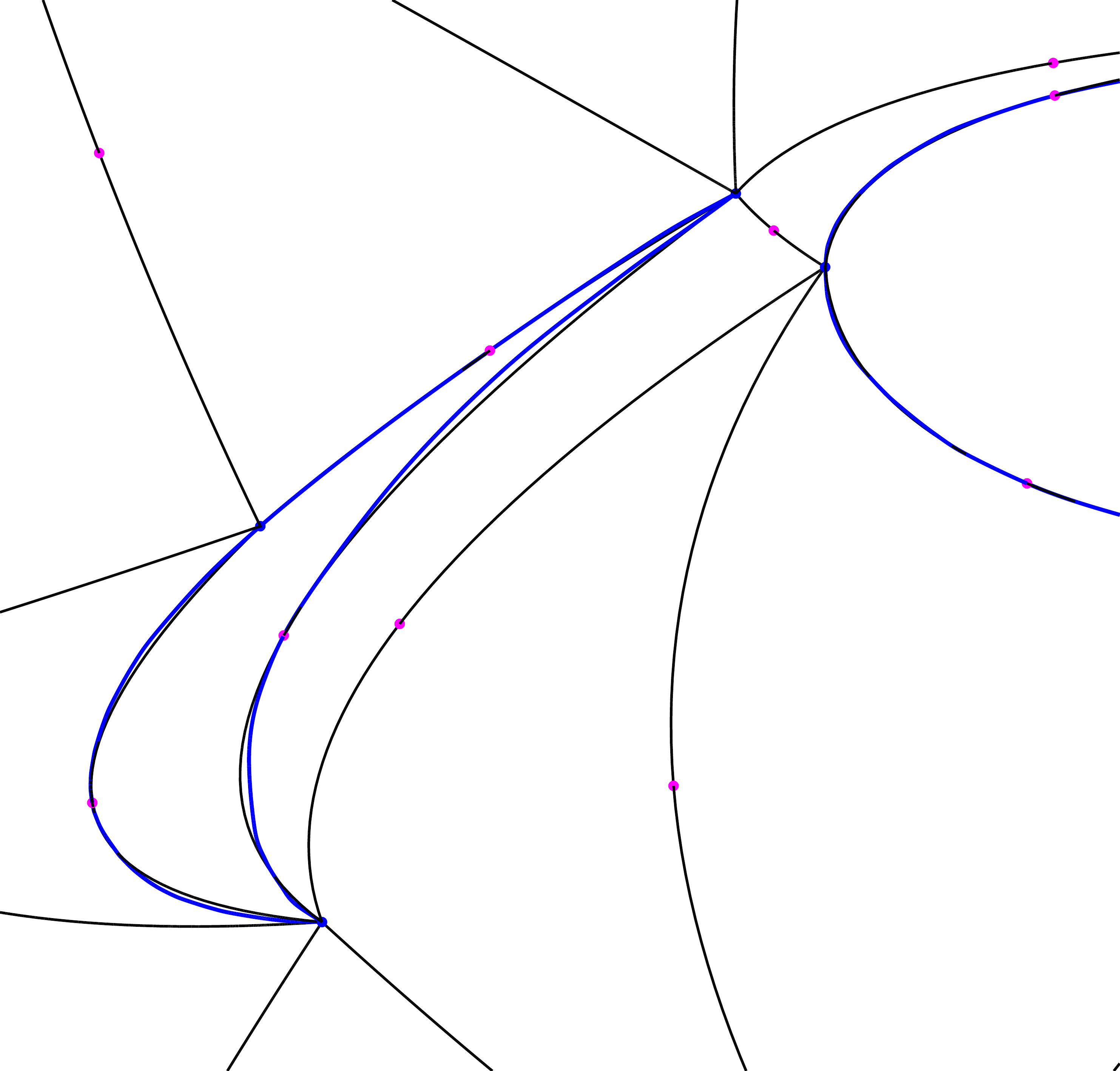}
\includegraphics[width=0.425\textwidth]{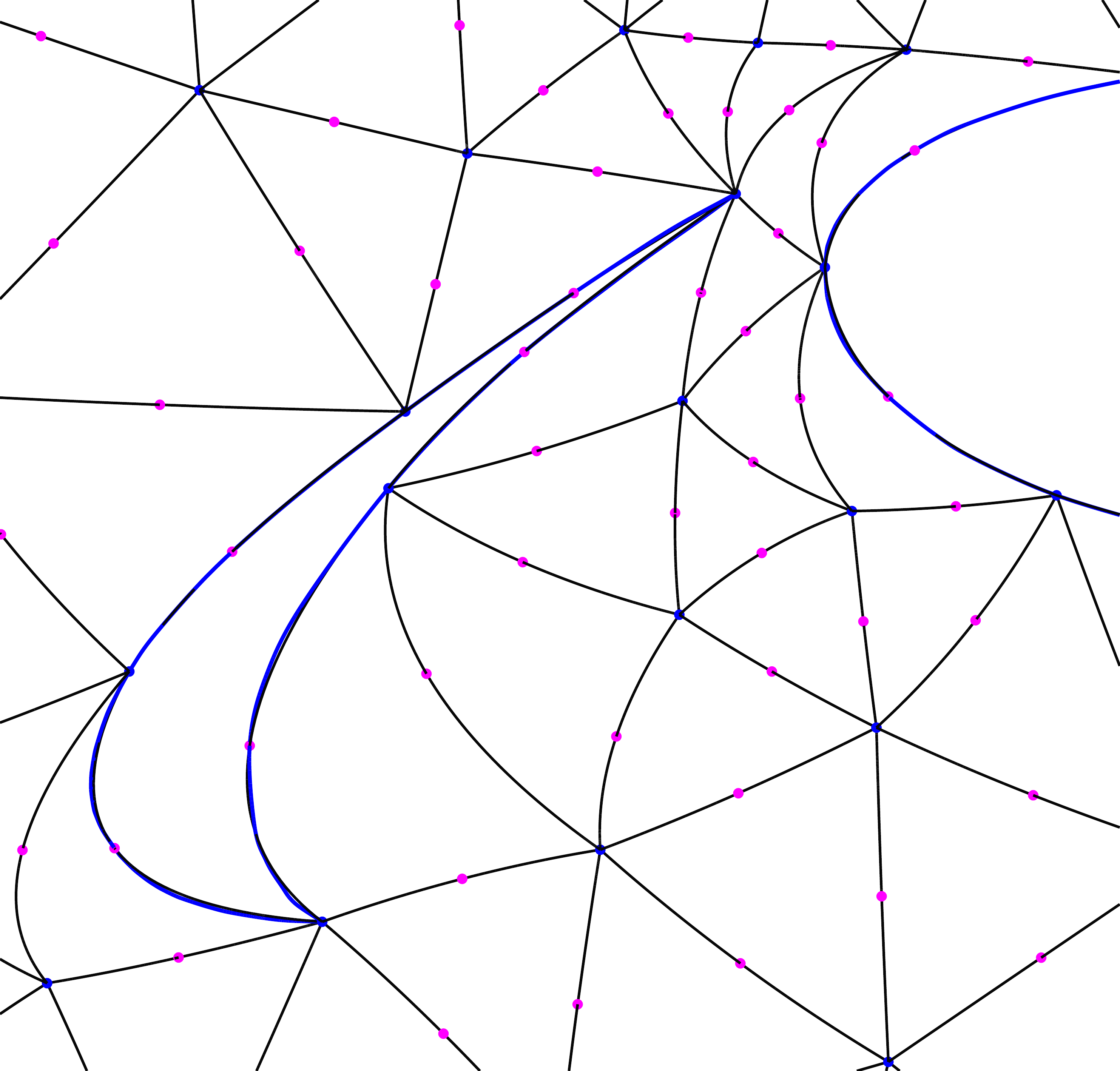}
\end{center}
\caption{Acoustic test case: details of the mesh around the slat and
the leading edge of the main component for the coarse meshes (left
column) and the medium-size meshes (right column). Linear meshes (top
row), quadratic meshes optimized for geometry only (middle row) and
quadratic meshes optimized for both validity and geometrical accuracy
(bottom row).
\label{fig:ra16sc1_comp-meshes}}
\end{figure}

Two sets of triangular meshes are generated, one composed of
medium-size meshes ($1617$ elements) and one composed of coarse
meshes ($296$ elements). Each set consists of a linear mesh,
the corresponding quadratic mesh optimized for element validity only
and the corresponding quadratic mesh optimized for both validity and
geometrical accuracy. Details of the meshes around the slat and the
leading edge of the main component, where the boundaries influence
most the acoustic field, are plotted in
Figure~\ref{fig:ra16sc1_comp-meshes}.

\begin{figure}
\begin{center}
\includegraphics[width=0.475\textwidth]{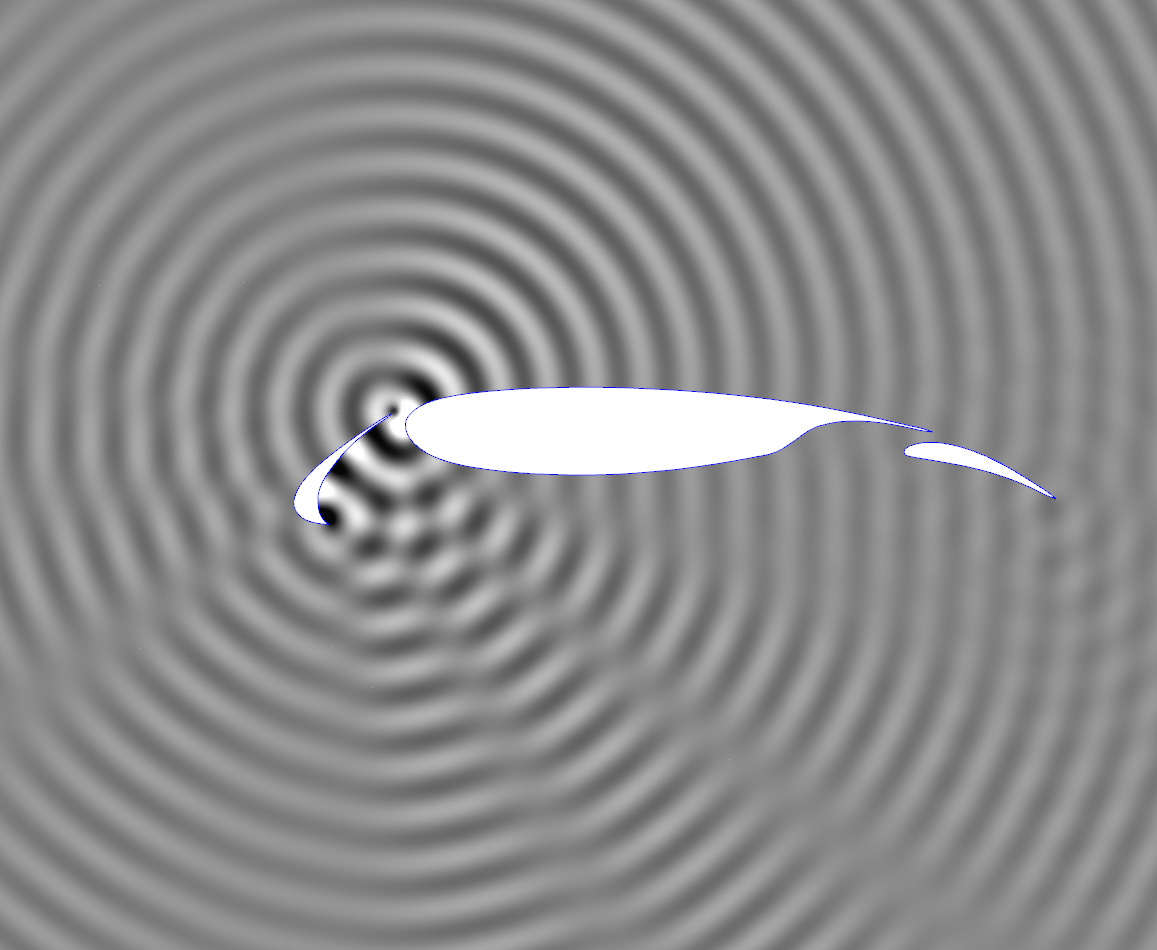}
\includegraphics[width=0.475\textwidth]{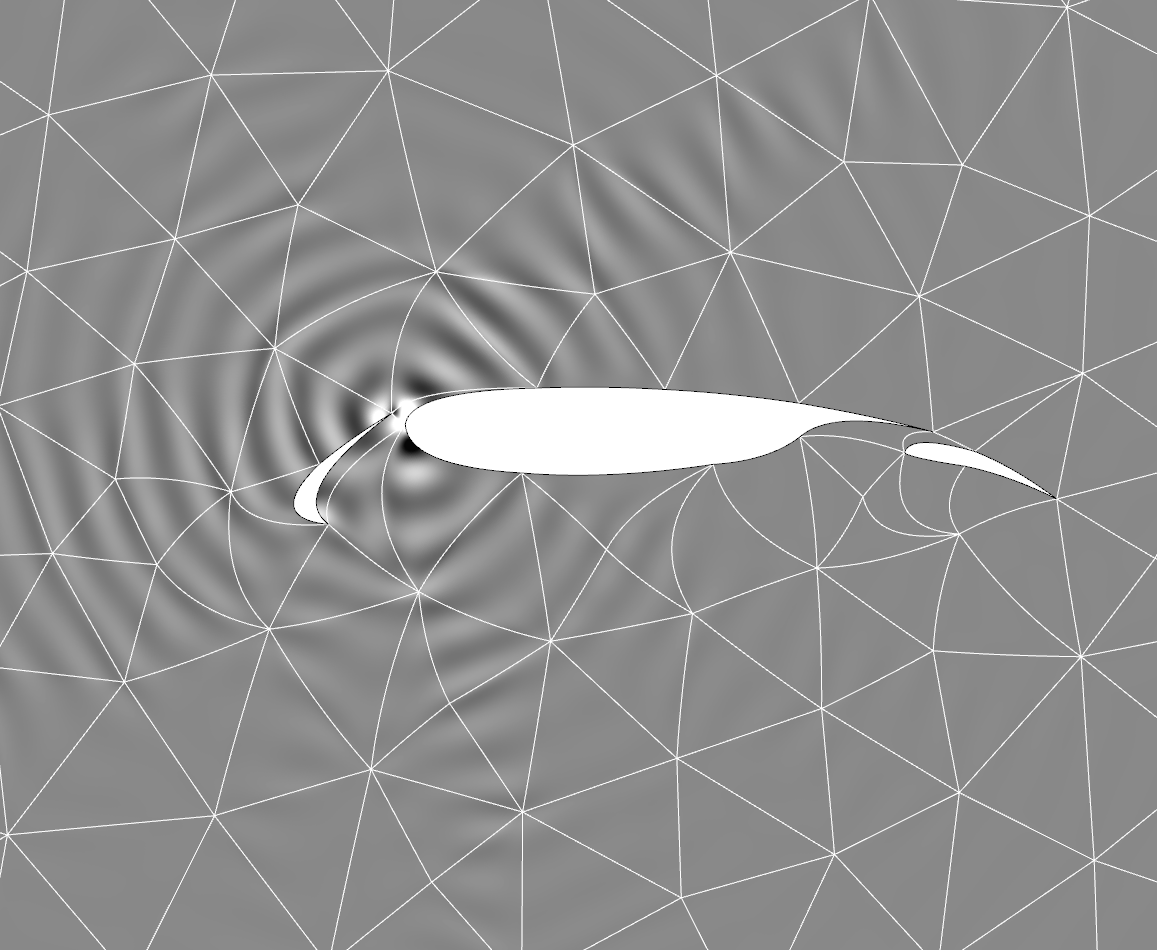}
\end{center}
\caption{Acoustics test case: acoustic pressure field at final time
for the reference solution (left) and the solution obtained on the
coarse quadratic mesh optimized for geometrical accuracy (right).
\label{fig:ra16sc1_sol}}
\end{figure}

It is obvious that the coarse quadratic mesh optimized for validity
only represents the model very poorly, while the minimization of
the geometrical error yields a fairly accurate and smooth approximation
of the airfoil. In the region of the computational domain shown in
Figure~\ref{fig:ra16sc1_comp-meshes}, the geometrical optimization
decreases the discrete model-to-mesh Hausdorff distance $\delta_H$ by
a factor 7 approximately ($\delta_H=9.3\,\textrm{mm}$ to
$\delta_H=1.3\,\textrm{mm}$), and the geometrical error $\delta_T$
drops by a factor 8 ($\delta_T=15.5\,\textrm{mm}$
to $\delta_T=1.8\,\textrm{mm}$). Above all, a close examination of
mesh optimized for validity only at the trailing edge of the slat
shows that the mesh edge on the lower side of the profile crosses the
edge representing the upper side: even though the mesh is valid in
the finite element sense, it is physically incorrect. On the
contrary, a simulation with the coarse quadratic mesh optimized for
geometrical accuracy can give an acceptable solution, provided that
the spatial discretization is of sufficiently high order. In the
present case, simulations with a 10$^\textrm{th}$-order discontinuous
Galerkin scheme were slightly more dissipative than the reference
computation on a fine mesh (see Figure~\ref{fig:ra16sc1_sol}).

\begin{figure}
\begin{center}
\includegraphics[width=1\textwidth]{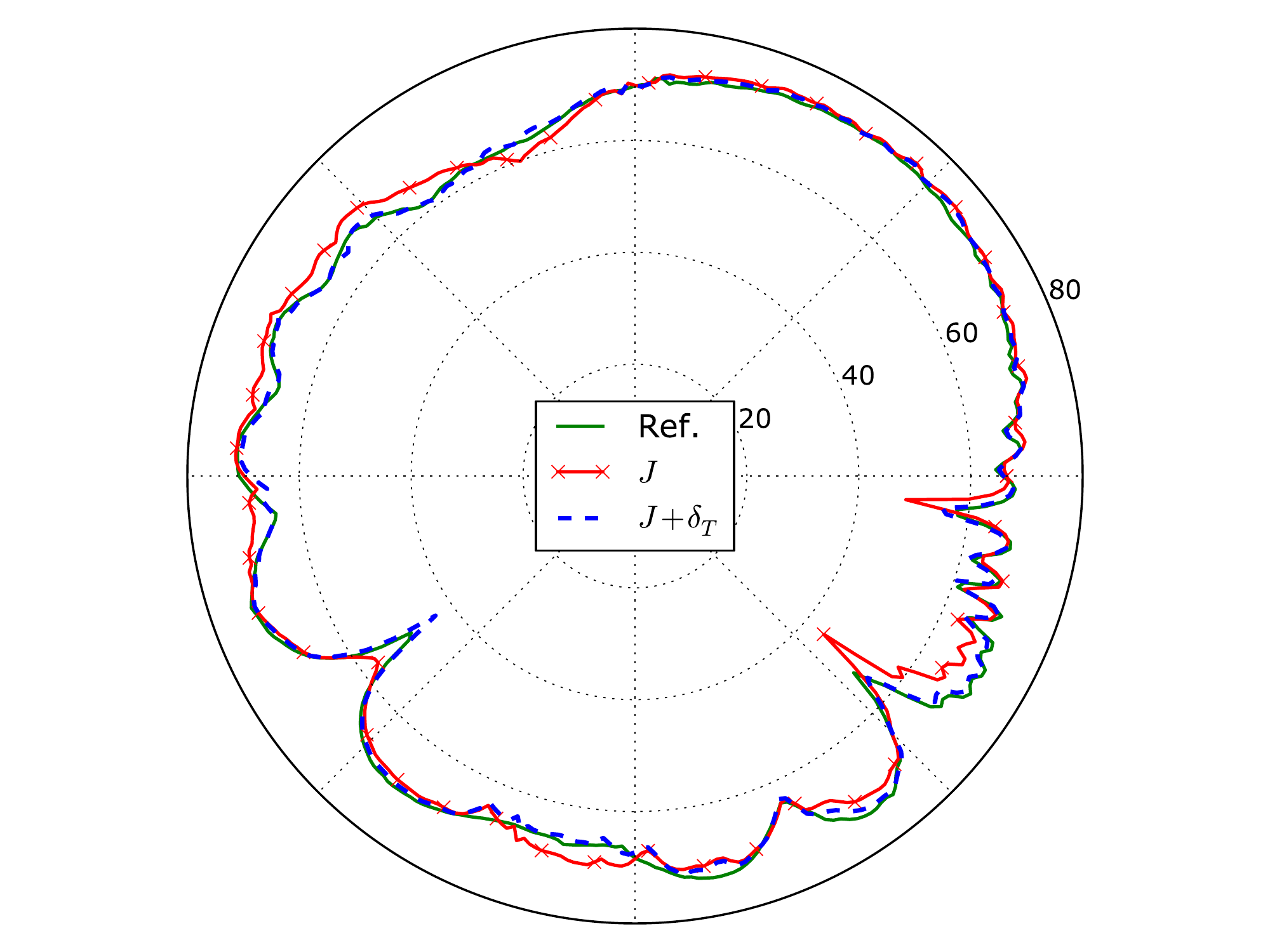}
\end{center}
\caption{Acoustics test case: sound directivity expressed in terms
of Sound Pressure Level (in dB) for the reference solution, as well as
the solutions obtained with quadratic medium-size meshes optimized for
validity only ($J$) and for both validity and geometrical accuracy
($J+\delta_T$).
\label{fig:ra16sc1_directivity}}
\end{figure}

The difference between both quadratic medium-size meshes is less
spectacular, as seen in Figure~\ref{fig:ra16sc1_comp-meshes}:
$\delta_H$ decreases by a factor 4 ($2.5\,\textrm{mm}$ to
$0.6\,\textrm{mm}$) and $\delta_T$ by a factor of 5 ($3.8\,\textrm{mm}$
to $0.8\,\textrm{mm}$) with geometrical optimization. Simulations are
run with these meshes, and the RMS acoustic pressure $p_\textrm{RMS}$
is measured over the last 6 oscillation periods along a circle of
radius $750\,\textrm{mm}$ centered at point P. The results are
expressed in terms of Sound Pressure Level as
$\textrm{SPL}=20\,\log(p_\textrm{RMS}/p_\textrm{ref})$, where
$p_\textrm{ref} = 2\cdot 10^{-5}\,\textrm{Pa}$.
Figure~\ref{fig:ra16sc1_directivity} shows that the effect of
the geometrical optimization impacts significantly the accuracy of
the sound directivity.

\section{Extension to three-dimensional meshes}\label{sec:3d}

In the same manner as described in Section~\ref{sec:distance} for
curves, it is possible to define a geometrical error for a surface
mesh element ${\mathcal S}_m$ approximating a model surface
${\mathcal S}$. At each of interpolation point $\mvx_i$ where both
surfaces coincide, a first-order estimation of the geometrical
error is:
\[
\delta^i_T = \left\| h (\mvn_m - \mvn)\right\|.
\]
where $\mvn_m$ represents the unit normal to ${\mathcal S}_m$ and
$\mvn$ represents the unit normal to ${\mathcal S}$. Here, $h$ is
proportional to the square root of a ``local surface element area'',
determined from the Jacobian of ${\mathcal S}_m$. The
geometrical error $\delta_T$ for the surface mesh element can then be
computed from all the $\delta^i_T$ in the element. It is then
possible to use the optimization process presented in
Section~\ref{sec:opti} in order to obtain a better representation of
the model surfaces by the boundary of a 3D volume mesh.

\begin{figure}
\begin{center}
\includegraphics[width=0.475\textwidth]{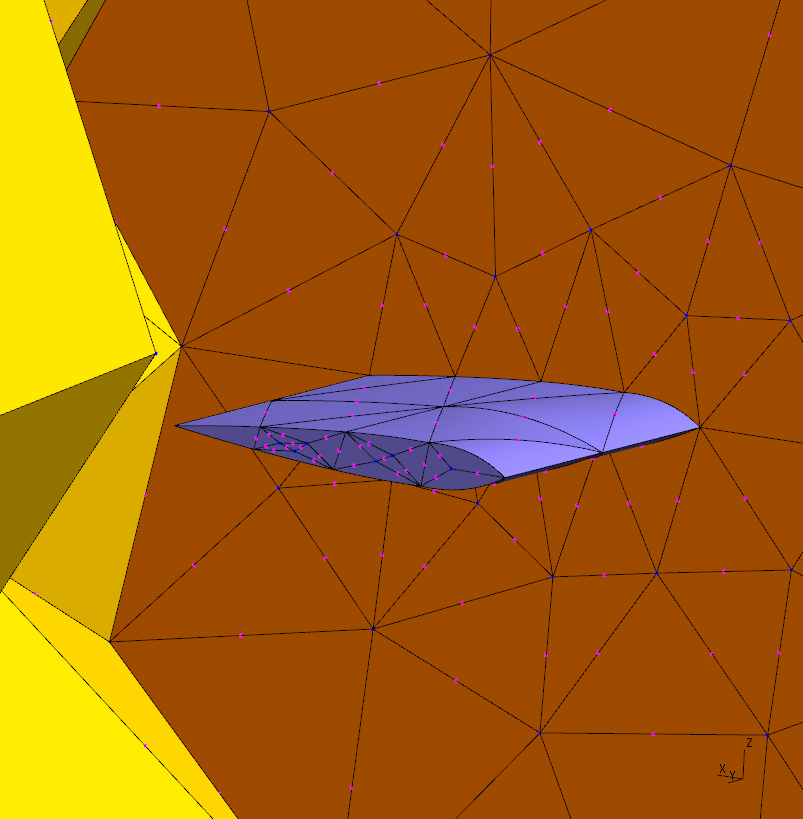}
\includegraphics[width=0.475\textwidth]{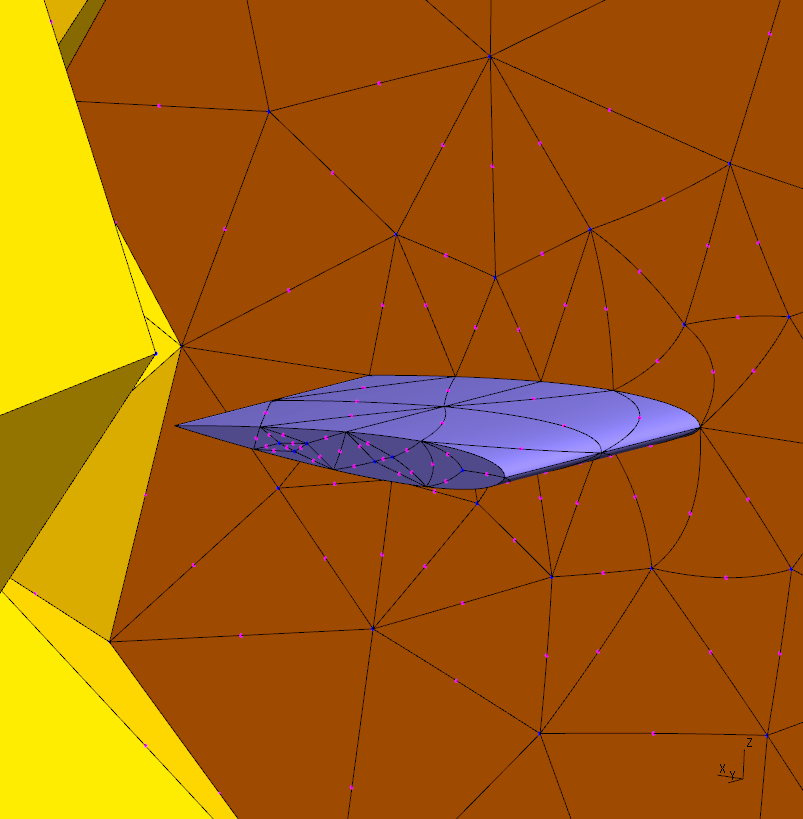}
\end{center}
\caption{Mesh of the 3D NACA0012 geometry optimized for validity only
(left) and for both validity and geometrical accuracy (right).
\label{fig:naca_3D}}
\end{figure}

In order to illustrate the potential of this approach for 3D meshes,
we apply the method to the case of a wing made from an extruded
NACA0012 profile of unit chord length. Fig.~\ref{fig:naca_3D} shows
a coarse mesh of $847$ tetrahedra in two versions, namely one optimized
for validity only and one optimized for both validity and geometrical
accuracy. As in the 2D case, the smoothness of the mesh boundary at the
leading edge is significantly improved by the minimization of $\delta_T$,
and the approximation of the airfoil seems to be more accurate. The
geometrical error $\delta_T$ is indeed decreased by more than an order
of magnitude ($0.16$ to $0.012$).

\begin{figure}
\begin{center}
\includegraphics[width=0.6\textwidth]{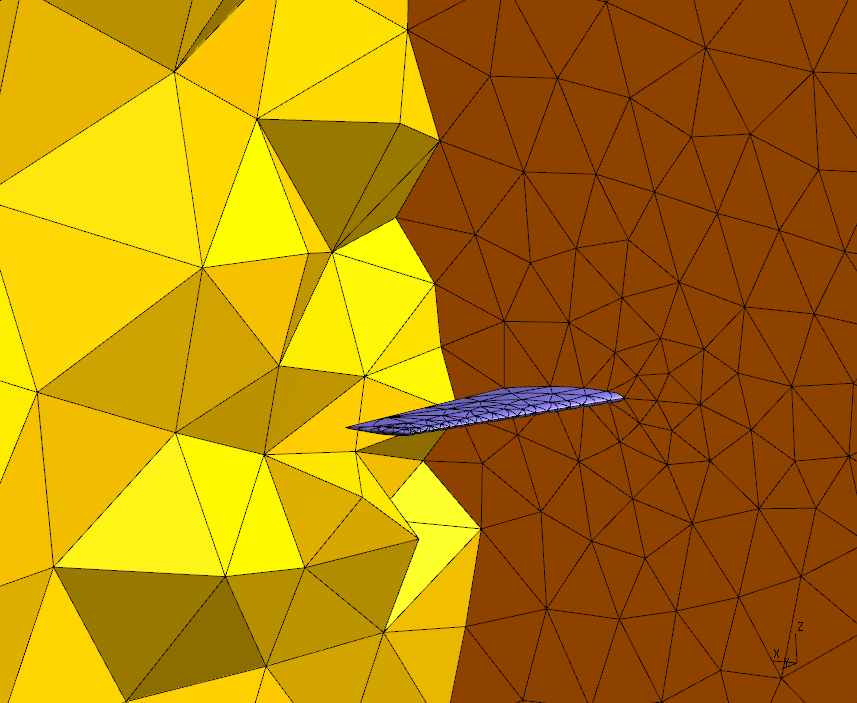}
\\[0.25\baselineskip]
\includegraphics[width=0.475\textwidth]{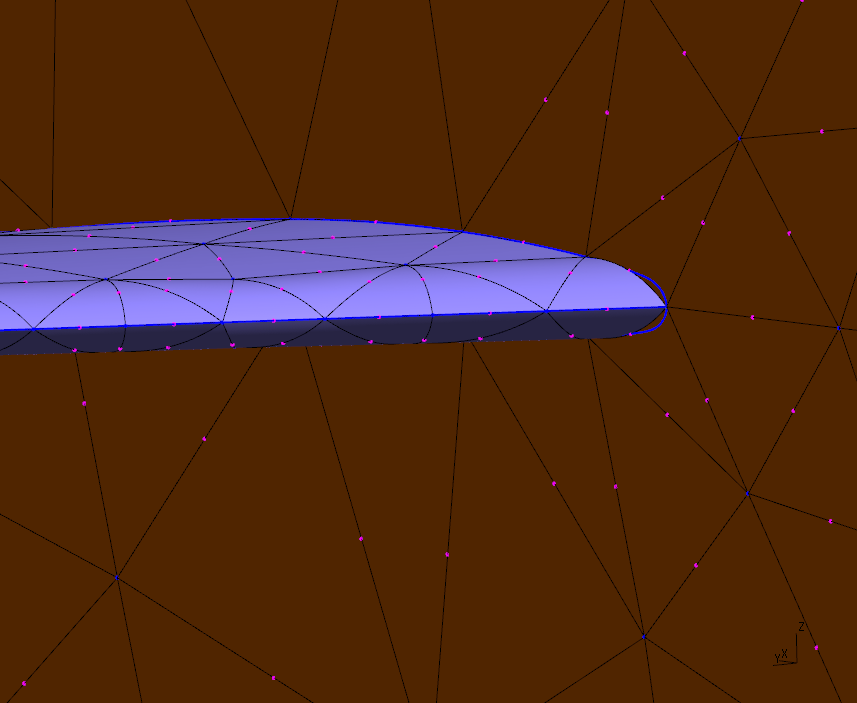}
\includegraphics[width=0.475\textwidth]{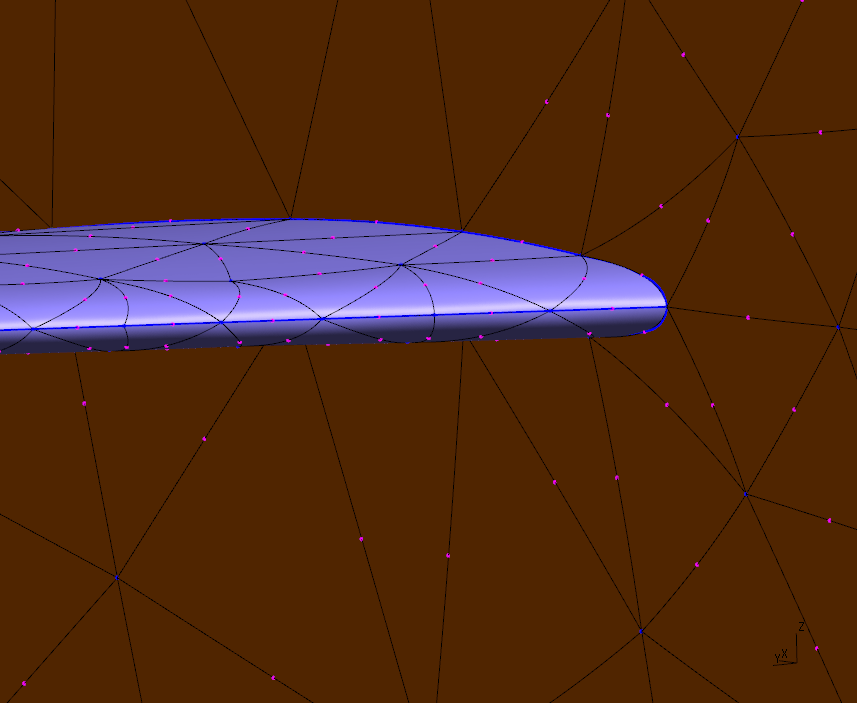}
\\[0.25\baselineskip]
\includegraphics[width=0.475\textwidth]{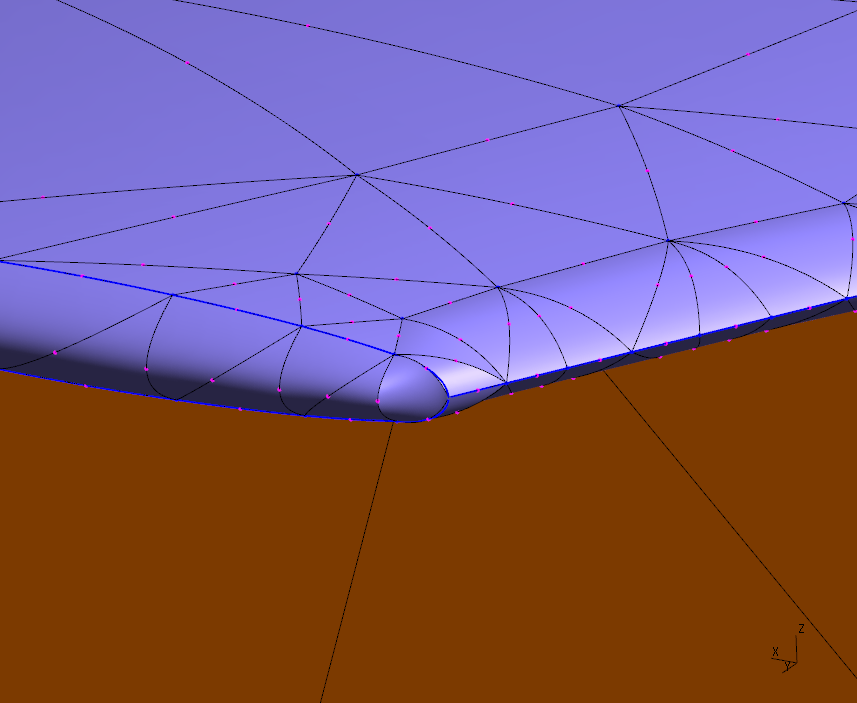}
\includegraphics[width=0.475\textwidth]{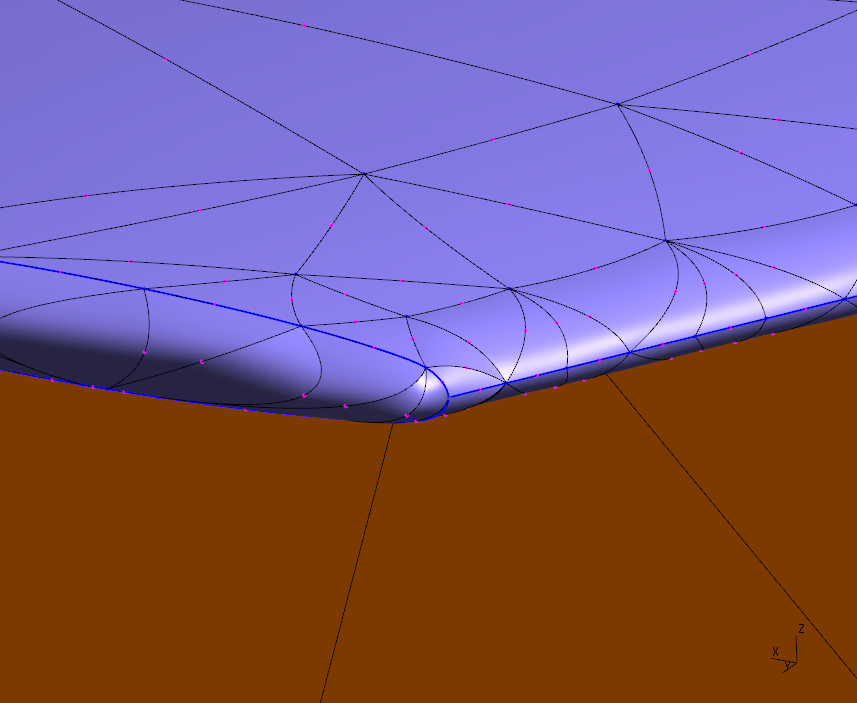}
\end{center}
\caption{ONERA M6 case: General view of the volume mesh and the wing (top),
details of the wing surface mesh optimized for validity only (center left
and bottom left) and for both validity and geometrical accuracy (center
right and bottom right).
\label{fig:oneraM6}}
\end{figure}

Another example is the case of an ONERA M6 wing of chord length of
$810$ at wing root, illustrated in Figure~\ref{fig:oneraM6}. A coarse
quadratic volume mesh of $11851$ terahedra is generated, then optimized
for validity only on one hand, and for both validity and geometrical
accuracy on the other hand. The geometrical error $\delta_T$ is reduced
from $40$ to $15$ by the geometric optimization. The representation of
the leading edge is clearly improved, particularly where it merges with
the tip surface of the wing.

\section{Conclusions}\label{sec:conclusions}

In this paper, we have presented methods to evaluate and improve the
geometrical representation of CAD models in high-order meshes.

The interest of formal distances in the plane for this purpose has been
assessed. The Fr\'echet and the Hausdorff distances between two curves
corresponding respectively to the model and to the mesh boundary have
been examined. A discrete version of these quantities, in particular
the Hausdorff distance, can be computed fast enough to assess the
quality of the geometrical model approximation in practical 2D meshes.
However, it is still computationally too costly to be employed in mesh
optimization algorithms.

To this end, a fast estimate of the geometrical error between the mesh
boundary and the CAD model is presented, which is based on a Taylor
development of each curve. It is then introduced in a pre-existing
optimization framework that guarantees the mesh validity. Several
examples show that minimizing this quantity significantly improves the
representation of the model: depending on the case, the model-to-mesh
Hausdorff distance is decreased by a factor $2$ to $35$, the gain being
larger for coarse meshes. An important aspect of the method lies in the
beneficial impact of the geometrical optimization on the mesh boundary
smoothness. As evidenced by several test cases, this effect is often
instrumental in obtaining accurate solutions from high-order
simulations. The approach is easily extended to 3D meshes, as
illustrated by two examples.

The method presented in this paper reduces the need to refine a
high-order mesh only for the purpose of representing the geometrical
model correctly. Therefore, it makes it easier to enjoy the
computational efficiency of very high order numerical schemes in
practical simulations. However, the constraints of element validity
and geometrical accuracy imposed on the mesh may lead to highly curved
elements inside the computational domain. The impact of the element
distortion on the accuracy, computational cost and robustness of the
simulation remains to be assessed and, if possible, controlled. This
topic is the subject of ongoing work.

\section*{Acknowledgements}

This work has been partly funded by the European Commission under the
FP7 grant ``IDIHOM'' (Industrialisation of High-Order Methods -- A
Top-Down Approach).

\bibliographystyle{abbrv}
\bibliography{biblio}

\begin{thebibliography}{10}

\bibitem{abgrall2012}
R.~Abgrall, C.~Dobrzynski, and A.~Froehly.
\newblock {A method for computing curved {2D} and {3D} meshes via the linear
  elasticity analogy: preliminary results}.
\newblock Rapport de recherche RR-8061, INRIA, Sept. 2012.

\bibitem{alt1995computing}
H.~Alt and M.~Godau.
\newblock Computing the {F}r\'echet distance between two polygonal curves.
\newblock {\em International Journal of Computational Geometry \&
  Applications}, 5(01n02):75--91, 1995.

\bibitem{arya1998optimal}
S.~Arya, D.~M. Mount, N.~S. Netanyahu, R.~Silverman, and A.~Y. Wu.
\newblock An optimal algorithm for approximate nearest neighbor searching fixed
  dimensions.
\newblock {\em Journal of the ACM (JACM)}, 45(6):891--923, 1998.

\bibitem{bassi}
F.~Bassi and S.~Rebay.
\newblock High-order accurate discontinuous finite element solution of the {2D}
  {E}uler equations.
\newblock {\em J. Comput. Phys.}, 138(2):251--285, 1997.

\bibitem{bernard}
P.-E. Bernard, J.-F. Remacle, and V.~Legat.
\newblock Boundary discretization for high-order discontinuous {G}alerkin
  computations of tidal flows around shallow water islands.
\newblock {\em International Journal for Numerical Methods in Fluids},
  59(5):535--557, 2009.

\bibitem{botti2012influence}
L.~Botti.
\newblock Influence of reference-to-physical frame mappings on approximation
  properties of discontinuous piecewise polynomial spaces.
\newblock {\em Journal of Scientific Computing}, 52(3):675--703, 2012.

\bibitem{dg:book}
B.~Cockburn, G.~Karniadakis, and C.-W. Shu, editors.
\newblock {\em Discontinuous {G}alerkin Methods}, volume~11 of {\em Lecture
  Notes in Computational Science and Engineering}, Berlin, 2000. Springer.

\bibitem{dey1999}
S.~Dey, R.~M. O'Bara, and M.~S. Shephard.
\newblock Curvilinear mesh generation in {3D}.
\newblock In {\em Proceedings of the 8th International Meshing Roundtable},
  pages 407--417. John Wiley \& Sons, 1999.

\bibitem{eiter1994computing}
T.~Eiter and H.~Mannila.
\newblock Computing discrete {F}r\'echet distance.
\newblock Technical report, Technische Universitat Wien, 1994.

\bibitem{gargallo2013}
A.~Gargallo-Peir\'o, X.~Roca, J.~Peraire, and J.~Sarrate.
\newblock High-order mesh generation on {CAD} geometries.
\newblock In J.~P.~M. de~Almeida, P.~D\'iez, C.~Tiago, and N.~Par\'es, editors,
  {\em Proceedings of the VI International Conference on Adaptive Modeling and
  Simulation (ADMOS 2013)}. International Center for Numerical Methods in
  Engineering (CIMNE), Barcelona, Spain, 2013.

\bibitem{held2001vroni}
M.~Held.
\newblock Vroni: An engineering approach to the reliable and efficient
  computation of {V}oronoi diagrams of points and line segments.
\newblock {\em Computational Geometry}, 18(2):95--123, 2001.

\bibitem{bounds-imr}
A.~Johnen, J.-F. Remacle, and C.~Geuzaine.
\newblock Geometrical validity of curvilinear finite elements.
\newblock In W.~R. Quadros, editor, {\em Proceedings of the 20th International
  Meshing Roundtable}, pages 255--271. Springer Berlin Heidelberg, 2012.

\bibitem{bounds-jcp}
A.~Johnen, J.-F. Remacle, and C.~Geuzaine.
\newblock Geometrical validity of curvilinear finite elements.
\newblock {\em Journal of Computational Physics}, 233:359--372, 2013.

\bibitem{adigma}
N.~Kroll, H.~Bieler, H.~Deconinck, V.~Couaillier, H.~Van Der~Ven, and
  K.~Sorensen.
\newblock {\em {ADIGMA} -- A European Initiative on the Development of Adaptive
  Higher-Order Variational Methods for Aerospace Applications: Results of a
  Collaborative Research Project Funded by the European Union, 2006-2009}.
\newblock Notes on Numerical Fluid Mechanics and Multidisciplinary Design.
  Springer, 2010.

\bibitem{idihom}
N.~Kroll, C.~Hirsch, F.~Bassi, C.~Johnston, and K.~Hillewaert, editors.
\newblock {\em IDIHOM: Industrialization of High-Order Methods - A Top-Down
  Approach}, volume 128 of {\em Notes on Numerical Fluid Mechanics and
  Multidisciplinary Design}.
\newblock Springer International Publishing, 2015.

\bibitem{luo2004automatic}
X.~Luo, M.~Shephard, R.~O'Bara, R.~Nastasia, and M.~Beall.
\newblock Automatic p-version mesh generation for curved domains.
\newblock {\em Engineering with Computers}, 20(3):273--285, 2004.

\bibitem{moxey2015}
D.~Moxey, M.~Green, S.~Sherwin, and J.~Peir\'o.
\newblock An isoparametric approach to high-order curvilinear boundary-layer
  meshing.
\newblock {\em Computer Methods in Applied Mechanics and Engineering},
  283:636--650, 2015.

\bibitem{perssonperaire}
P.-O. Persson and J.~Peraire.
\newblock Curved mesh generation and mesh refinement using lagrangian solid
  mechanics.
\newblock In {\em Proceedings of the 47th AIAA Aerospace Sciences Meeting and
  Exhibit, Orlando (FL), USA, 5-9 January 2009}, 2009.

\bibitem{rote1991}
G.~Rote.
\newblock Computing the minimum hausdorff distance between two point sets on a
  line under translation.
\newblock {\em Information Processing Letters}, 38(3):123--127, 1991.

\bibitem{sahni2010}
O.~Sahni, X.~Luo, K.~Jansen, and M.~Shephard.
\newblock Curved boundary layer meshing for adaptive viscous flow simulations.
\newblock {\em Finite Elements in Analysis and Design}, 46(1-2):132--139, 2010.

\bibitem{sherwin2002}
S.~J. Sherwin and J.~Peir\'o.
\newblock Mesh generation in curvilinear domains using high-order elements.
\newblock {\em International Journal for Numerical Methods in Engineering},
  53(1):207--223, 2002.

\bibitem{toulorge}
T.~Toulorge and W.~Desmet.
\newblock Curved boundary treatments for the discontinuous {G}alerkin method
  applied to aeroacoustic propagation.
\newblock {\em AIAA J.}, 48(2):479--489, 2010.

\bibitem{toulorge2}
T.~Toulorge and W.~Desmet.
\newblock Spectral properties of discontinuous {G}alerkin space operators on
  curved meshes.
\newblock In J.~S. Hesthaven, E.~M. Ronquist, T.~J. Barth, M.~Griebel, D.~E.
  Keyes, R.~M. Nieminen, D.~Roose, and T.~Schlick, editors, {\em Spectral and
  High Order Methods for Partial Differential Equations}, volume~76 of {\em
  Lecture Notes in Computational Science and Engineering}, pages 495--502.
  Springer Berlin Heidelberg, 2011.

\bibitem{jcp2013}
T.~Toulorge, C.~Geuzaine, J.-F. Remacle, and J.~Lambrechts.
\newblock Robust untangling of curvilinear meshes.
\newblock {\em J. Comput. Phys.}, 254:8--26, 2013.

\bibitem{wolanski1994}
E.~Wolanski.
\newblock {\em Physical {O}ceanographic {P}rocesses of the {G}reat {B}arrier
  {R}eef}.
\newblock CRC Press, Boca Raton, Florida, 1994.

\bibitem{xie}
Z.~Q. Xie, R.~Sevilla, O.~Hassan, and K.~Morgan.
\newblock The generation of arbitrary order curved meshes for {3D} finite
  element analysis.
\newblock {\em Computational Mechanics}, 51(3):361--374, 2013.

\end{thebibliography}

\end{document}